\documentclass[twocolumn]{aastex631}

\shorttitle{Quantum machine learning for anomaly detection}
\shortauthors{Kawamuro et al.}
\graphicspath{{./}{figures/}}

\usepackage{listings}
\usepackage{amssymb}
\usepackage{braket}
\usepackage{amsmath}

\usepackage{xcolor}

\definecolor{codeblue}{rgb}{0.1, 0.1, 0.6}
\definecolor{codegreen}{rgb}{0, 0.6, 0}
\definecolor{codegray}{rgb}{0.5, 0.5, 0.5}
\definecolor{codepurple}{rgb}{0.58, 0, 0.82}
\definecolor{codebg}{rgb}{0.95, 0.95, 0.95}

\lstdefinestyle{python}{
    language=Python,
    basicstyle=\ttfamily\small,
    keywordstyle=\color{codeblue}\bfseries,
    stringstyle=\color{codepurple},
    commentstyle=\color{codegreen}\itshape,
    identifierstyle=\color{black},
    showstringspaces=false,
    numberstyle=\tiny\color{codegray},
    numbers=left,
    frame=single,
    breaklines=true,
    captionpos=b,
    xleftmargin=1em,
    framexleftmargin=0.1em,
    backgroundcolor=\color{codebg} 
}

\begin{document}

\title{Quantum Machine Learning for Identifying Transient Events in X-ray Light Curves}

\correspondingauthor{Taiki Kawamuro}
\email{kawamuro@ess.sci.osaka-u.ac.jp}

\author[0000-0002-6808-2052]{Taiki Kawamuro}
\affiliation{Department of Earth and Space Science, Osaka University, 1-1 Machikaneyama, Toyonaka 560-0043, Osaka, Japan}
\affiliation{RIKEN Cluster for Pioneering Research, 2-1 Hirosawa, Wako, Saitama, Saitama 351-0198, Japan}

\author[0000-0003-4808-893X]{Shinya Yamada}
\affiliation{Department of Physics, Rikkyo University, 3-34-1 Nishi Ikebukuro, Toshima-ku, Tokyo 171-8501, Japan}

\author[0000-0002-7025-284X]{Shigehiro Nagataki}
\affiliation{Astrophysical Big Bang Laboratory (ABBL), RIKEN Pioneering Research Institute (PRI), 2-1 Hirosawa, Wako, Saitama 351-0198, Japan}
\affiliation{RIKEN Center for Interdisciplinary Theoretical \& Mathematical Science (iTHEMS), 2-1 Hirosawa, Wako, Saitama 351-0198, Japan}
\affiliation{Astrophysical Big Bang Group (ABBG), Okinawa Institute of Science and Technology (OIST), 1919-1 Tancha, Onna-son, Kunigami-gun, Okinawa 904-0495,
7 Japan}

\author{Shunji Matsuura}
\affiliation{RIKEN Center for Interdisciplinary Theoretical \& Mathematical Science (iTHEMS), 2-1 Hirosawa, Wako, Saitama 351-0198, Japan}
\affiliation{Department of Electrical and Computer Engineering, University of British Columbia, Vancouver, BC V6T 1Z4, Canada}
\affiliation{Department of Physics, University of Guelph, ON N1G 1Y2, Canada}
\affiliation{Center for Mathematical Science and Advanced Technology, Japan Agency for Marine-Earth Science and Technology, Yokohama 236-0001, Japan}

\author[0000-0002-5809-3516]{Yusuke Sakai}
\affiliation{Department of Physics, Rikkyo University, 3-34-1 Nishi Ikebukuro, Toshima-ku, Tokyo 171-8501, Japan}

\author[0000-0002-9754-3081]{Satoshi Yamada}
\affiliation{Frontier Research Institute for Interdisciplinary Sciences, Tohoku University, Sendai 980-8578, Japan}
\affiliation{Astronomical Institute, Graduate School of Science, Tohoku University, Sendai 980-8578, Japan}
\affiliation{RIKEN Cluster for Pioneering Research, 2-1 Hirosawa, Wako, Saitama, Saitama 351-0198, Japan}

\begin{abstract}


We investigate whether a novel method of quantum machine learning (QML) can identify anomalous events in X-ray light curves as transient events and apply it to detect such events from the XMM-Newton 4XMM-DR14 catalog. 
The architecture we adopt is a quantum version of the long-short term memory (LSTM) where some fully connected layers are replaced with quantum circuits. The LSTM, making predictions based on preceding data, allows identification of anomalies by comparing predicted and actual time-series data.
The necessary training data are generated by simulating active galactic nucleus-like light curves as the species would be a significant population in the XMM-Newton catalog. 
Additional anomaly data used to assess trained quantum LSTM (QLSTM) models are produced by adding flares like quasi-periodic eruptions to the training data. 
Comparing various aspects of the performances of the quantum and classical LSTM models, we find that QLSTM models incorporating quantum superposition and entanglement  slightly outperform the classical LSTM (CLSTM) model in expressive power, accuracy, and true-positive rate. 
The highest-performance QLSTM model is then used to identify transient events in  4XMM-DR14. Out of 40154 light curves in the 0.2--12 keV band, we detect 113 light curves with anomalies, or transient event candidates.
This number is $\approx$ 1.3 times that of anomalies detectable with the CLSTM model.  
By utilizing SIMBAD and four wide-field survey catalogs made by ROSAT, SkyMapper, Pan-STARRS, and WISE, no possible counterparts are found for 12 detected anomalies.

\end{abstract}

\keywords{Surveys(1671) --- Astronomy data analysis(1858) --- X-ray astronomy(1810)}

\section{Introduction}\label{sec:intro}

To date, numerous wide-field surveys have been conducted at various wavelengths \citep[e.g.,][]{Becker1995ApJ...450..559B,Planck2014A&A...571A..28P,Moshir1992ifss.book.....M,Wri10,Abazajian2003AJ....126.2081A,Bianchi2014AdSpR..53..900B,Voges1999A&A...349..389V,Kawamuro2018ApJS..238...32K,Tueller2008ApJ...681..113T}, 
and forthcoming surveys covering wider fields, achieving higher sensitivities, and/or being executed at higher cadences  
will usher in a new era for time-domain astronomy. 
For example, notable surveys are expected to be performed by the Square Kilometer Array (SKA), 
Large Synoptic Survey Telescope (LSST), 
Nancy Grace Roman Space Telescope, UltraViolet EXplorer (UVEX), Ultraviolet Transient Astronomy Satellite (ULTRASAT), NewAthena, Hyper-Kamiokande, Cosmic Explorer, and Einstein telescope  \cite[e.g.,][]{Fender2015aska.confE..51F,Kulkarni2021arXiv211115608K,Haiman2023arXiv230614990H,Shvartzvald2024ApJ...964...74S,HK2018arXiv180504163H,Maggiore2020JCAP...03..050M,Nandra2013arXiv1306.2307N,Ivezic2019ApJ...873..111I}. 
Therefore, a sizable amount of time-series data will become available in the new era, and aid in improving our understanding of astronomical phenomena, particularly from the perspective of time variability. 
In addition, these data will facilitate the identification of various transient events such as stellar flares \citep[e.g.,][]{Davenport2016ApJ...829...23D}, 
outbursts from binary systems \citep[e.g.,][]{Yan2015ApJ...805...87Y,Lin2019ApJ...870..126L}, fast blue optical transients \citep[e.g.,][]{Drout2014ApJ...794...23D,Ho2023ApJ...949..120H}, gamma-ray bursts \citep[e.g.,][]{Lien2016ApJ...829....7L,vonKienlin2020ApJ...893...46V,Poolakkil2021ApJ...913...60P}, changing-look active galactic nuclei (AGNs; e.g., \citealt{Panda2024ApJS..272...13P}), and tidal disruption events (TDEs; e.g., \citealt{Hammerstein2023ApJ...942....9H}; \citealt{Kaw16a}).  
Similarly, there would be new types of transient objects, awaiting discovery. These include those expected but little unexplored, such as electromagnetic counterparts to multi-messenger objects (e.g., \citealt{Metzger2017LRR....20....3M}).

In the new era, due to the anticipated large data size, the development of 
automatic methods for identifying variable sources and new transients has become one of the most important research topics (e.g., \citealt{Muthukrishna2022MNRAS.517..393M}). 
Machine learning (ML) is recognized as an efficient approach; indeed, the ML has extracted features from big data. 
A prevalent ML application in astronomical time-series data is 
the classification of light curves and color evolution into different groups (e.g., \citealt{Tranin2022A&A...657A.138T}; \citealt{Kovacevic2022A&A...659A..66K}; \citealt{Perez-Diaz2024MNRAS.528.4852P}). 
Additionally, ML is occasionally used to determine whether a new transient candidate is genuine or bogus without extensive human examination  (e.g., \citealt{Duev2019MNRAS.489.3582D}; \citealt{Makhlouf2022A&A...664A..81M}). 
Furthermore, 
ML plays a crucial role in identifying new classes of transients as outliers (e.g., \citealt{Nun2014ApJ...793...23N}; \citealt{Villar2021ApJS..255...24V}; \citealt{Sanchez-Saez2021AJ....162..206S}). 
Typically, this approach involves deriving characteristics from observed data (e.g., amplitude, skewness, and duration) and constructing feature spaces to distinguish between normal and anomalous behaviors. 

\begin{figure}
 \begin{center}
  \includegraphics[width=4cm]{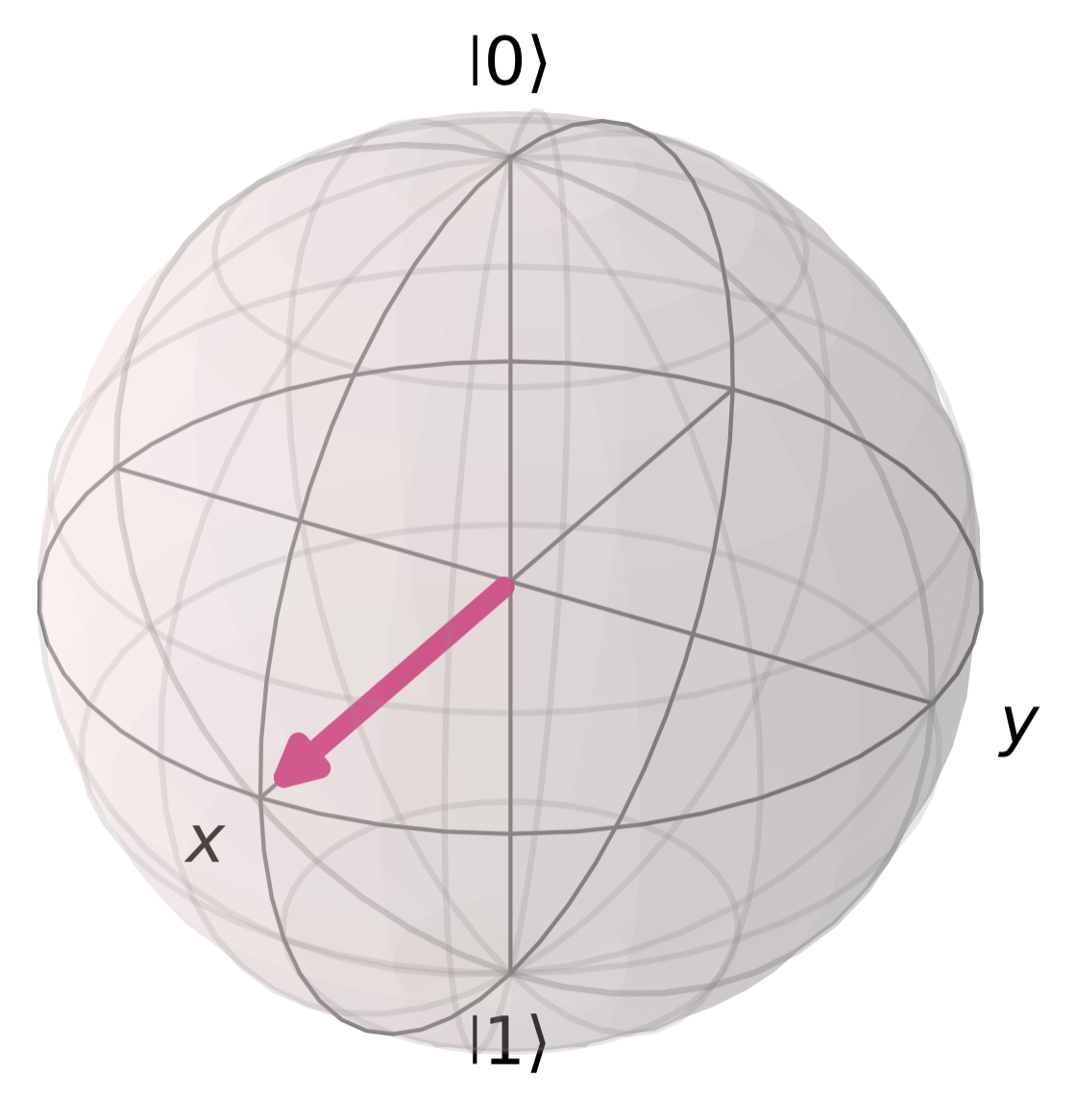} 
 \end{center}
 \caption{ 
    Example of the state of a qubit, which is the fundamental unit in quantum computing, on a complex sphere or the Bloch sphere. 
    The magenta arrow indicates the state $|\Phi\rangle = 1/\sqrt{2} |0\rangle + 1/\sqrt{2} |1\rangle$. The initial state of $|0\rangle$ corresponds to the arrow pointing straight upward. 
 }
 \label{fig:bloch_h} 
\end{figure}

While ML has solidified its position in astronomy by performing various tasks (e.g., \citealt{Aniyan2017ApJS..230...20A}; \citealt{Tanaka2022PASJ...74....1T}; \citealt{Shimakawa2022PASJ...74..612S}; \citealt{Parker2022MNRAS.514.4061P}), 
future advancements in ML involve leveraging quantum computers
(e.g., \citealt{Schuld2015ConPh..56..172S}; \citealt{PeralGarcia2022arXiv220104093P}; \citealt{Corli2024arXiv240811047C}). The ML combined with quantum computers is generally referred to as quantum ML (QML). 
As detailed in Appendix~\ref{app:quantum}, quantum computing uses 
a fundamental unit called ``qubit". 
The qubit is a two-level system, and various quantum systems are utilized to implement it\footnote{For instance, superconducting qubits, developed by  companies, including IBM, are based on superconducting resonant circuits. 
These circuits can quantize energy under specific conditions (i.e., high charges and low temperatures).
Although the energy levels are equal in the conditions, the introduction of  a Josephson junction into the circuit 
creates unequal energy levels and isolate the two energy states, 
crucial to prevent identical photon inputs from causing energy transitions between multiple neighboring states.
Typically, the ground state and the next higher state are used to implement a qubit.}. 
A distinguishing feature of the qubit is its ability to exist in a superposition of two states, represented as $|\Phi\rangle = \alpha |0\rangle + \beta|1\rangle$, where $\alpha$ and $\beta$ are complex numbers. Unlike a  classical bit, which must be in one of two states at a time, a qubit can simultaneously exist in both. 
Given the condition of $|\alpha|^2 + |\beta|^2 = 1$, the state of a qubit can be visualized as a vector on the complex sphere (Figure~\ref{fig:bloch_h}), or the Bloch sphere in the context of quantum computing (Appendix~\ref{app:quantum} describes more details). 
This expression simplifies the understanding of the state of a qubit at any time and is widely used. 
The advantage of quantum superposition is that the number of possible states represented by qubits increases exponentially with the number of qubits; for instance, two qubits can represent four states ($2\times2$), and $n$ qubits can represent $2^n$ states.
In addition to superposition, two notable features are entanglement and interference.
Entanglement allows different qubits to be correlated, and thus the state of one qubit determines another. 
Interference means that the probability of one state affects another, even within a single qubit. 
These quantum features are believed to accelerate computation and represent complex feature spaces (e.g., \citealt{Havlicek2019Natur.567..209H}; \citealt{Huang2021NatCo..12.2631H}).
For example, \citet{Abbas2020arXiv201100027A} showed that quantum neural networks (NNs) can express data more effectively than classical networks. 
However, computations in currently available noisy intermediate-scale quantum (NISQ) computers are considerably affected by quantum noise (see Appendix~\ref{app:quantum_current}). Additionally,
the superiority of QML over classical ML in solving real-world problems remains unclear and is being actively explored. 

In this study, 
we evaluate the feasibility of using a novel QML method to identify anomalous events in X-ray light curves as transient events 
and detect actual transients from a large set of light curves in the 
XMM-Newton Serendipitous Source Catalog, Fourteenth Data Release (4XMM-DR14; \citealt{Webb2020A&A...641A.136W}; \citealt{Traulsen2020A&A...641A.137T}). 
To the best of our knowledge, this is the first attempt to use the 
QML method for transient event detection. 
XMM-Newton is a space telescope with three detectors retaining large effective areas in the soft X-ray band (0.2--12\,keV) and large 30-arcmin diameter field-of-views \citep{Jan01}. 
Thus, the telescope has efficiently accumulated X-ray data, making the 4XMM catalog one of the largest X-ray-band database.

Our QML models are constructed by following
a quantum version of the long short-term memory (LSTM) architecture (\citealt{hochreiter1997long}) introduced by \citet{Chen2020arXiv200901783Y}, and are trained and assessed with normal and anomaly datasets. 
The LSTM, a type of recurrent neural network (RNN), can maintain long-term information by updating dedicated memory, unlike the most basic RNN, while 
determining short-term memory to be forwarded. 
This capability allows it to learn sequential data effectively and make predictions from preceding data, useful in tasks like stock price forecasting, speech and text interpretation, and translation. 
Consequently, the LSTM can detect transient events by identifying significant deviations from its predictions. 
Among the two necessary datasets, the normal one, mainly used for training LSTM models, is prepared by generating AGN-like light curves, as 
AGNs are a significant population in the XMM-Newton catalog (e.g., \citealt{Lin2012ApJ...756...27L}). 
For the anomaly dataset, we produce light curves by adding 
flares like quasi-periodic eruptions (QPEs) to AGN-like light curves. 
We focus on QPEs owing to their recent identification and significant interest (e.g., \citealt{Arcodia2021Natur.592..704A}; \citealt{Bao2022MNRAS.509.3504B}; \citealt{Webbe2023RASTI...2..238W}). 

The paper is organized as follows. 
Section~\ref{sec:lstm} details the construction of QLSTM models and also their classical counterpart, i.e., classical LSTM (CLSTM), for performance comparison. 
Section~\ref{sec:data} describes our normal  (AGN-like light curves) and anomalous datasets (AGN-like light curves with QPE-like flares) and 
explains the real X-ray light curves from the 4XMM-DR14 catalog used to 
detect transient events. 
Section~\ref{sec:results} presents 
the training results and performance assessment of the LSTM models. 
Section~\ref{sec:app2xmm} discusses 
the application of a QLSTM model to the XMM-Newton light curves.
Finally, Section~\ref{sec:summary} summarizes our results, and 
Section~\ref{sec:prospect} describes prospects. 

\begin{figure*}
 \begin{center}
  \includegraphics[width=8cm]{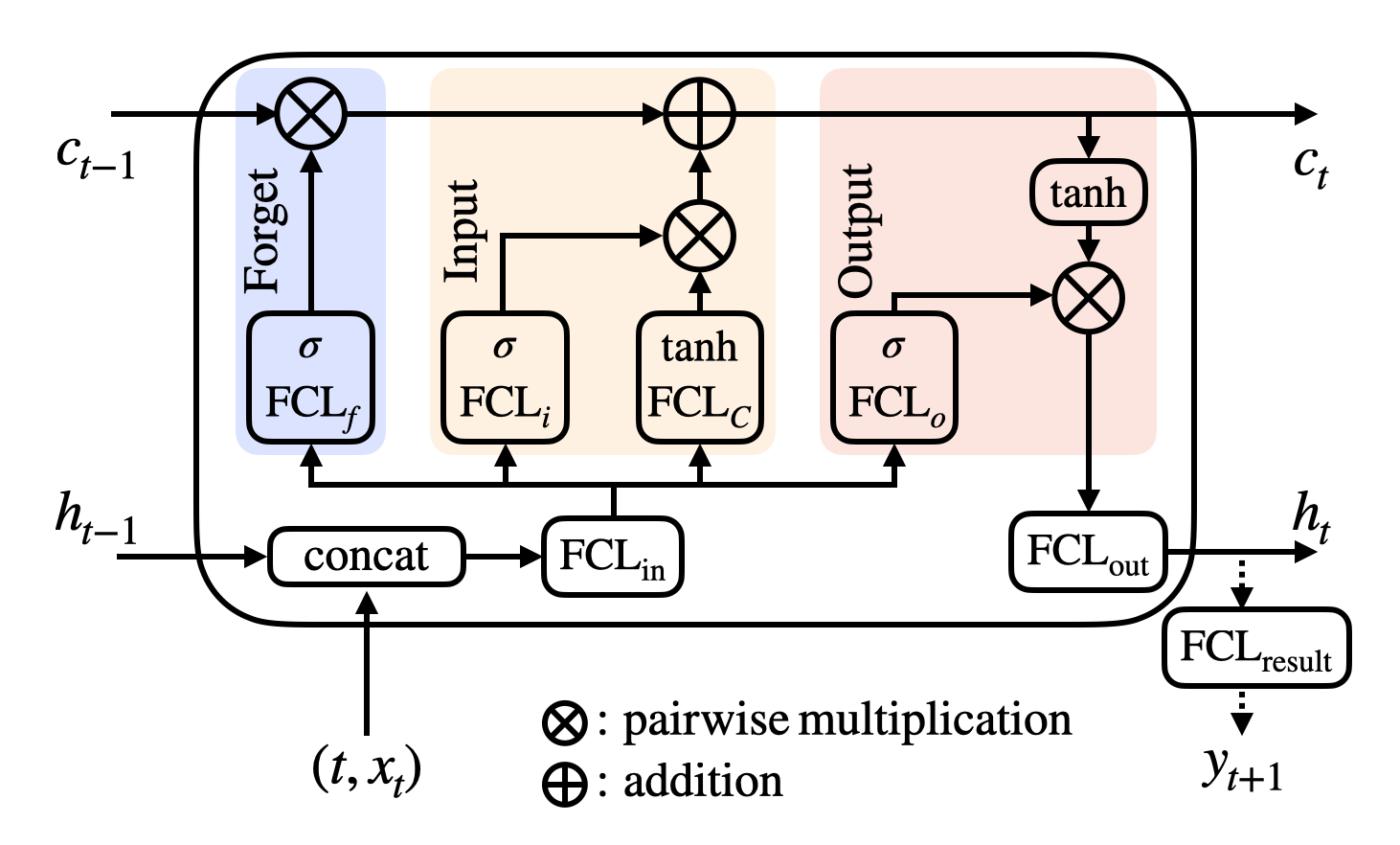} 
  \includegraphics[width=8cm]{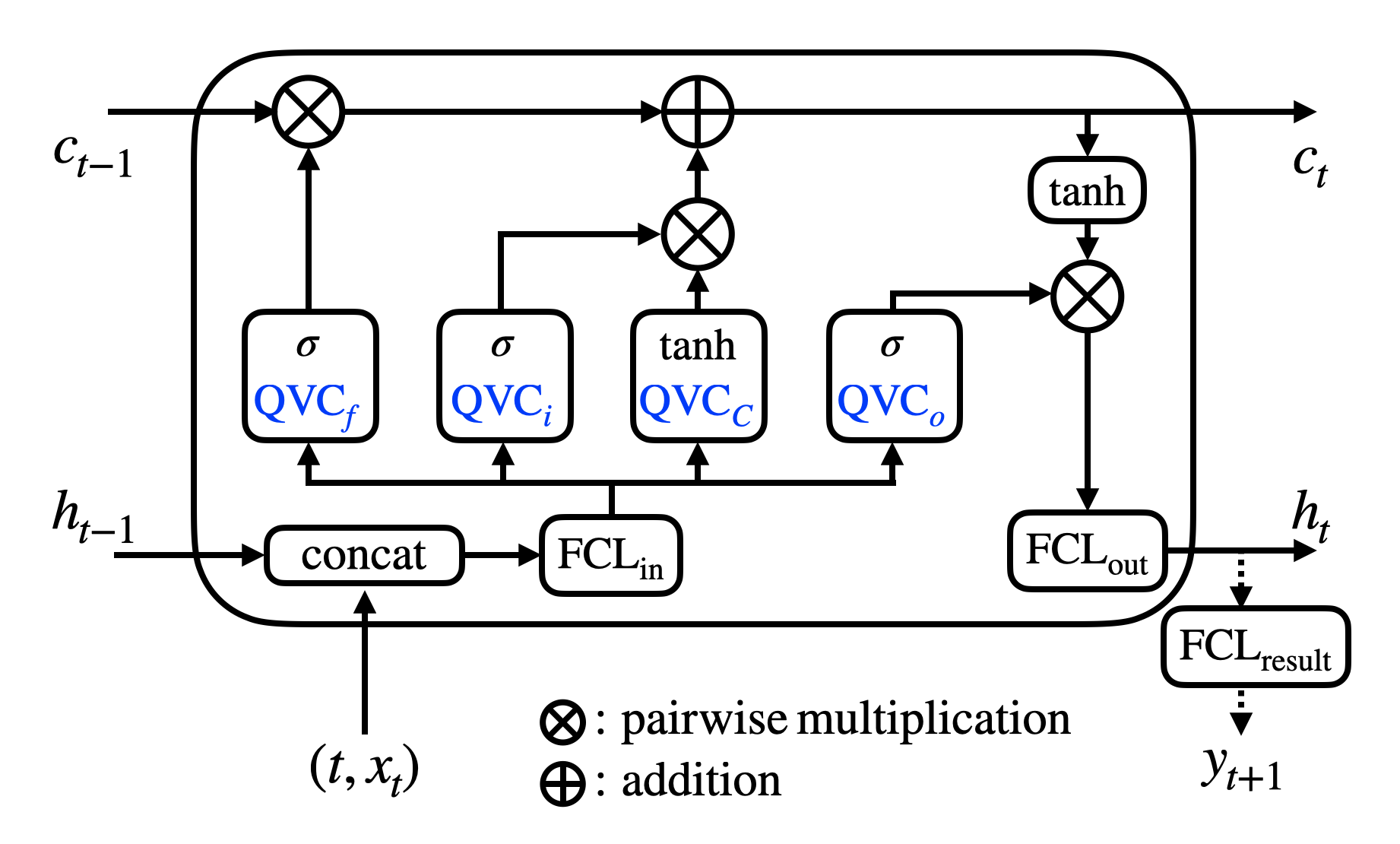}   
 \end{center}
 \caption{ 
    Left: Schematic picture of a single CLSTM layer at a time step of $t$. 
    The layer takes four inputs of $c_{t-1}$, $h_{t-1}$, $t$, and $x_t$, which are
    the cell state (long-term memory), hidden state (shot-term memory), time, and a vector of input values at the time, respectively. The FCL denotes a fully connected layer. 
    Right: the schematic picture of a single QLSTM layer at a time step of $t$.
    The FCLs of the gates in the CLSTM layer are replaced with QVCs.  
 }
 \label{fig:lstm_layer} 
\end{figure*}

\section{Construction of CLSTM and QLSTM Models}\label{sec:lstm} 

\subsection{CLSTM model}\label{sec:clstm}  

The left panel of Figure~\ref{fig:lstm_layer} presents a schematic picture of a CLSTM model. Unlike a basic RNN, which has only a hidden state ($h_t$, a vector carrying information) updated at each step to predict values, the LSTM holds two states: the hidden state and cell state ($c_t$). 
The cell state is responsible for retaining long-term memory, enabling the LSTM to track long-term trends and make more accurate predictions compared to an RNN.
The LSTM model comprises three gates: 
forget, input, and output gates (the left panel of Figure~\ref{fig:lstm_layer}). 
The forget gate reduces information from the 
cell state (long-term memory) based on concatenated input data ([$x_t$, $h_{t-1}$], with $x_t$ representing observed data). 
The input gate has two components; 
one uses a \texttt{sigmoid} function to determine the degree of information to be included, and the other employs a hyperbolic tangent function (\texttt{tanh}) to determine the actual information to add. 
The cell state is updated by the results of the forget and input gates. 
The output gate then literally extracts information from the updated cell state as the output. 
The output information, or the hidden state ($h_t$), is derived by first applying the hyperbolic tangent function to the cell state and then determining the degree of the obtained result. 
Typically, the LSTM layer is repeated, sequentially updating information. 
In the final LSTM layer, 
the hidden state is used as the prediction. At this point, if necessary, a fully connected layer is occasionally used to adjust the dimension of the hidden state to what is desired for comparison with observed data.
We implement the actual code of the CLSTM using the LSTM class in \texttt{PyTorch}.

\subsection{QLSTM models}\label{sec:qlstm}  

\begin{figure}
 \begin{center}
  \includegraphics[width=8.5cm]{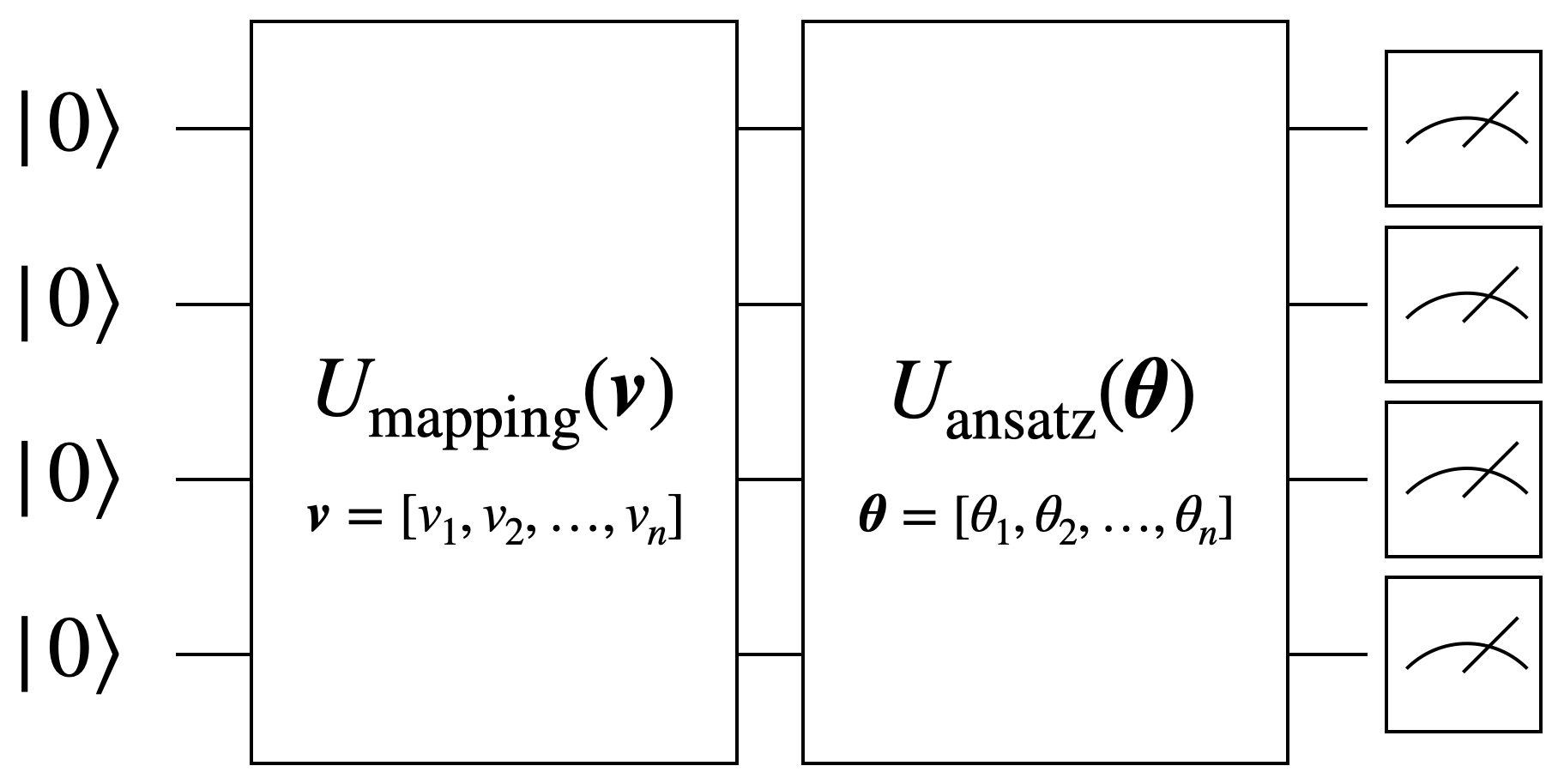}
 \end{center}
 \caption{ 
    Structure of the QVC for four qubits. The QVC consists of two layers.
    The first layer, represented by $U_{\rm mapping}$, encodes inputs ($v$) into the qubits. The second layer, called ansatz, has trainable parameters ($\theta$) to fit the input values to desired ones.
    The four squares denote measurements. 
 }
 \label{fig:circ_concept} 
\end{figure}

We construct QLSTM models, following \citet{Chen2020arXiv200901783Y}; 
while the overall CLSTM structure is maintained, each fully connected layer in the three gates is replaced with a quantum circuit (the right panel of Figure~\ref{fig:lstm_layer}), so called the quantum variational circuit (QVC). 
The conceptual structure of the QVC is shown in Figure~\ref{fig:circ_concept}. 
The QVC consists of two main layers. 
the first one is the feature-mapping layer, which encodes input data into qubits initially set to the ground state $\ket{0}$. 
The second layer, called ``ansatz", alters qubit states 
based on trainable parameters. 
After feature mapping and ansatz, qubits are measured using operators, often  employing the Pauli-Z operator to observe its eigenvalues $1$ or $-1$. 
By repeating the entire process from the feature mapping to the measurement multiple times, probabilities of obtaining eigenvalues are estimated. 
With the probabilities, expected values are also calculable. 
These derived values are then compared with the observed values to optimize trainable parameters. 
In this study, we derive the expected value of each qubit for 
the Pauli-Z operator. 

\begin{figure*}
 \begin{center}
  \includegraphics[width=17.5cm]{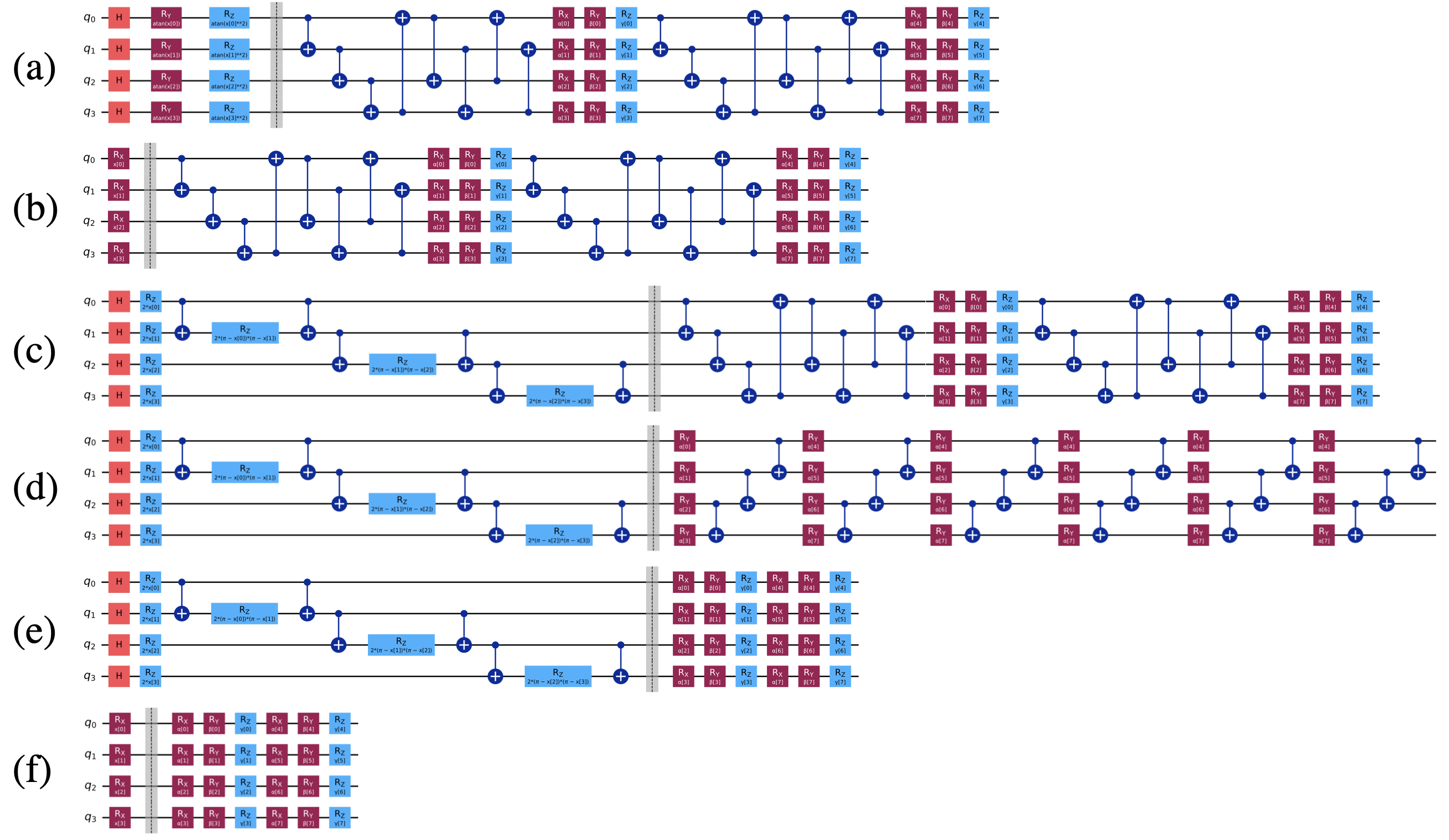}
 \end{center}
 \caption{ 
    From top to bottom, six types of QVCs are presented in the order of Table~\ref{tab:qvcs}: 
    (Feature Map, Ansatz) = 
    (C20, C20), (AngleX, C20), 
    (ZZFeatureMap, C20), 
    (ZZFeatureMap, C20),
    (ZZFeatureMap, C20NoCNOT), and 
    (AngleX, C20). 
    In each QVC, the feature-map and ansatz sections are divided by a dotted line with a gray band. 
    Light red squares labeled $H$ denote Hadamard gates, which create a superposition state. 
    Squares labeled $R_X$, $R_Y$, or $R_Z$ indicate gates rotating the relevant qubit by a specific angle around the respective axis.
    The remaining gates are CNOT gates, 
    each represented by a line connecting the control qubit (small circle) to the target qubit (cross). 
 }
 \label{fig:qvcs} 
\end{figure*}

Regarding the feature map, we consider three types. 
(1) The first one is what was adopted in a QLSTM study by \citet{Chen2020arXiv200901783Y}. 
We refer it to as C20. 
As shown in (a) of Figure~\ref{fig:qvcs}, 
the C20 feature map begins with a Hadamard gate.  
This gate effectively puts the initial state into a superposition with equal probabilities for $\ket{0}$ and $\ket{1}$ (Figure~\ref{fig:bloch_h}; see Appendix~\ref{app:quantum} for more detail). 
Then, input vector elements ($v_i$) are encoded in the relevant qubit by applying $\arctan(v_i)$ and $\arctan(v_i^2)$ and then rotating the state vector around the $Y$ and $Z$ axes by the two obtained values, respectively. 
(2) The second feature map is ZZFeatureMap, proposed by \citet{Havlicek2019Natur.567..209H}. 
The circuits (c), (d), and (e) of Figure~\ref{fig:qvcs} adopts the ZZFeatureMap. 
The map creates superposition states using the Hadmard gate and rotates  
qubit state vectors around the $Z$ axis based on input values, while occasionally creating entanglements between qubits via CNOT gates (see Appendix~\ref{app:quantum} for the detail of CNOT).  
This map is noted for its ability to achieve mappings difficult to reproduce classically, utilizing quantum properties. 
(3) We lastly use a very simple feature map which we define as AngleX. 
In Figure~\ref{fig:qvcs}, (e) adopts it.
The feature map encodes input data only by rotating qubit state vectors around the $X$-axes according to input values, without entanglement. 
This mapping is used to evaluate the impact of the quantum entanglement by comparing results from AngleX with those from the other maps.

Like the feature maps, we consider three types of ansatzes. 
(1) The first ansatz, as used in \citet{Chen2020arXiv200901783Y},  
entangles each qubit with an adjacent one using the CNOT gate and rotate  all qubits around the $X$, $Y$, and $Z$ axes (see (a), (b), and (c) of Figure~\ref{fig:qvcs}). 
(2) The second ansatz is RealAmplitudes, which has been heuristically applied in QML for chemistry and classification tasks. 
It rotates the state vector of each qubit around the $Y$-axis axis, and uses a CNOT gate for entanglement with an adjacent qubit (see (d) in Figure~\ref{fig:qvcs}). 
The term RealAmplitudes indicates that the complex component is always zero. 
We use RealAmplitudes to assess how largely changes in ansatz affect results while preserving quantum characteristics like CNOT gate entanglements. 
(3) The final ansatz we adopt is a modified version of the C20 ansatz, labeled C20NoCNOT, wherein CNOT gates are removed. 
In the ansatz, no entanglement exists. 
This version allows us to evaluate the significance of quantum entanglement by comparing its results with those from the other ansatzes, which create entanglement.

Based on the three feature maps and the three ansatzes, we construct six QVCs (Table~\ref{tab:qvcs}). 
We consider the QVC presented by \citet{Chen2020arXiv200901783Y} as the foundation, and adopt the combinations of the C20 ansatz with the other two feature maps as well. 
Then, to discuss the effectiveness of ZZFeatureMap, we additionally adopt the combinations of ZZFeatureMap with the two remaining ansatzes of  RealAmplitudes and C20NoCNOT. 
Finally, to discuss the impact of the absence of quantum entanglement, we take the QVC that consists of AngleX and C20NoCNOT into consideration.

The actual code to train the QLSTM models are created using the Python-compatible libraries Pennylane and Qiskit, developed by Xanadu and IBM, respectively.
Pennylane facilitates rapid quantum circuit simulation, whereas we use Qiskit to execute realistic NISQ-era simulations affected by quantum noise. 
Additionally, \texttt{PyTorch} is employed to implement fully connected layers, necessary to adjust the dimensions of input and output data, and optimize trainable parameters. 
An example of quantum circuit simulation code is provided in  Appendix~\ref{app:code}, especially for those new to quantum computing.

\begin{deluxetable}{ccccc}
\tablecaption{QVC setups\label{tab:qvcs}}
\tablewidth{0pt}
\tablehead{
\colhead{ID} & \colhead{Feature map} & \colhead{Ansatz} & \colhead{\# of rep.} & \colhead{\# of pars.} 
}
\decimalcolnumbers
\startdata
      (a) & C20 & C20 & 2 & 129 \\
      (b) & AngleX & C20 & 2 &  129 \\
      (c) & ZZFeatureMap & C20 & 2 & 129 \\
      (d) & ZZFeatureMap & RealAmp. & 6 & 129 \\
      (e) & ZZFeatureMap & C20NoCNOT & 2 & 129 \\
      (f) & AngleX & C20NoCNOT & 2 & 129 \\      
\enddata
\tablecomments{
    (1) Identification. Figure~\ref{fig:qvcs} uses the identifications. 
    (2) Name of the feature map that encodes the input data into qubits. 
    (3) Name of an ansatz.  
    (4) Number of ansatz repetitions in QVC. 
    (5) The total number of trainable parameters, including the parameters in a single LSTM layer and those in fully connected layers, necessary to adjust the input and output dimensions to appropriate ones. 
}
\end{deluxetable}

\section{Time-series Data}\label{sec:data} 

\subsection{Simulated Training and Validation Data}\label{sec:train_test_data}


As detailed in the following paragraphs, we simulate AGN-like light curves using the method proposed by \citet{Timmer1995A&A...300..707T}, and define these as our normal data, part of which are used for training. Anomalous data are then created by superimposing QPE-like flares onto these AGN-like light curves, based on parameters informed by observational studies.
This approach allows us to construct a sufficiently large and diverse dataset for both training and validation, encompassing a broad range of realistic AGN and QPE variability behaviors. 
In particular, given the scarcity of confirmed QPE events in current X-ray observations \citep[e.g.,][]{Webbe2023RASTI...2..238W}, the simulation is crucial. 
Moreover, the simulation offers the key advantage of full control over the characteristics of normal and anomalous data, allowing for a systematic evaluation of the detection capabilities of models.

Regarding the normal data, we assume appropriate power spectrum densities (PSDs) for the AGN and then perform an inverse Fourier transformation \citep{Timmer1995A&A...300..707T}. 
The PSD shape is set as a power-law function. 
PSD slopes are sampled from a Gaussian distribution with a mean of $-2$ and a standard deviation of $0.3$. This Gaussian distribution is adopted based on an XMM-Newton study of nearby AGNs ($z < 0.4$) by \citet{Gonzalez-Martin2012A&A...544A..80G}. 
Their AGN sample was created by selecting AGNs whose XMM-Newton data were publicly available as of February 2012. Thus, this may represent AGNs with good quality data in the 4XMM-DR14 catalog, but as an additional effort, 
we also investigated the representative possible values of the slope by including more recent data. 
We create our sample by referring to \citet{Tortosa2023MNRAS.526.1687T}. The authors constructed a sample of 151 un-obscured AGNs which were detected in the Swift/BAT 70-month catalog and had publicly available XMM-Newton data as of December 2022. 
The BAT selection enables the choice of bright objects suitable for constructing PSDs. 
For simplicity, in cases where multiple observations are available for a single object, we 
analyze only the data with the longest exposure time. 
By following the methodology of \citet{Gonzalez-Martin2012A&A...544A..80G} (e.g., limiting to AGNs having data with exposure times longer than 40 ks and using maximum likelihood methods for PSD fitting) and checking the quality of the PSD fits, the PSD slopes of 30 AGNs are constrained. 
Fitting the histogram of the 30 slopes with a Gaussian distribution yields a mean of $-2$ with a variance of 0.2, confirming consistency with the results of \citet{Gonzalez-Martin2012A&A...544A..80G}.
The normalization of each PSD, or variability strength, is then determined based on an excess variance. For the variance, we again refer to \citet{Tortosa2023MNRAS.526.1687T}. 
The authors derived the excess variances for AGNs detected with Swift/BAT and observed with XMM-Newton. The reported excess variances of the AGNs range roughly from $-5$ to $-1$. We thus draw excess variances from a uniform distribution within the range. 


Our actual simulation procedure is summarized as follows. 
(1) The slope of a PSD is randomly determined from the Gaussian distribution, and, with a tentative normalization value, a light curve is generated via the inverse Fourier transformation. The exposure time and bin size are set to 100 ks and 1 ks, respectively. The exposure time is reasonable, given that it falls between the shortest and longest exposures of the 4XMM-DR14 light curves that we finally use to detect anomalies (Section~\ref{sec:xmm_data}): i.e., $\sim$ 30 ks and $\sim$ 130 ks. 
(2) An appropriate PSD normalization is then incorporated into the light curve based on an excess variance, sampled from the uniform distribution. 
(3) Lastly, the light curve is standardized, as is often done in ML tasks. 

Anomaly light curves are generated by first simulating AGN-like light curves in the way described above and then superimposing QPE-like flares on them. 
Actual QPE flares were systematically identified and characterized by \citet{Webbe2023RASTI...2..238W}, who determined their durations, amplitudes, and duty cycles. 
The authors then fitted exponentially modified Gaussian distributions to their frequency histograms. 
These distributions are accessible on their GitHub page\footnote{https://github.com/robbie-webbe/ML\_QPEs} and 
can be used to generate QPE-like flares by randomly determining parameters. 
To investigate the detectability of individual QPE-like flares, 
we superimpose one flare onto an AGN-like light curve for simplicity. 
Thus, we do not use the histogram of the duty cycle. 
Regarding the amplitude, it significantly affects the detectability of  flares, and we produce four anomaly datasets with fixed amplitudes of 1.1, 1.5, 2.0, and 3.0. An amplitude of 1.1 roughly corresponds to the lowest observed QPE amplitude (\citealt{Webbe2023RASTI...2..238W}).  
The remaining duration parameter is randomly sampled from the distribution provided by \citet{Webbe2023RASTI...2..238W}. 

We prepare three sets of time-series data, each consisting of normal and anomalous datasets: Sets A, B, and C. 
In Set A, 100 light curves are generated per dataset type. 
The normal dataset in Set A is used to train the LSTM models.
As detailed in Section~\ref{sec:training}, because we conduct the learning every 5 ksec, the actual number of training instances is larger than the number of the normal light curves. 
Specifically, 70\% of each 100-ks exposure light curve is used for  training, resulting in 6500 training data points. 
The remaining 30\% is used for performance evaluation after each training epoch. 
We do not increase the dataset size due to the high computational cost of quantum simulations.
The second set, Set B, consists of 1000 normal and 1000 anomalous light curves. 
This set is designed to thoroughly assess the performance of the trained LSTM models. 
Set C is a smaller set with nine light curves per dataset type and is used to visualize the accuracy of the LSTM models in reproducing light curves. 
These nine light curves are generated from nine parameter sets derived from three excess variances  ($10^{-5}, 10^{-3},$ and $10^{-1}$) and three PSD slopes ($-1.7, -2.0,$ and $-2.3$) to cover the parameter spaces. 

\subsection{Real Light Curves in the XMM-Newton Serendipitous Source Catalog}\label{sec:xmm_data}

Our goal is to detect transient events by using a trained QLSTM model on a large set of light curves in the 4XMM-DR14 catalog. 
The catalog includes 1035832 detections with 372313 detections having 
light curves. 
By applying the summary flag SUM\_FLAG of 1 or less to ensure detection quality, 297803 reliable detections remain.  
Although Metal Oxide Semi-conductor (MOS) and pn light curves are available, we focus only on pn light curves due to their generally higher signal-to-noise ratios compared to MOS light curves. 
As low count rates (i.e., low signal-to-noise ratios) and short exposures would make it difficult to detect anomalous behaviors, or transient events, significantly, we limit our sample to those that were identified in observations with exposures longer than 30 ksec and average count rates above 0.01 in the 0.2--12 keV band (PN\_ONTIME $> 30000$ and PN\_8\_RATE $> 0.01$). 
By furthermore excluding observations with the exposure flag ``X", meaning ``not applicable" in the catalog, we downloaded the remaining light curves from the 
XMM-Newton Science Archive Server.
We then subtract background from the source light curves. As probably non-celestial anomalous behavior appears near the end of the exposure time in some light curves, particularly in the final few ks,  
we uniformly exclude the final 5 ks from all light curves. 
Similar issues have been reported in \cite{Webbe2023RASTI...2..238W}. 
Finally, 40154 light curves with exposure times of $> 30$ ks compose our final XMM-Newton sample. These light curves were produced in the 0.2--12 keV band, and we bin them using a time bin size of 1 ks.

\section{Setup and General Performance}\label{sec:learn_perass} 

\subsection{Setup}\label{sec:learn_perass} 

In the CLSTM model, the parameter to be set is the size of the hidden state, and we set it to four, resulting in 133 trainable parameters. 
The breakdown is as follows. 
The input is two-dimensional ($t$, $x_t$), and the hidden state ($h_{t-1}$) is four-dimensional. To compress the six-dimensional data ($t$, $x_t$, $h_{t-1}$) into a four-dimensional space, a fully connected layer with 28 parameters (i.e., $4\times6+4$) is needed. 
Each of the three gates also requires a fully connected layer with four inputs and four outputs, adding 20 parameters per gate (i.e., $4\times4+4$). 
As the input gate requires an additional fully connected layer of the same type, the three gates have 80 trainable parameters in total. 
Finally, two fully connected layers are needed.
One extracts the hidden state, or short-term memory, to be forwarded (20 parameters, i.e., $4\times4+4$), and the other converts the hidden state into a one-dimensional output for comparison with the actual value (five parameters, i.e., $1\times4+1$). 
Consequently, the total number of parameters becomes 133. 
If the hidden state size is set to three (five), the number of trainable parameters becomes 88 (186). 

The number of trainable parameters in the QLSTM model depends on the number of ansatz repetitions, the number of qubits, and the size of the hidden layer. 
Among the various combinations of these three factors, for each QLSTM model, we select the combination in which the number of trainable parameters matches that of the CLSTM model as much as possible. 
When we use an ansatz other than RealAmplitudes, we repeat the ansatz twice within a QVC and set the size of the hidden state and the number of qubits to two and four, respectively. 
Under these conditions, the number of trainable parameters becomes 129. 
Among these, 20 parameters are required to implement, like the CLSTM, a fully connected layer that processes the concatenated vector of ($t$, $x_t$, $h_{t-1}$) before its encoding into the four qubits. 
Then, because each of the four qubits in one QVC is rotated by using six free parameters and there are four QVCs, we need 96 parameters related to the QVCs (i.e., $6 \times 4 \times 4$). 
After passing the QVCs, we need one fully connected layer to determine the next hidden state, and thus, 10 parameters are required (i.e., $2 \times 4 + 2$).
Finally, to produce a one-dimensional output for comparison with the observed value, three parameters are necessary (i.e., $1 \times 2 + 1$). 
In the case of RealAmplitudes, while maintaining the same hidden layer size and number of qubits, we repeat the ansatz six times and, as a result, maintain the same number of trainable parameters.

\subsection{Performance Indicator: Effective Dimension}\label{sec:effdim}

\begin{figure}
 \begin{center}
  \includegraphics[width=8.4cm]{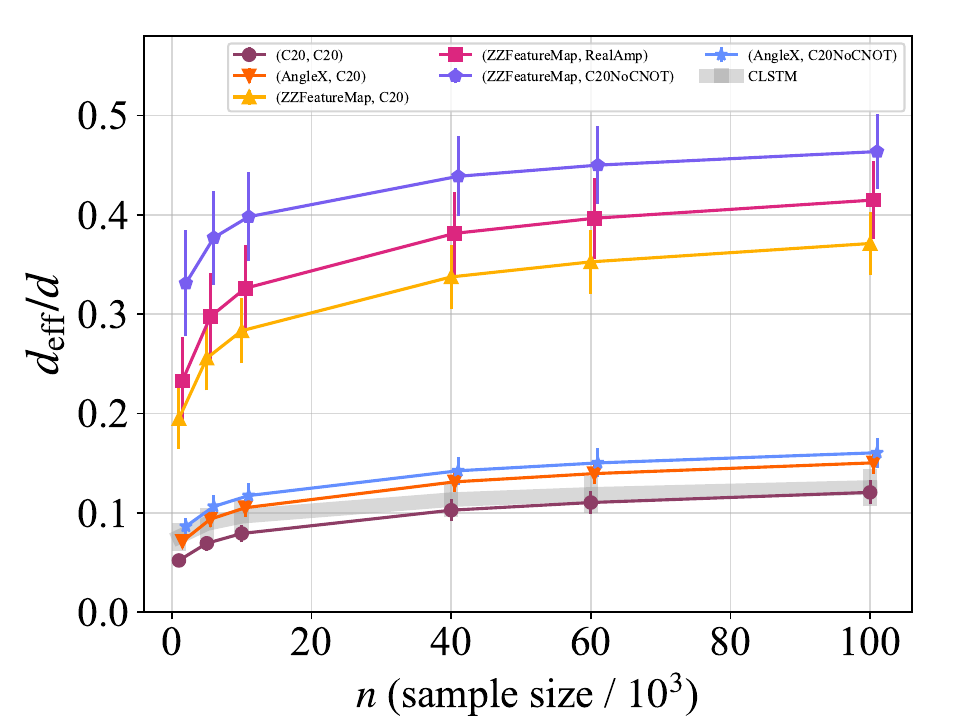} 
 \end{center}
 \caption{ 
    Effective dimension divided by the actual dimension as a function of sample size. 
    The data points and error bars indicate the average values and standard deviations of multiple calculations, respectively. 
    For each QLSTM model, we show the result of 10 calculations, while the result of the CLSTM model is based on 100 calculations. 
    The model for each curve is denoted in the legend. 
 }
 \label{fig:eff} 
\end{figure}

To evaluate and compare the performance of the CLSTM and six QLSTM models, we derive effective dimensions, which measure 
the number of parameters actively contributing to the performance of a model. 
We adopt the definition proposed by \citet{Abbas2020arXiv201100027A} and \citet{Berezniuk2020arXiv200110872B}.
The important quantity in their definition is complexity, representing how well a model can fit diverse data, and the effective dimension indicates the number of parameters that significantly contribute to the quantity. 
Briefly, the significant parameters are identified by 
assessing the extent to which changes in parameters affect the model output. 
A notable advantage of this effective dimension is its consideration of sample size in the calculation (\citealt{Abbas2020arXiv201100027A}).
For a detailed explanation, one can refer to Appendix~\ref{app:eff} or the original papers (\citealt{Abbas2020arXiv201100027A}; \citealt{Berezniuk2020arXiv200110872B}).

We compute the effective dimension for the CLSTM and QLSTM models by referring to the program provided by \citet{Abbas2020arXiv201100027A}\footnote{https://github.com/amyami187/effective\_dimension}. 
As the original program was designed for NNs, we slightly modify it for the LSTM models. 
Each model is composed of five LSTM layers (Figure~\ref{fig:lstm_layer}), where each layer is designed to receive two-dimensional data, such as time and a value at the time, as input. 
In addition, whereas our LSTM models are designed to predict a possible future value for anomaly detection, we simplify the problem by considering  binary classification as in  \citet{Abbas2020arXiv201100027A}.
Because the effective dimensions depends on input data and parameters, 
we randomly generate 100 input datasets and 100 parameter sets, resulting in a total of 10000 sets, and use them to derive the average effective dimensions. 
To account for statistical fluctuations owing to random sampling, we perform 10 calculations for each QLSTM model and 100 calculations for the CLSTM model. For each model, we compute the average and standard deviation.

The estimated effective dimensions as functions of the number of available training data are shown in Figure~\ref{fig:eff}. We find that the effective dimensions of the three QLSTM models implemented with ZZFeatureMap are significantly higher than that of the CLSTM model. 
Thus, the feature map ZZFeatureMap seems to enhance the performance of the LSTM model. 
Regarding the choice of an ansatz, we find that C20NoCNOT, which has no entanglement operation, achieves the highest effective dimension.
However, the effective dimension is only $\sim$ 20\% higher than the second highest one, and as discussed later in Section~\ref{sec:training}, the QLSTM models with quantum entanglements, that is, (ZZFeatureMap, C20) and (ZZFeatureMap, RealAmplitudes), can achieve higher prediction accuracies. 
Considering the effective dimension and prediction accuracy (Section~\ref{sec:training}), we 
adopt the QVC consisting of (ZZFeatureMap, C20) 
as the pivotal point for further discussion. 
Although the QVC constructed with ZZFeatureMap and RealAmplitudes has a higher effective dimension that the adopted one, we prioritize the combination of ZZFeatureMap and C20 because it achieves the highest prediction accuracy.

\begin{figure}
 \begin{center}
  \includegraphics[width=8.6cm]{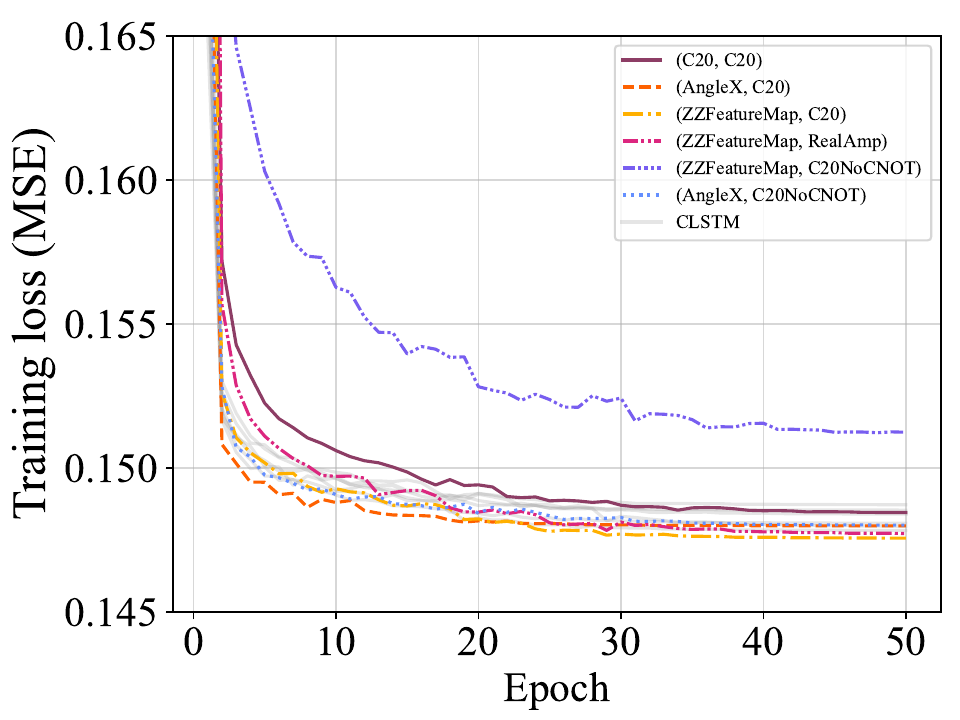} 
  \includegraphics[width=8.6cm]{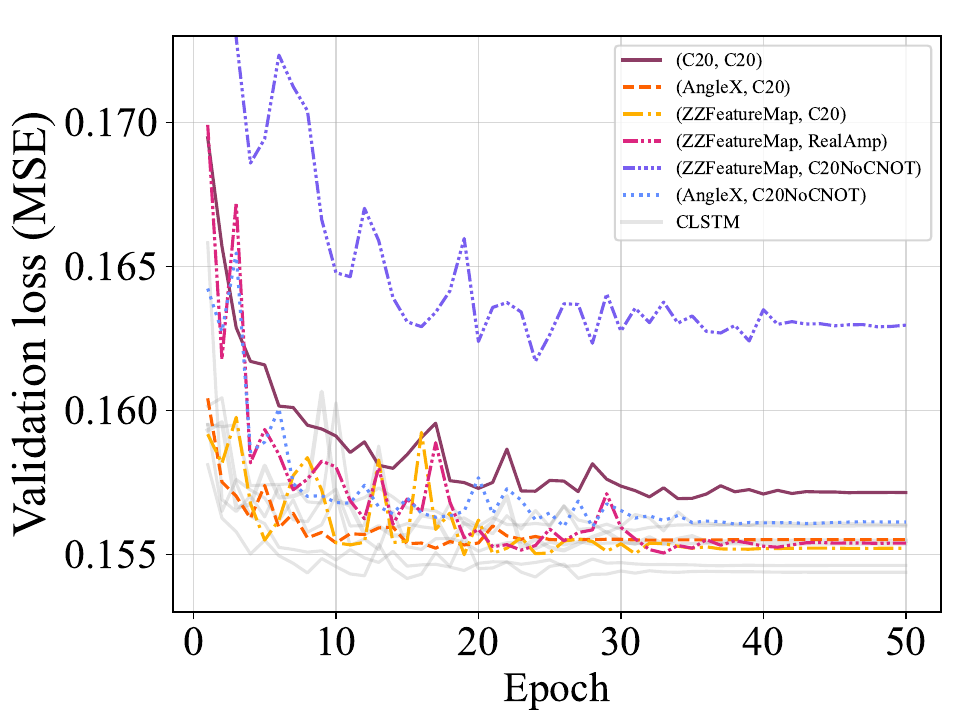}   
 \end{center}
 \caption{
    Loss histories for the training and validation datasets, shown in the top and bottom figures, respectively. We use the Set A data, providing 100 normal light curves (see Section~\ref{sec:train_test_data}). 
    For each line, the used feature map and ansatz are presented in the legend. 
 }
 \label{fig:loss_hist} 
\end{figure}

\section{Training and Performance}\label{sec:results} 

\subsection{Training}\label{sec:training}

Using Set A (100 normal light curves and 100 anomaly curves; see Section~\ref{sec:train_test_data}), we train both the CLSTM model and the six QLSTM models (Table~\ref{tab:qvcs}). 
We allocate 70\% of the normal dataset for training and the remaining 30\% for validation. 
In each training process, time-series data within 5 ks (i.e., five data points such as $(t_0, x_0)$, $(t_1, x_1)$, ..., $(t_4, x_4)$, where $t_i$ and $x_i$ are the time and count rate, respectively), are used to predict the subsequent count rate. Then, the trainable parameters are optimized to reduce the mean squared error (MSE) between the predicted and actual values. 
The optimization is based on the \texttt{Adagrad} algorithm. 
By examining different optimization algorithms, such as \texttt{Adam}, we find that \texttt{Adagrad} seems to be one of the best options for rapidly  reducing loss. 
The initial learning rate is set to 0.05, and the rate is reduced by half if the decrease in MSE is less than 0.001 for three consecutive epochs. 
The total number of training epochs is set to 50.

Regarding the training of the QLSTM models, more things need to be described. 
We simulate quantum circuits in the QLSTM models under two conditions; one is free from quantum noise, and the other is affected by the noise. 
Simulations in noisy conditions, in addition to the ideal case, are important, 
because the currently available quantum computers are generally noisy. 
When simulating noise-free quantum circuits, we utilize the \texttt{lightning.qubit} simulator, provided by Pennylane. 
For the noisy case, real computers are usable but quite busy. We therefore simulate noisy circuits by using a suitable simulator of \texttt{AerSimulator} and a noise model provided by Qiskit. 
The noise model is the one created based on the "ibm\_brisbane" quantum computer. 
As particularly the simulation of noise-involved training is extremely time-consuming, we limit the number of training epochs to three. We, however,  
initialize the parameters with values determined from noise-free training, allowing the QLSTM models to achieve low loss values within a few steps. 

\begin{figure*}
 \begin{center}
  \includegraphics[width=18cm]{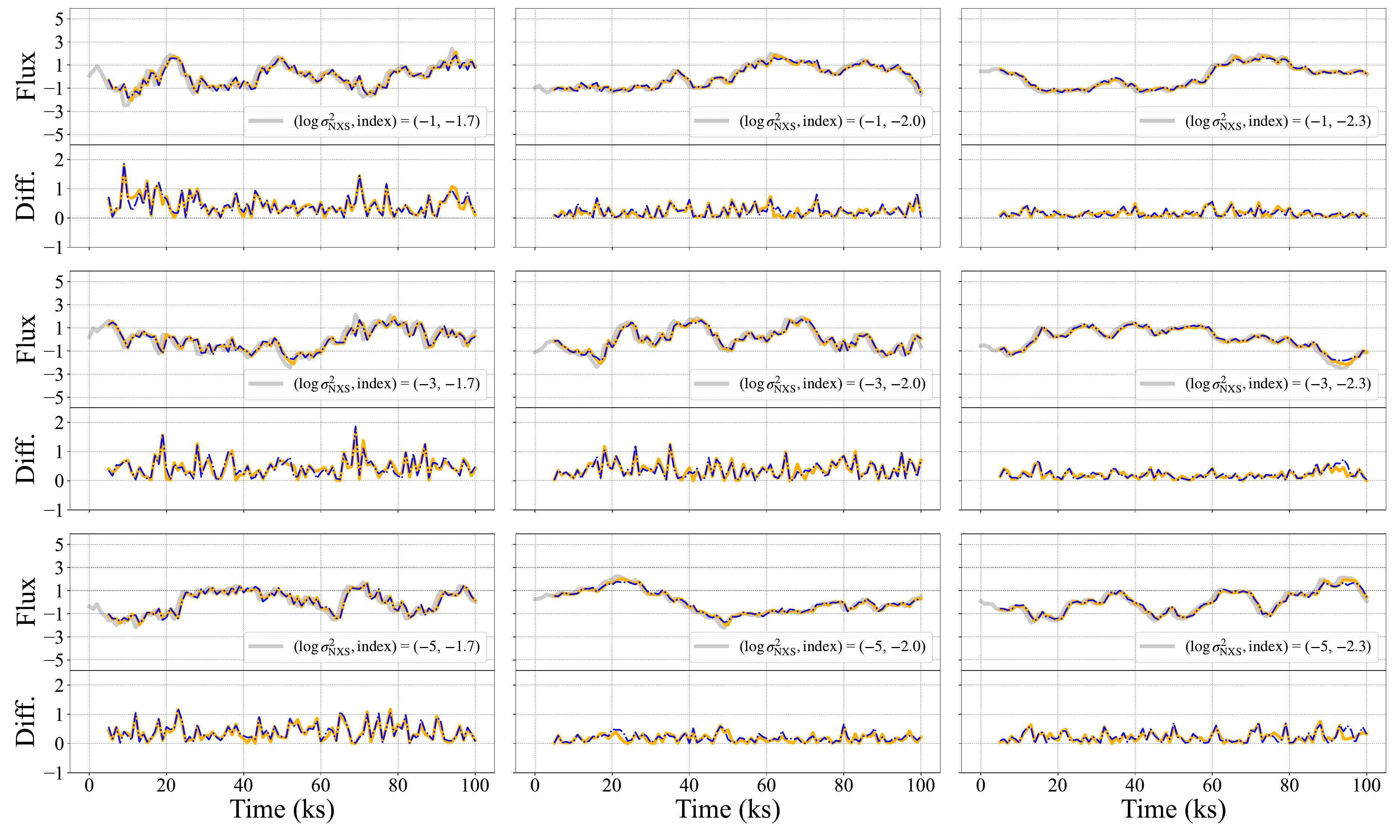} 
 \end{center}
 \vspace{-.5cm}
 \caption{
    QML predictions for nine representative, simulated AGN-like light curves in Set C (Section~\ref{sec:train_test_data}). 
    Results are based on the pivotal QML model constructed with (ZZFeatureMap, C20). 
    The upper sub-panel of each panel shows 
    the normal light curve (gray solid line), the noise-free prediction (orange solid line), and the noisy prediction (blue dashed-dotted line). 
    Fluxes are in arbitrary units. 
    Each lower sub-panel displays the absolute difference between the AGN-like light curve and the predictions; 
    the orange solid and blue dashed-dotted lines corresponds to the results under the noise-free and noisy situations, respectively. 
 }
 \label{fig:demo_normal} 
\end{figure*}

\begin{figure*}
 \begin{center}
  \includegraphics[width=18cm]{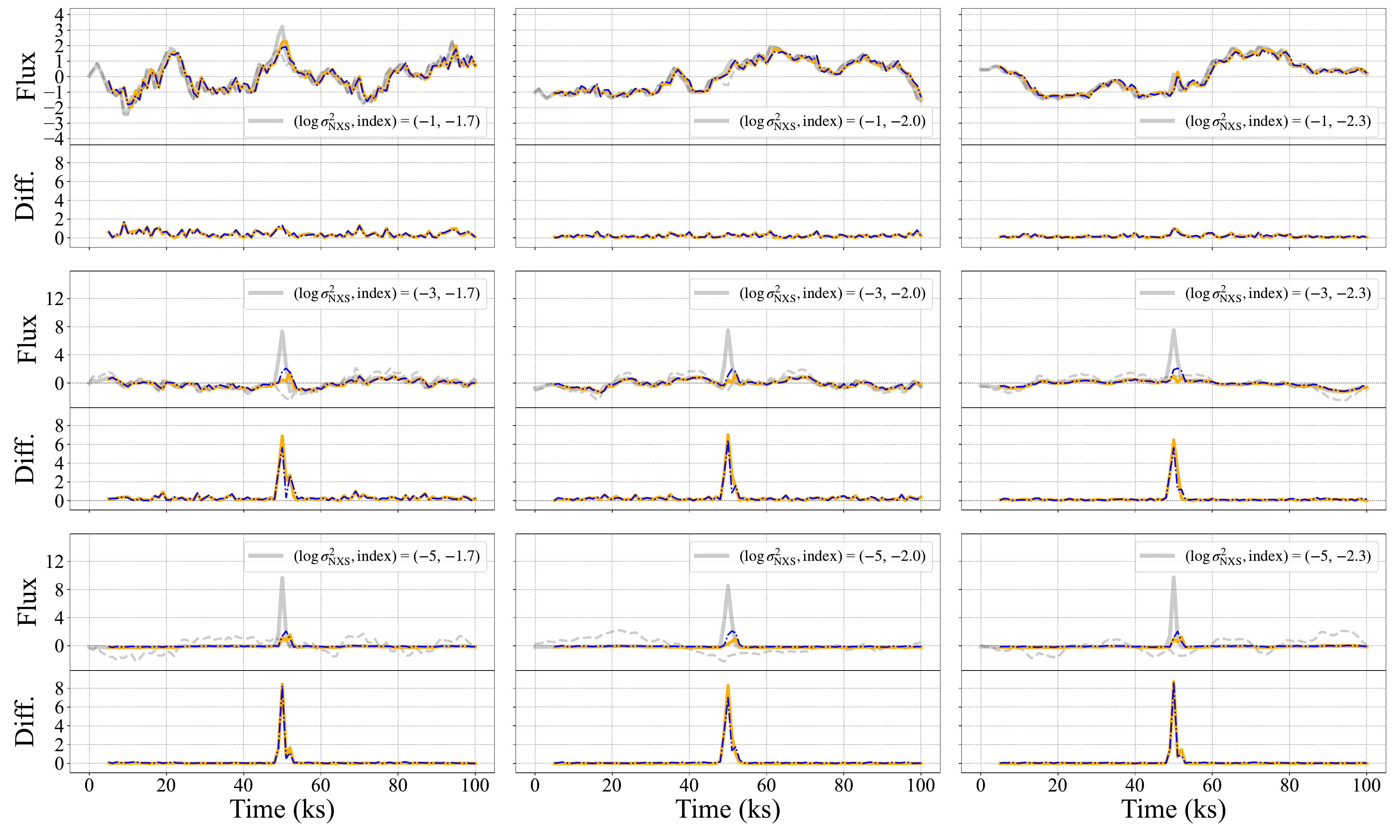}   
 \end{center}
 \vspace{-.5cm} 
 \caption{
    Same as Figure~\ref{fig:demo_normal}, but for anomaly light curves simulated with a fixed amplitude of 1.5. All anomaly flares occur around 50 ks. 
    In each upper sub-panel, the dashed gray line, not present in Figure~\ref{fig:demo_normal}, represents the normal AGN-like light curve, serving as a base for the anomaly curve (solid gray line). 
    We note that anomaly flares appear weaker with increasing $\sigma^2_{\rm NXS}$ because a higher $\sigma^2_{\rm NXS}$ results in a more variable light curve,  making the anomaly flare less noticeable. 
 }
 \label{fig:demo_anomaly} 
\end{figure*}

The loss histories for the six QLSTM and one CLSTM models are shown in Figure~\ref{fig:loss_hist}, where 
the results of the QLSTM models are what are obtained under the noise-free condition. 
Although the QLSTM models using AngleX, the simplest feature map, exhibit the fastest initial loss reduction, the combinations of ZZFeatureMap with ansatzes creating quantum entanglement achieve the lowest losses. 
This result may suggest that the quantum entanglement can enhance prediction accuracy. 
However, this advantage is not seen in the validation data loss histories when compared to the LSTM model. 
Regarding the accuracy under the noisy condition, the lowest losses of 0.155 and 0.158 are obtained for the QLSTM models with (ZZFeatureMap, C20) and (ZZFeatureMap, RealAmplitudes), respectively. 
The losses of the other combinations exceed 0.16. 
Due to the quantum noise and the fewer training steps, the accuracy is lowered, but in practice, such a decrease in performance about 0.01 is not a significant issue for detecting anomalies, or transient events.

\begin{figure}
 \begin{center}
  \includegraphics[width=8.6cm]{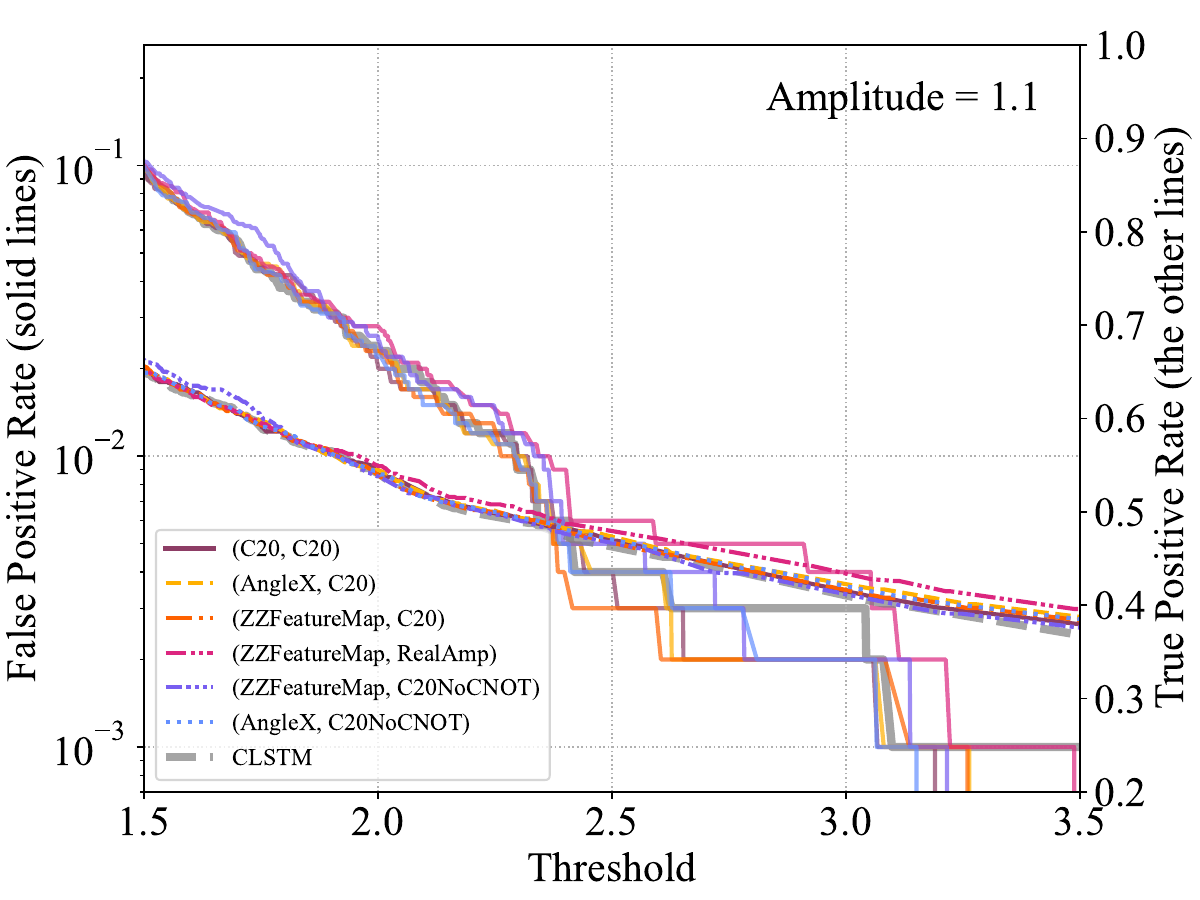} 
  \includegraphics[width=8.6cm]{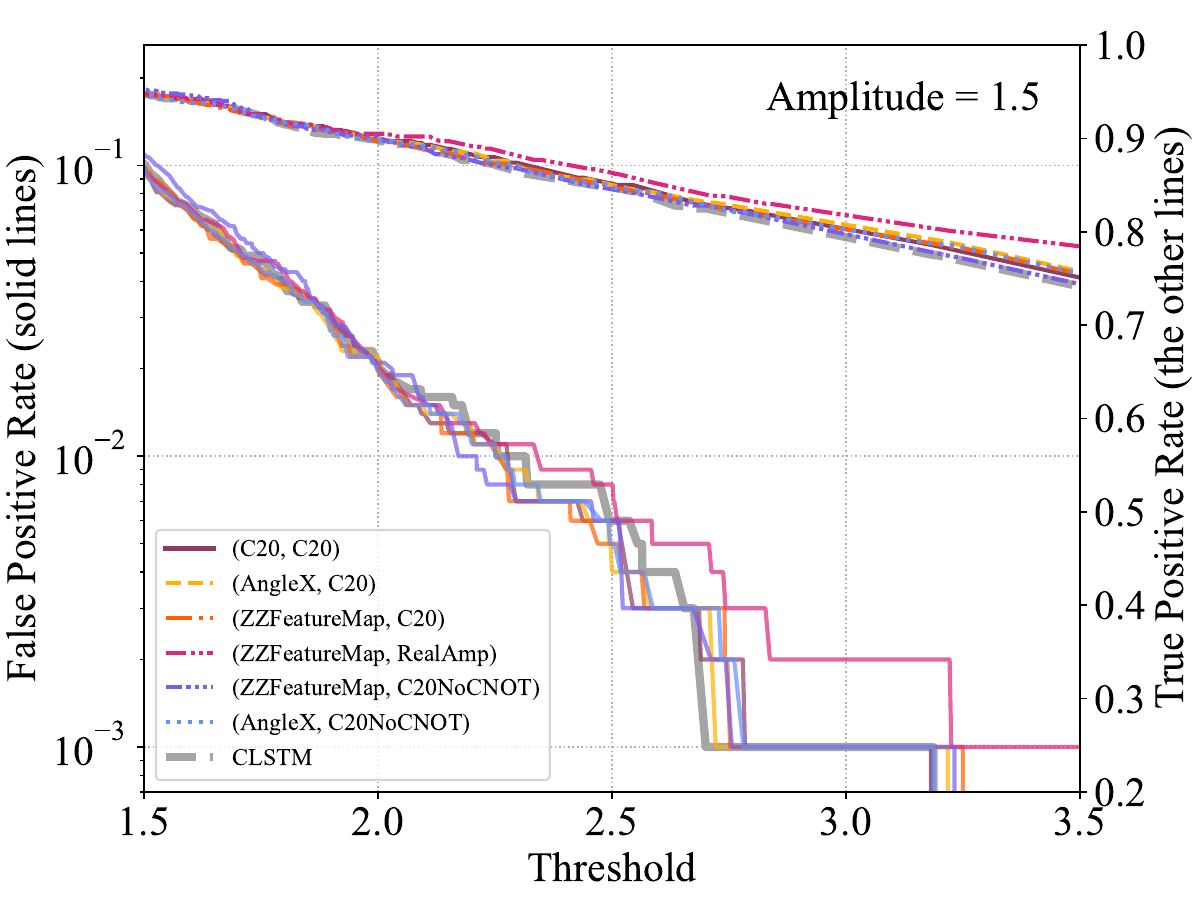}
  \includegraphics[width=8.6cm]{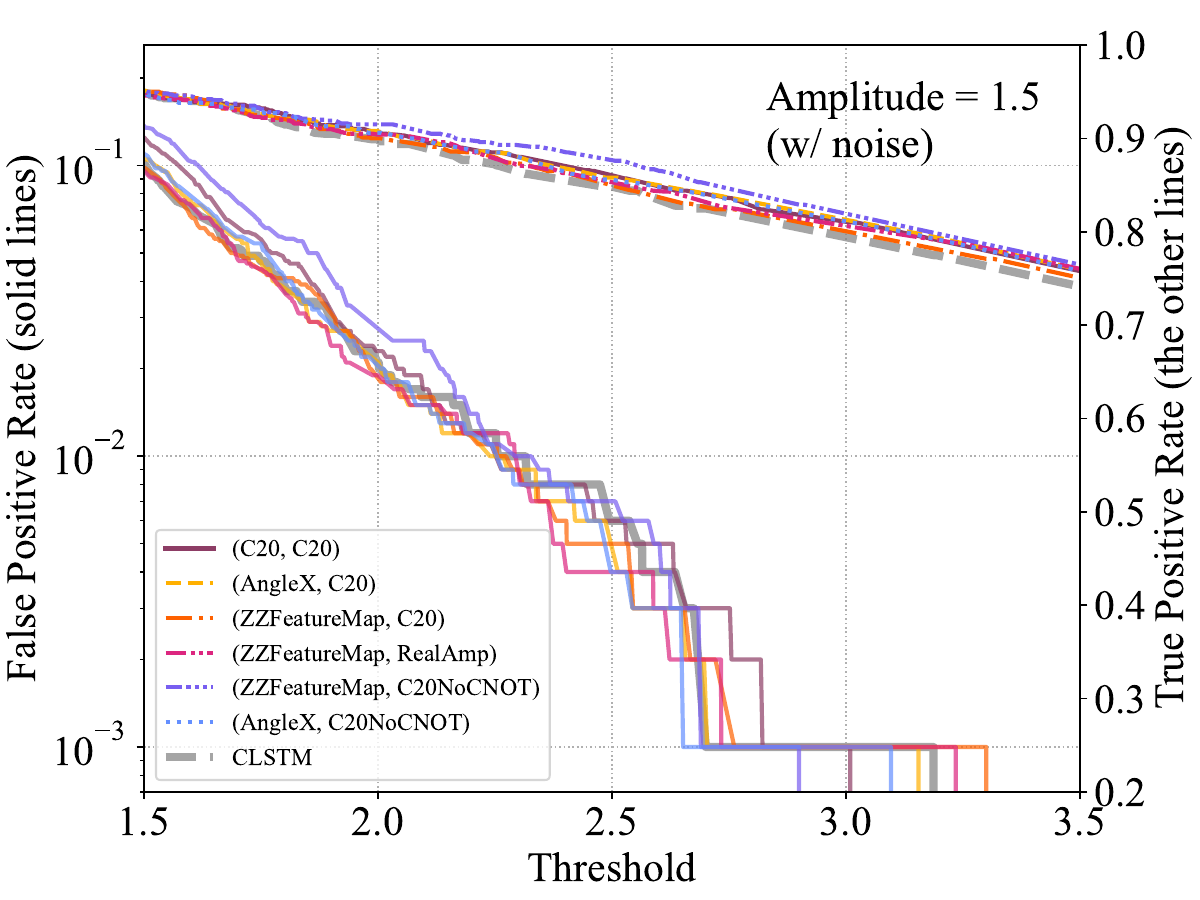}  
 \end{center}
 \caption{
    False-positive and true-positive rates as a function of the threshold for distinguishing between normal and anomalous events. The top and middle panels display the results for the amplitudes of 1.1 and 1.5, respectively. The bottom panel replicates the middle one but includes quantum noise in the simulations. 
 }
 \label{fig:fpr_tpr} 
\end{figure}

\subsection{Predictions and Anomaly Detection}\label{sec:result_of_training}

In Figure~\ref{fig:demo_normal}, 
we demonstrate the predictions for unseen normal data in Set C (i.e., nine representative AGN-like light curves) by the fiducial QLSTM model using (ZZFeatureMap, C20). 
Both results in the noise-free and noisy conditions are shown. 
The predicted light curves closely fit the prepared ones, with differences of $\sim$ 1.5 at most. 
Figure~\ref{fig:demo_anomaly} shows the results obtained by applying the QLSTM model to the anomaly data with an amplitude of 1.5 in Set C. 
It demonstrates that the model generally cannot explain flaring events, thereby identifying anomaly flares. 
We note that, when the excess variance used for generating an underlying light curve is high (e.g., $10^{-1}$), the curve shows strong variability and can obscure flares. 
This effect can explain the challenge in identifying anomalies in the light curves  with $\sigma^2_{\rm NXS} = 10^{-1}$ (the top panels in Figure~\ref{fig:demo_anomaly}). 
Additional results from the other QLSTM and CLSTM models are provided in 
Appendix~\ref{app:demos}.

\subsection{Performance Assessment and Determination of Optimal Threshold}\label{sec:performance}   

We quantify and compare the performances of the QLSTM and CLSTM models by calculating their false- and true-positive rates. 
As anomaly flares are identified by comparing actual data with LSTM model predictions and finding differences exceeding a threshold, the rates of interest depend on the threshold. Thus, by varying the threshold from 1.5 to 3.5, we calculate the rates.
The top and middle panels of Figure~\ref{fig:fpr_tpr} 
show the results for the amplitudes of 1.1 and 1.5, respectively. 
The large datasets of Set B (i.e., 1000 normal and 1000 anomaly data) are used to reduce the statistical uncertainty. 
The figure shows no clear differences between the false-positive rates of the QLSTM and CLSTM models regardless of the adopted amplitudes of 1.1 and 1.5, except for the combination of ZZFeatureMap and RealAmplitudes. The particular combination exhibits the worst performance. 
Regarding the true-positive rate, all QLSTM models generally outperform the CLSTM model. 
Even if quantum noise is present (the bottom panel of Figure~\ref{fig:fpr_tpr}), the same trend is observed, except that  the combination of ZZFeatureMap and RealAmplitudes shows a better false-positive rate than that in the noise-free situation. 
As a result, we find that, from the point of view of the true-positive rate, QLSTM seems to have the capability to outperform CLSTM. 

We determine the optimal threshold based on the calculated false-positive and true-positive rates. If the amplitudes of flares are 1.1 (the top panel of Figure~\ref{fig:fpr_tpr}), it is difficult to simultaneously achieve 
a low false-positive rate and a high true-positive rate at any threshold. 
Thus, we focus only on anomalous flares with amplitudes greater than 1.5.
Based on the middle panel of Figure~\ref{fig:fpr_tpr}, we adopt 3.0 as our optimal threshold. 
This threshold would be a reasonable option, as we can achieve a low false-positive rate of $\approx 10^{-3}$ and a relatively high true-positive rate of $\approx$ 0.8 in any QLSTM model. The determined threshold is used to detect anomaly candidates from the large set of XMM-Newton light curves (Section~\ref{sec:app2xmm}).

\section{Detection of Anomaly Light Curves from XMM-Newton Catalog}
\label{sec:app2xmm}

\begin{figure*}
 \begin{center}
  \includegraphics[width=5.6cm]{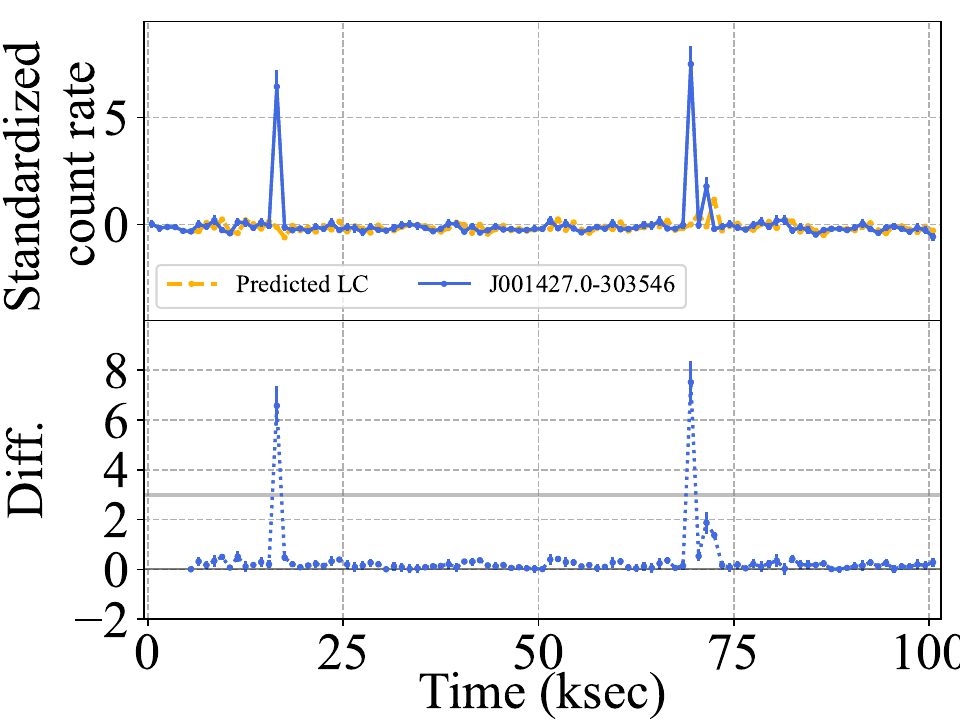}
  \includegraphics[width=5.6cm]{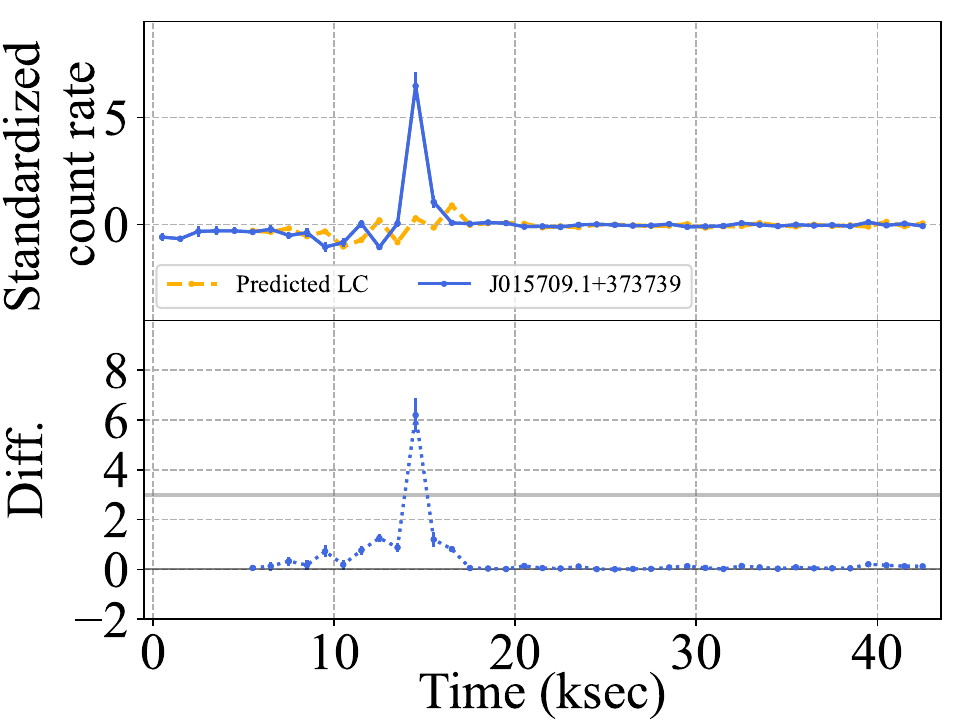}
  \includegraphics[width=5.6cm]{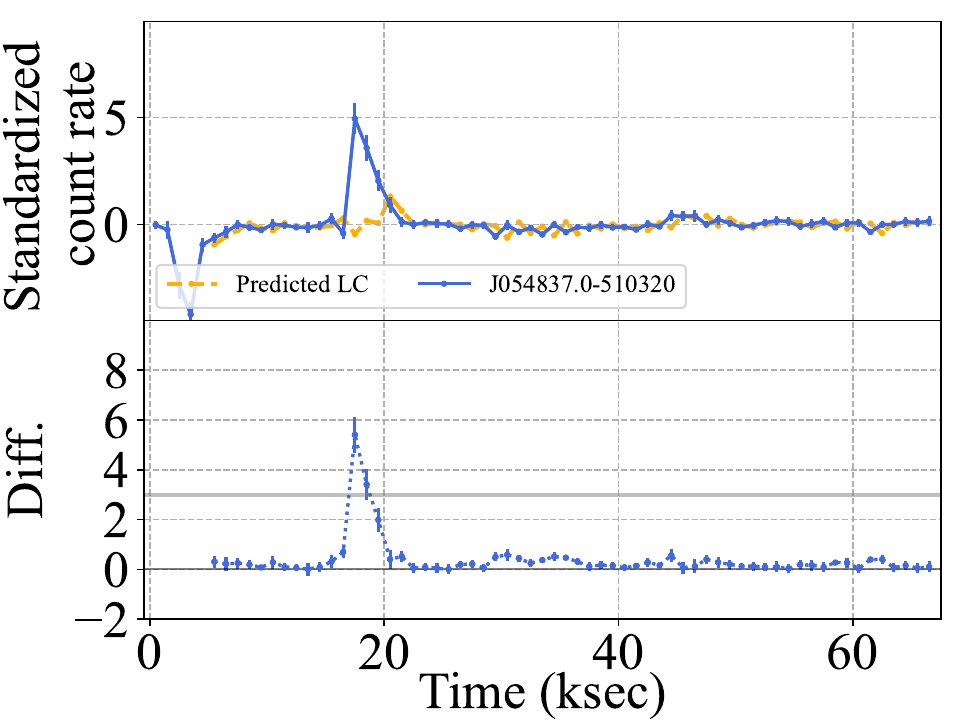}
  \includegraphics[width=5.6cm]{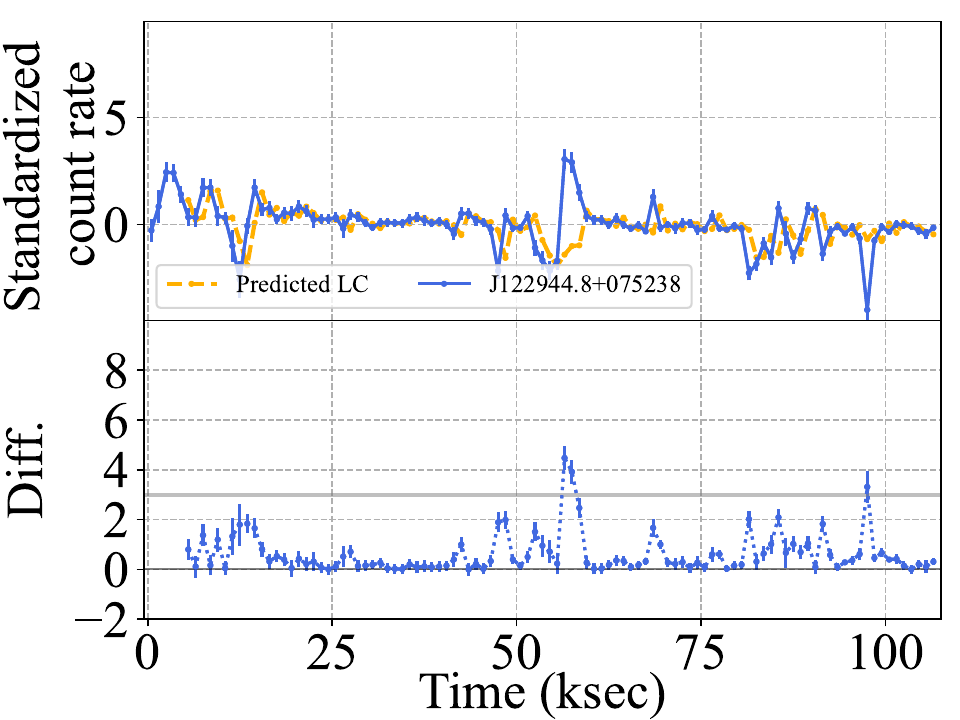}
  \includegraphics[width=5.6cm]{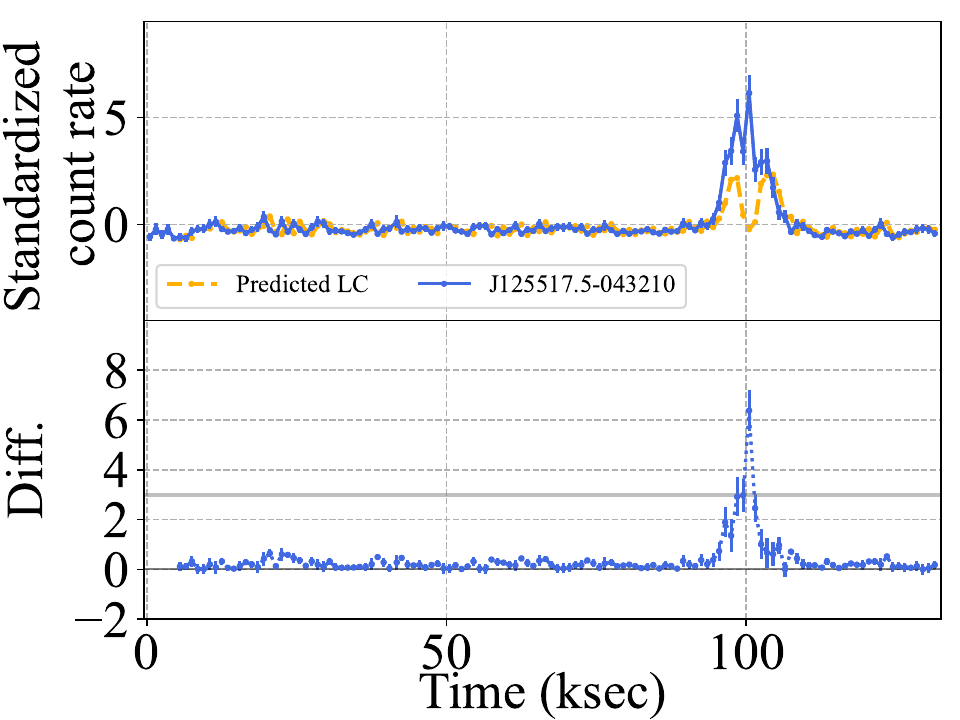}
  \includegraphics[width=5.6cm]{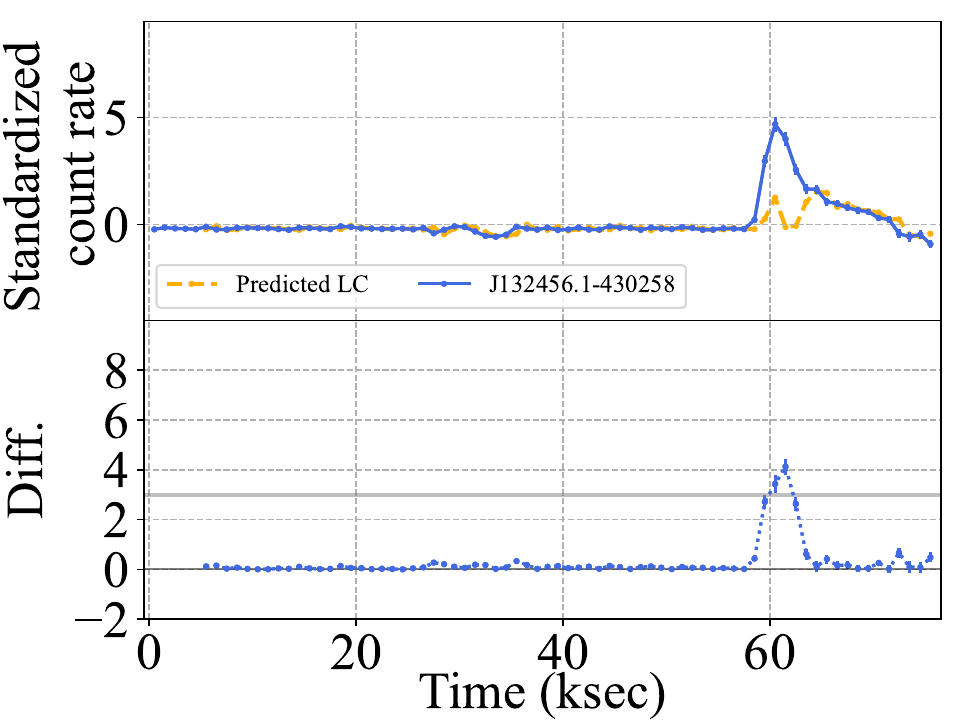}
  \includegraphics[width=5.6cm]{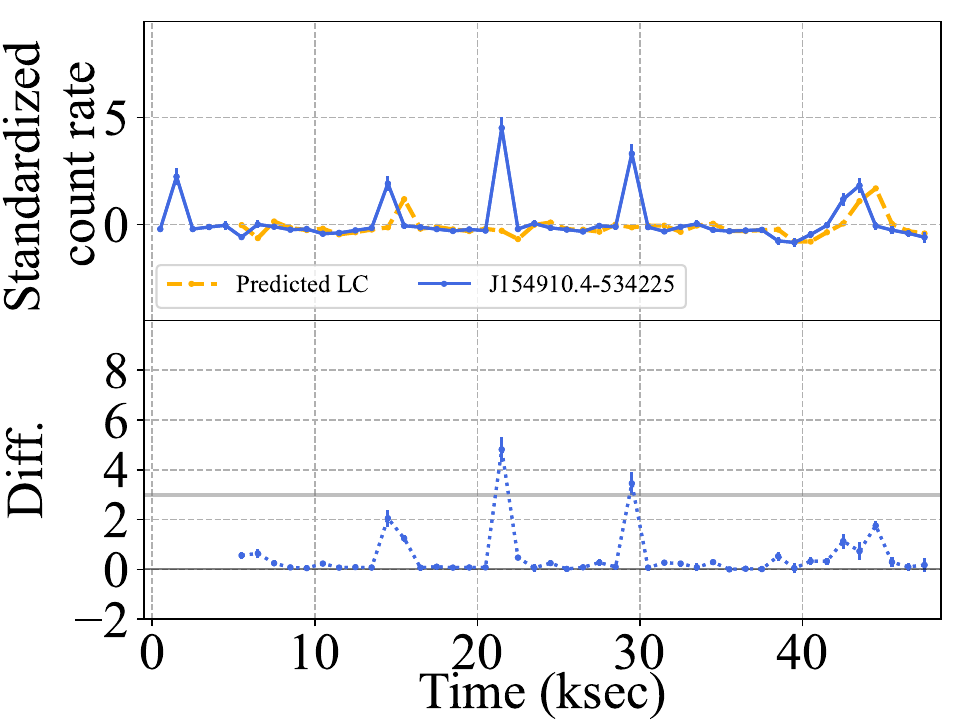}
  \includegraphics[width=5.6cm]{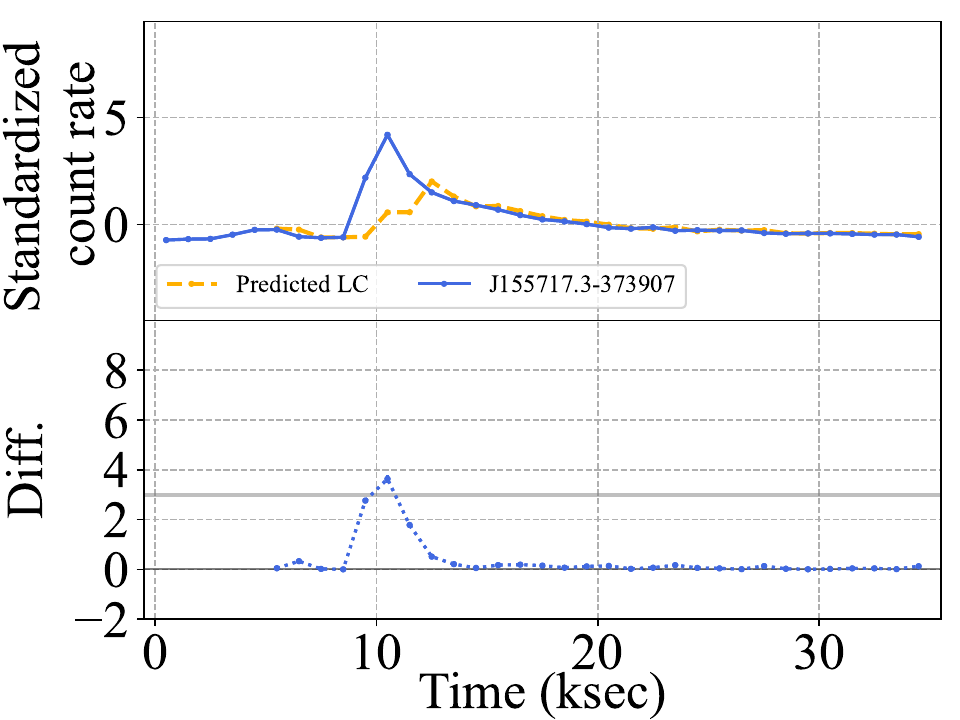}
  \includegraphics[width=5.6cm]{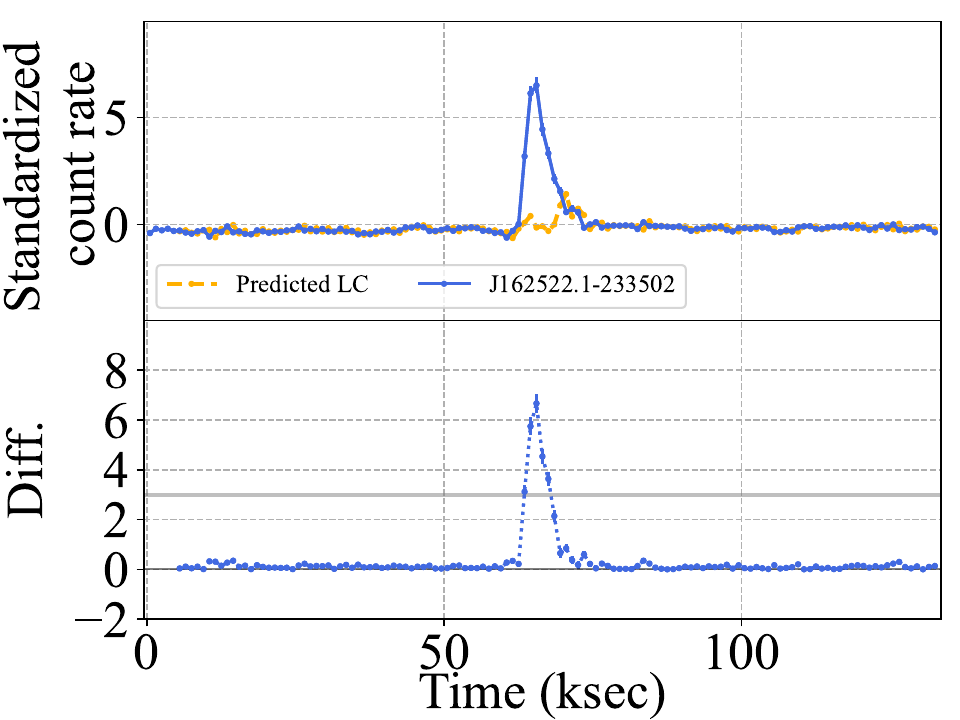}
  \includegraphics[width=5.6cm]{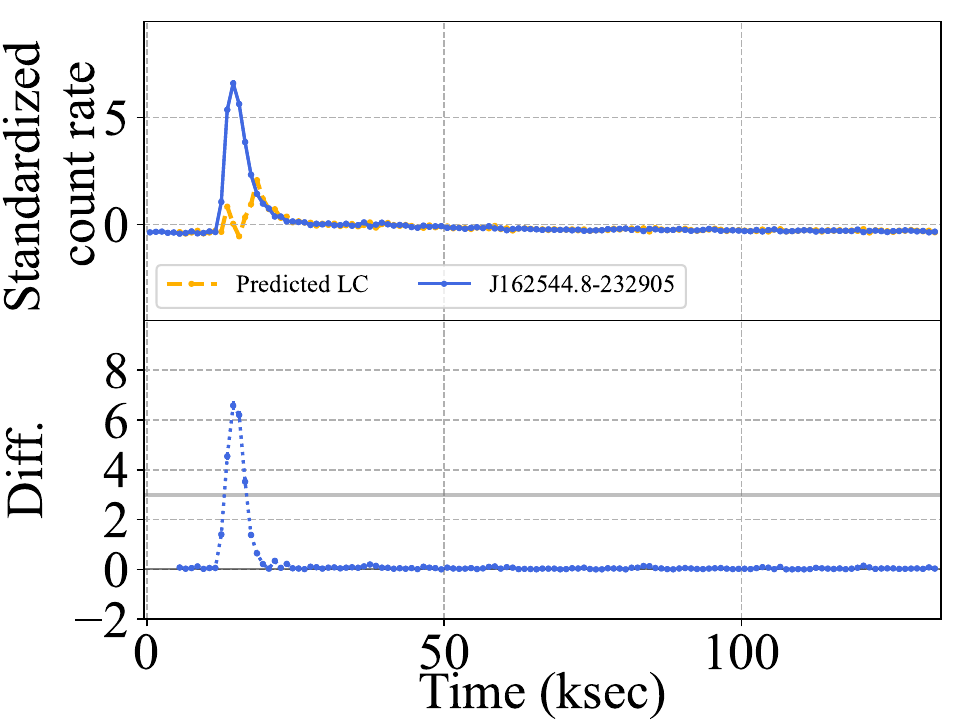}
  \includegraphics[width=5.6cm]{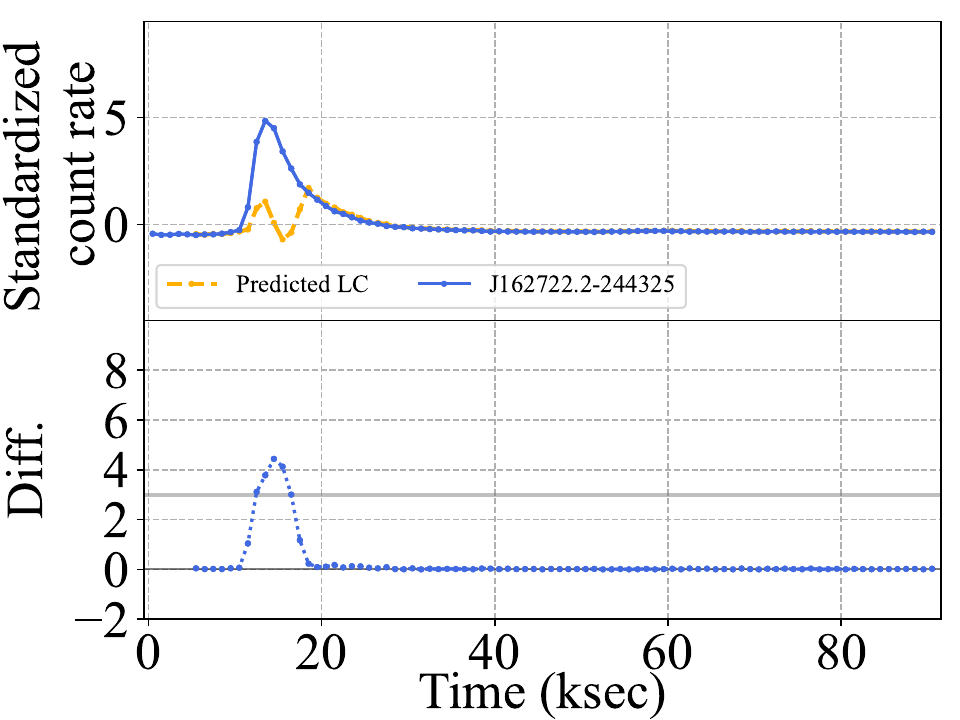}
  \includegraphics[width=5.6cm]{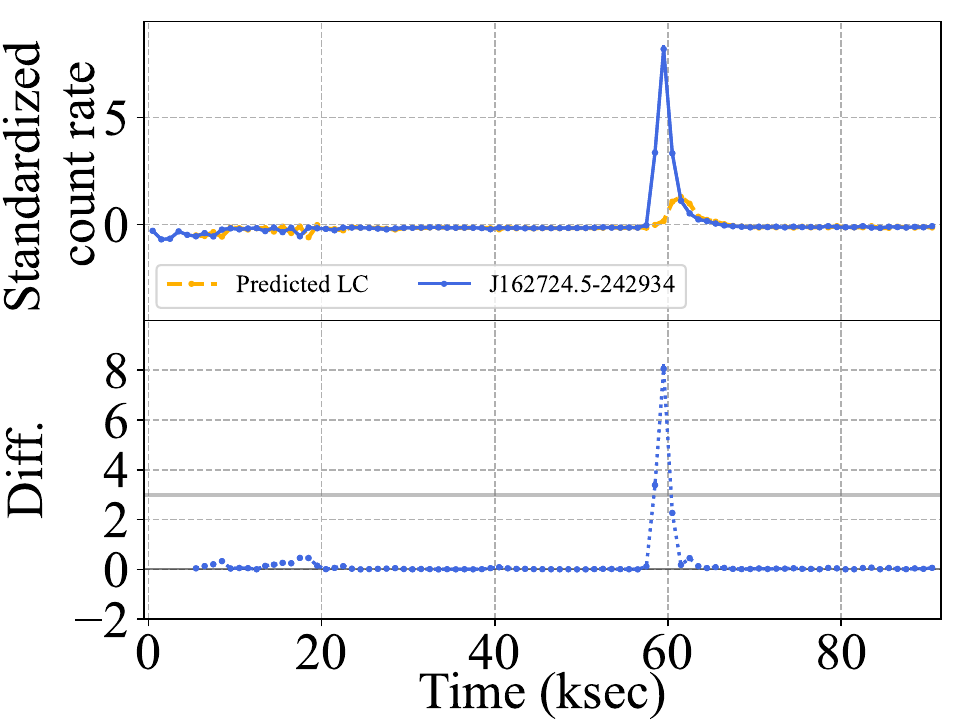}
  \includegraphics[width=5.6cm]{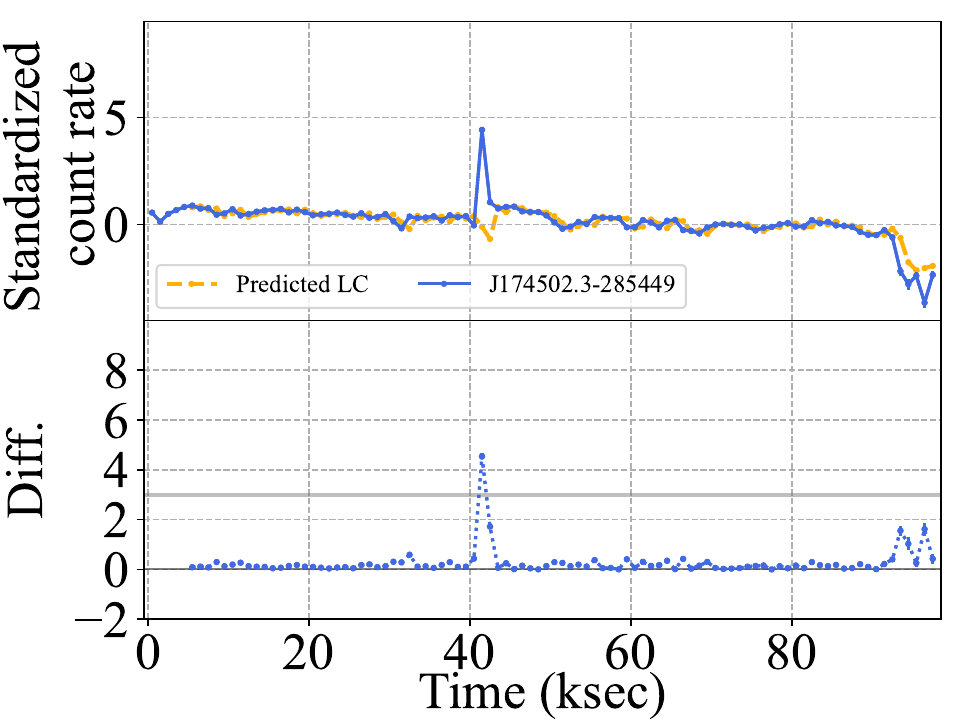}
  \includegraphics[width=5.6cm]{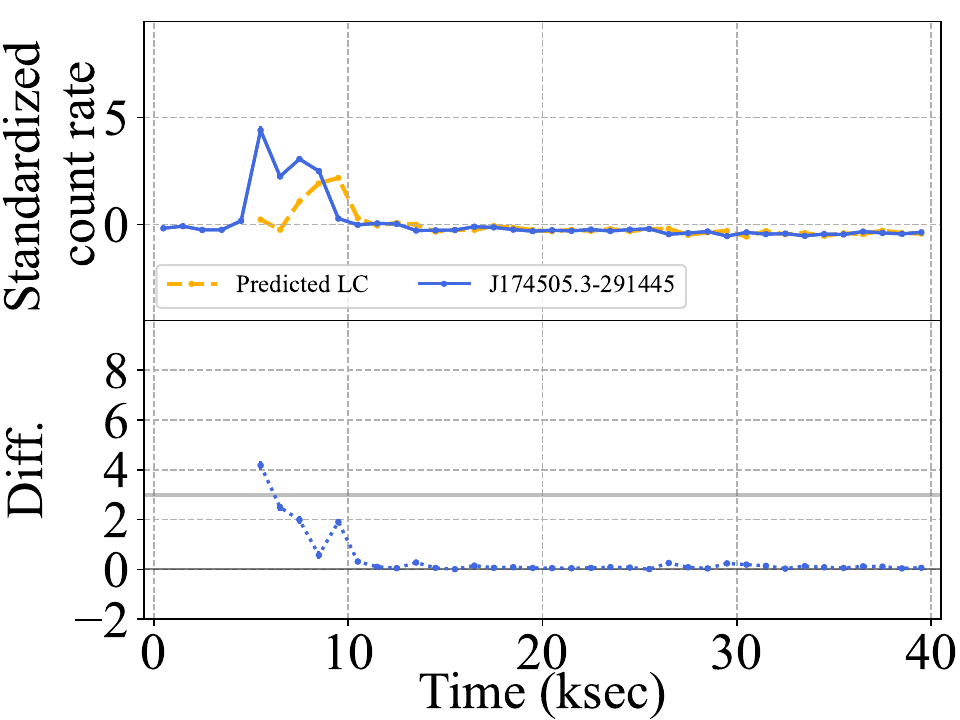}
 \end{center}
 \caption{
        Standardized XMM-Newton light curves in the 0.2--12 keV band (blue) and QLSTM model predictions (orange) using ZZFeatureMap and C20. 
        These light curves lack possible counterparts in the referenced databases. 
        Lower panels show the 
        absolute differences between the real and predicted values. 
 }
 \label{fig:xmm_anom} 
\end{figure*}

We attempt to identify anomaly flares from the XMM-Newton catalog by applying the QLSTM model constructed with the combination of (ZZFeatureMap, C20) and setting the identification threshold to 3.0 (Section~\ref{sec:performance}). 
The QLSTM model is adopted as it achieves a relatively high effective dimension (Figure~\ref{fig:eff}),  one of the lowest loss values (Figure~\ref{fig:loss_hist}), and the fairly good false-positive and true-positive rates (Figure~\ref{fig:fpr_tpr}). 
As a result, of the 40154 detections (Section~\ref{sec:xmm_data}), 
we successfully identify 113 detections, showing anomalous flares in their light curves. 
The table in Appendix~\ref{app:catalog}
summarizes their basic information. 
Ten examples of our anomaly detections are shown in Figure~\ref{fig:xmm_anom}. 
Each panel shows an XMM-Newton light curve, a predicted curve, and the absolute differences between the two curves. 

\begin{figure}
 \begin{center}
  \includegraphics[width=8.cm]{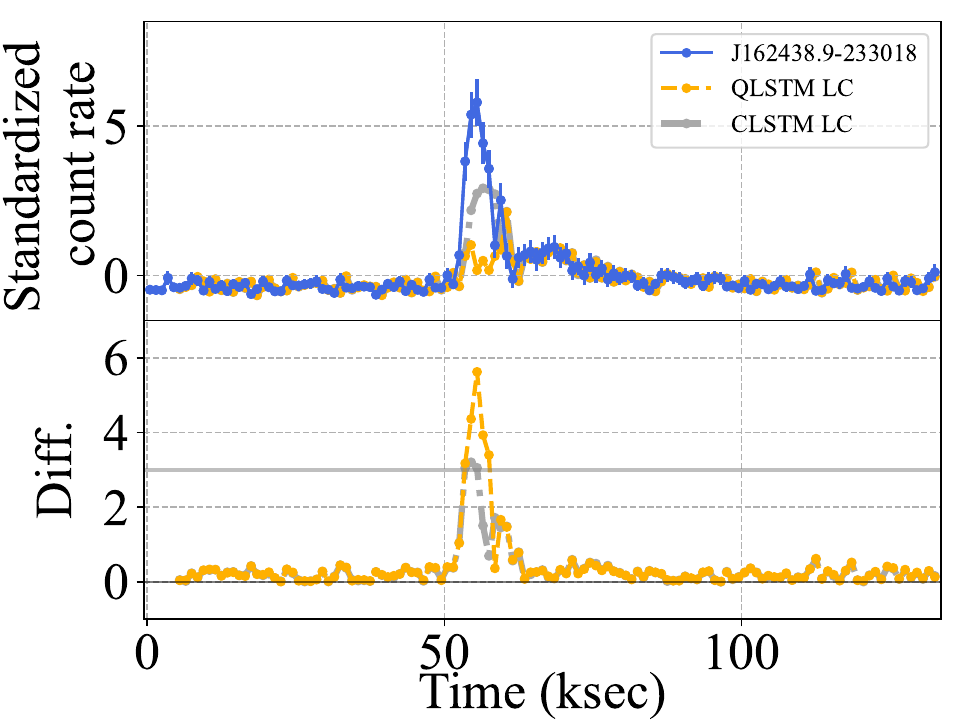}
 \end{center}
 \caption{
        Standardized XMM-Newton light curve in the 0.2--12 keV band (blue), and light curves predicted by the best QLSTM model and the CLSTM one. 
        The lower panel shows 
        the absolute differences between the real and predicted values, with 
        orange dashed and gray dot-dashed lines representing the results obtained with the QLSTM and CLSTM models, respectively.  
        The horizontal gray line indicates our detection threshold, with errors omitted for clarity. 
 }
 \label{fig:clstm_qlstm} 
\end{figure}

We conduct the same anomaly search with the CLSTM model as done with the QLSTM model and find 85 anomalies. 
This result is consistent with 
the higher true-positive rate of the QLSTM model than that of the CLSTM model.
As an example, Figure~\ref{fig:clstm_qlstm}
shows a comparison of the predictions made by the QLSTM and CLSTM models for an anomaly mioverlooked  by the CLSTM model.  
The CLSTM model tends to over-fit a possible flare, resulting in reduced flare signal.

To identify a possible celestial object responsible for each anomaly, we cross-match our anomaly catalog (Table~\ref{tab:anom_catalog}) with the SIMBAD database using the CDS xMatch service\footnote{http://cdsxmatch.u-strasbg.fr/xmatch/doc/index.html}. 
The database includes $\sim 2\times10^7$ objects with known positions. 
Also, using the same service, we search four catalogs created by the wide-field surveys of ROSAT, SkyMapper, Pan-STARRS, and WISE for possible counterparts. 
The actual catalogs are Second ROSAT All-Sky Survey Point Source Catalog (2RXS), SkyMapper DR4, Pan-STARRS DR1, and AllWISE 
(\citealt{Boller2016A&A...588A.103B}; \citealt{Wolf2018PASA...35...10W}; \citealt{Flewelling2020ApJS..251....7F}; \citealt{Wri10}; \citealt{Mai11}).  
The 2RXS catalog provides all-sky X-ray data, while 
we use the SkyMapper and Pan-STARRS catalogs to cover all sky in optical bands, and  the all-sky WISE catalog allows for searches less affected by dust extinction. 
For each anomaly, we consider an object spatially consistent with the XMM-Newton position within 3-sigma errors as a candidate. The cross-match accounts for errors in both XMM-Newton and referenced catalogs. 
When multiple candidates are found, we simply list the nearest one. 
As listed in Table~\ref{tab:anom_catalog} in Appendix~\ref{app:catalog}, 
possible counterparts exist for 101 anomalies; stellar activities seem to be dominant.

We examine how many of the detected anomalies correspond to previously confirmed QPE events. We refer to the list of known QPE sources compiled by \citet{Webbe2023RASTI...2..238W}, which includes five confirmed QPE objects 
(GSN 069, RX J1301.9$+$2747, XMMSL1 J024916.6$-$041244, eRASSU J023147.2$-$102010, and eRASSU J023448.9$-$441931) 
and 12 associated XMM-Newton light curves. 
Out of the light curves, five light curves are left after applying our data filtering criteria, selecting observations with exposure flags ``S" or ``U" and exposure times longer than 30 ks. 
Among these, our model has successfully detected anomalous variability in two light curves of RX J1301.9$+$2747, which is 4XMM J130200.1$+$274657. 
The remaining three light curves include two observations of eRASSU J023448.9-441931, whose QPEs are characterized by relatively slow variability, possibly making them more difficult to detect. The other corresponds to the one of GSN 069 where the flare amplitude was relatively small and anomalous flares did not exceed the threshold of the model ($=$ 3).
These results suggest that the QLSTM model is capable of recovering a subset of previously reported QPEs, while also indicating the importance of flare characteristics (e.g., variability timescale and amplitude) in determining detection sensitivity.

For the remaining 12 sources, we manually search for counterparts. 
4XMM J015709.1$+$373739 was identified as a shock breakout candidate by \citet{Alp2020ApJ...896...39A}. 
Although no counterpart is found for 4XMM J154910.4--534225, its light curve shows 
flares resembling QPEs, suggesting that it might be a new QPE object. 
We however note that the durations of the first three flares ($\lesssim$ 300 sec) are shorter than those typical for already identified QPEs ($\gtrsim 700$ sec; \citealt{Webbe2023RASTI...2..238W}). Thus, some non-celestial events may produce some, or all, flares. 

Although we have cross-matched the detected anomalies with SIMBAD, the four survey catalogs, the confirmed QPE list from \citet{Webbe2023RASTI...2..238W}, and literature, we have not systematically included catalogs that are dedicated to galactic or extragalactic transient events, such as stellar flares or gamma-ray bursts. 
As such, unmatched cases are not interpreted as new phenomena, but are treated as just unidentified anomaly candidates. Their classification requires follow-up observations and extended cross-matching. 


\section{Summary}\label{sec:summary} 

In the forthcoming era of time-domain astronomy, ML techniques will be crucial for analyzing continuously growing big data, and we have explored the potential of QML, which leverages quantum computers, as an advanced technique. 
In this study, we have investigated whether the quantum LSTM architecture, QLSTM, can detect anomalous events in X-ray light curves as transient events (Section~\ref{sec:results}). 
Additionally, we have used a trained QLSTM model to detect actual transient events from the 4XMM-DR14 catalog (\citealt{Webb2020A&A...641A.136W}; \citealt{Traulsen2020A&A...641A.137T})  (Section~\ref{sec:app2xmm}). 
Our investigations are summarized as follows:

\begin{itemize}
    \item Six QLSTM models were constructed by considering three feature maps and three ansatzes (Table~\ref{tab:qvcs}). Training and performance verification were performed with normal and anomaly data; the normal data were generated by simulating AGN-like light curves, while the anomaly ones were created by adding QPE-like flares to the normal data (Section~\ref{sec:data}). 
    
    \item Effective dimensions were calculated for all LSTM models by following the approach developed by \citet{Abbas2020arXiv201100027A} (Section~\ref{sec:effdim}). 
    The effective dimension indicates the number of parameters that actively contribute to data expression among the available parameters. 
    Our results reveal that the QLSTM models with  ZZFeatureMap, which enables encoding patterns difficult for classical methods, can achieve significantly higher effective dimensions than the other LSTM models, including the CLSTM one. 
    
    \item Training the QLSTM models to minimize MSEs, we found that the two models with the highest effective dimensions and quantum entanglement in their ansatzes attained the lowest MSEs. 
    This suggests that the selection of an optimal feature map and an ansatz can enhance prediction performance. 
    Compared to the CLSTM model, some QLSTM models achieved lower MSEs for the training data (the left panel of Figure~\ref{fig:loss_hist}), but, for the unseen validation data, no significant differences are found (the right panel of Figure~\ref{fig:loss_hist}). 
    Although we found that quantum noise reduces the performance of the QLSTM models, 
    the effect on anomaly detection would not be significant. 
    
    \item We calculated false-positive and true-positive rates by varying the threshold to detect anomalies based on the difference between observed and predicted values (Section~\ref{sec:performance}). 
    The QLSTM and CLSTM models exhibited similar false-positive rates, but 
    the QLSTM models achieved higher true-positive rates than the CLSTM one (Figure~\ref{fig:fpr_tpr}). 
    This result suggests the superior performance of the QLSTM models. 
    We note that the quantum noise generally did not affect these results. 
\end{itemize}
We then summarize our attempt to detect anomalies from the actual XMM-Newton data below: 
\begin{itemize}
    \item Using the best QLSTM model with ZZFeatureMap and C20 and our optimal threshold of 3.0 (Section~\ref{sec:performance}), we aimed to detect transient events from 4XMM-DR14. Among the 40154 detections, we found transient event candidates from the 113 detections (Table~\ref{tab:anom_catalog}). 
    This detection number is about 1.3 times higher than that by the CLTM model, suggesting an outperformance of the QLSTM model. 
    Based on SIMBAD and four catalogs of 2RXS, SkyMapper DR4, Pan-STARRS DR1, and AllWISE, it is found that many of the transient event candidates may be due to  stellar activities. Notably, no possible counterparts are found for 12 candidates. 
\end{itemize}

\section{Prospect}\label{sec:prospect} 

For future developments, we mention a technical insight we obtained through this research.
An LSTM-like model that frequently switches between classical and quantum computers results in substantial queue time for accessing a real quantum computer. This makes the entire process highly time-consuming and impractical for actual quantum computation.
Even if computations are completed, the benefits may be limited.
Therefore, in the context of quantum models, it may be more advantageous to explore fundamentally different architectures.

Despite the technical and methodological challenges discussed above, the exploration of quantum computing in astronomy remains a promising frontier.
As quantum computers continue to evolve with advancements in speed, scalability, fault tolerance, and accessibility, current limitations are gradually diminishing. 
Our research marks an initial step in understanding the practical applications and boundaries of quantum computing in astrophysical studies, potentially inspiring innovative approaches in the future. 
By transforming classical frameworks, quantum computing would shed light on the hidden dynamics of the universe.

Separately from the quantum computing aspect, we lastly comment on the possible strengths and limitations of LSTM models compared to classical regression models and  future works. 
While the LSTM models are often considered black boxes and it is challenging to fully understand their internal processes, 
the models may offer an advantage in the context of our study. They are designed to learn temporal dependencies directly from data, making them effective for identifying deviations from normal variability in complex time series. In contrast, classical regression models, which are more interpretable, rely on the assumption of a specific functional form. This assumption may limit their ability to represent the full diversity of normal behaviors. Also, such models may inadvertently fit transient events too well, thereby failing to recognize them as anomalies. Since our goal is to detect unexpected deviations, not to reproduce them, this may pose a limitation for regression-based approaches. While interpretability is a valid concern, we think that the detection capability and adaptability of LSTM-based models would be essential in our work. That said, we recognize the importance of model transparency, and as part of future work, it would be worthwhile to apply explainable AI techniques (e.g., SHAP values, attention mechanisms) to gain insight into the decision-making process of models. In addition, it is worthwhile, as a future work, to explore other modeling frameworks that naturally incorporate uncertainty estimation.
One such candidate is Gaussian Process Regression, which provides both predictions and associated uncertainties in time-series modeling.\\

We are grateful to the anonymous referee for their insightful comments, which helped to improve the paper.
We thank UTokyo Quantum Initiative for their support, which provided us with the opportunity to execute quantum circuits on quantum computers and allowed us to gain valuable insights by visiting an IBM quantum computer. T.K. and S.Y. are grateful for the support from the RIKEN Special Postdoctoral Researcher Program. T.K., S.N., and S.Y. are supported by JSPS KAKENHI grant numbers 23K13153/24K00673, JP25H00675, and 23K13154, respectively. 
S.N. is supported by JST ASPIRE Program "RIKEN-Berkeley mathematical quantum science initiative".

This work was supported by the iTHEMS, RIKEN TRIP initiative (RIKEN Quantum) and UTokyo Quantum Initiative.

A part of this work was conducted with the supercomputer HOKUSAI BigWaterfall2, maintained by RIKEN. 

This research used data obtained from the 4XMM XMM-Newton serendipitous source catalogue compiled by the XMM-Newton Survey Science Centre consortium.

This research made use of the cross-match service provided by CDS, Strasbourg. 

This research has made use of the SIMBAD database, operated at CDS, Strasbourg, France. 

The national facility capability for SkyMapper has been funded through ARC LIEF grant LE130100104 from the Australian Research Council, awarded to the University of Sydney, the Australian National University, Swinburne University of Technology, the University of Queensland, the University of Western Australia, the University of Melbourne, Curtin University of Technology, Monash University and the Australian Astronomical Observatory. SkyMapper is owned and operated by The Australian National University's Research School of Astronomy and Astrophysics. The survey data were processed and provided by the SkyMapper Team at ANU. The SkyMapper node of the All-Sky Virtual Observatory (ASVO) is hosted at the National Computational Infrastructure (NCI). Development and support of the SkyMapper node of the ASVO has been funded in part by Astronomy Australia Limited (AAL) and the Australian Government through the Commonwealth's Education Investment Fund (EIF) and National Collaborative Research Infrastructure Strategy (NCRIS), particularly the National eResearch Collaboration Tools and Resources (NeCTAR) and the Australian National Data Service Projects (ANDS).

The Pan-STARRS1 Surveys (PS1) and the PS1 public science archive have been made possible through contributions by the Institute for Astronomy, the University of Hawaii, the Pan-STARRS Project Office, the Max-Planck Society and its participating institutes, the Max Planck Institute for Astronomy, Heidelberg and the Max Planck Institute for Extraterrestrial Physics, Garching, The Johns Hopkins University, Durham University, the University of Edinburgh, the Queen's University Belfast, the Harvard-Smithsonian Center for Astrophysics, the Las Cumbres Observatory Global Telescope Network Incorporated, the National Central University of Taiwan, the Space Telescope Science Institute, the National Aeronautics and Space Administration under Grant No. NNX08AR22G issued through the Planetary Science Division of the NASA Science Mission Directorate, the National Science Foundation Grant No. AST-1238877, the University of Maryland, Eotvos Lorand University (ELTE), the Los Alamos National Laboratory, and the Gordon and Betty Moore Foundation.

This publication makes use of data products from the Wide-field Infrared Survey Explorer, which is a joint project of the University of California, Los Angeles, and the Jet Propulsion Laboratory/California Institute of Technology, and NEOWISE, which is a project of the Jet Propulsion Laboratory/California Institute of Technology. WISE and NEOWISE are funded by the National Aeronautics and Space Administration."

\vspace{5mm}
\facilities{XMM (pn)}

\software{
NumPy \citep{Har20}, 
SciPy \citep{Vir20},
Pandas \citep{Tea22}, 
Matplotlib \citep{Hun07}, 
Astropy \citep{Ast13,Ast18}, 
PyTorch \citep{Paszke2019arXiv191201703P}, 
}

\appendix

\section{Brief Introduction to Qubits and Quantum Gates}\label{app:quantum}

\begin{figure*}
 \begin{center}
  \includegraphics[width=16cm]{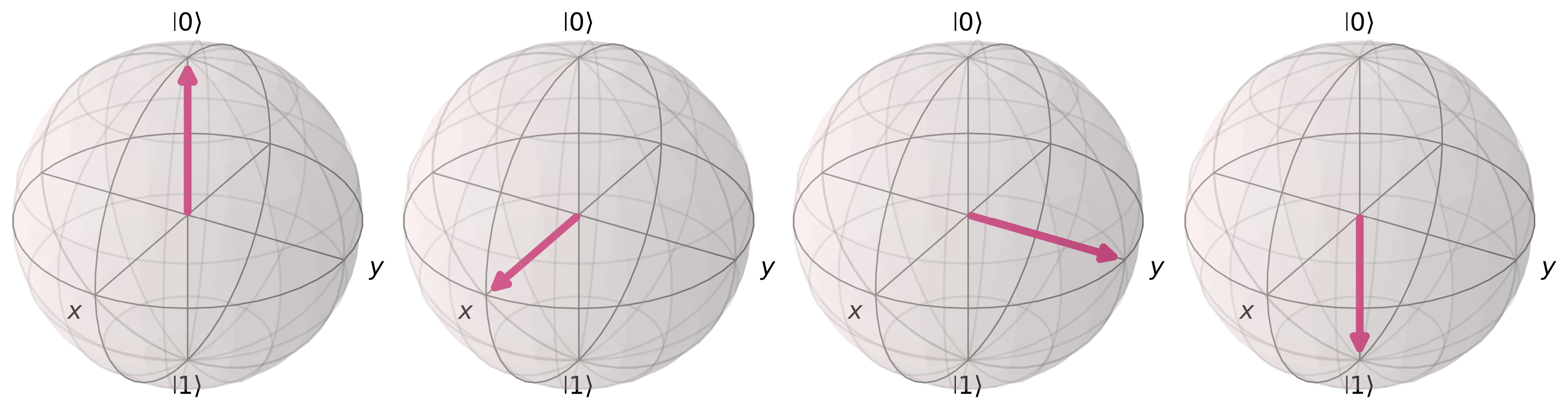} 
 \end{center}
 \caption{
    The leftmost sphere indicates the $|0\rangle$ state of a qubit in the Bloch sphere. 
    The arrows in the other spheres from left to right show the results after applying the $R_{\rm Y}(\pi/2)$,
    $R_{\rm Z}(\pi/2)$, and $R_{\rm X}(-\pi/2)$ gates to the qubit of $|0\rangle$ sequentially. 
    The final state of the qubit shown in the rightmost sphere corresponds to $|1\rangle$. 
 }
 \label{fig:sphere2} 
\end{figure*}

The qubit is the fundamental unit of information in quantum computing. This is analogous to the classical bit; however, unlike the bit, which can be either 0 or 1, the qubit can exist in a superposition of both states. The state of a qubit can be represented by the following equation: 
\begin{equation}
|\psi\rangle = \alpha |0\rangle + \beta |1\rangle, 
\end{equation}
where $\alpha$ and $\beta$ are complex numbers satisfying \(|\alpha|^2 + |\beta|^2 = 1\). 
In vector notation, the above equation can be expressed as 
\begin{equation}
    \begin{pmatrix} \alpha \\ \beta \end{pmatrix} = 
    \alpha 
    \begin{pmatrix}  1 \\ 0 \end{pmatrix} + 
    \beta 
    \begin{pmatrix} 0 \\ 1 \end{pmatrix}. 
\end{equation}
Furthermore, by considering a spherical coordinate, where $\theta$ and $\phi$ denote the polar and azimuthal angles, respectively, a different form of the equation is 
\begin{equation}
    |\psi\rangle = e^{i\delta} \left(
        \cos\left(\frac{\theta}{2}\right) |0\rangle + e^{i\phi} \sin\left(\frac{\theta}{2}\right) |1\rangle  \right).
\end{equation}
The term $e^{i\delta}$ is called the global phase, and 
may be ignored because it does not affect the probability of observing any state. 
As inferred from the equation, 
a state can be realized as a vector pointing to a point on a sphere, which is usually called the Bloch sphere. 
We note that the states of $|0\rangle$ and $|1\rangle$ are conventionally set so that the vectors point toward the north and south poles, respectively. 
The representation of the state in the Bloch sphere is convenient, because the states of the qubit and quantum operations can be understood as rotations in the sphere. It is noteworthy that the global phase does not appear in the sphere.

The states of qubits can be changed by using quantum gates. 
The fundamental gates are the Pauli gates of Pauli-$X$, Pauli-$Y$, and Pauli-$Z$. 
The Pauli-$X$ gate is analogous to the classical NOT gate, and 
swaps the amplitudes of the $|0\rangle$ and $|1\rangle$ states. 
In the matrix notation, the Pauli-$X$ is expressed as 
\begin{equation}
    X = \begin{pmatrix} 0 & 1 \\ 1 & 0 \end{pmatrix}    
\end{equation}
The applications of the Pauli-X gate to the basis states result in 
\begin{equation}
    X|0\rangle = |1\rangle, \quad X|1\rangle = |0\rangle
\end{equation}
As inferred from these transformations, the Pauli-X gate corresponds to a rotation around the $X$-axis by $\pi$ radians.  
Similarly, the Pauli-$Y$ gate flips the state of the qubit but introduces a phase factor of $i$ (the imaginary unit). The matrix representation is as follows: 
\begin{equation}
    Y = \begin{pmatrix} 0 & -i \\ i & 0 \end{pmatrix}    
\end{equation}
Accordingly, the Pauli-$Y$ gate transforms the $|0\rangle$ and $|1\rangle$ states as 
\begin{equation}
    Y|0\rangle = i|1\rangle, \quad Y|1\rangle = -i|0\rangle. 
\end{equation}
Like the Pauli-$X$ gate, a rotation around the $Y$-axis by $\pi$ radians can be realized by the Pauli $Y$-gate. 
Finally, the Pauli-$Z$ gate is a phase-flip gate that leaves the $|0\rangle$ state unchanged but flips the phase of the $|1\rangle$ state:
\begin{equation}
    Z = \begin{pmatrix} 1 & 0 \\ 0 & -1 \end{pmatrix}    
\end{equation}
The transformations performed by the Pauli-Z gate are as follows. 
\begin{equation}
Z|0\rangle = |0\rangle, \quad Z|1\rangle = -|1\rangle
\end{equation}
The Pauli-$Z$ gate corresponds to the $\pi$-radian rotation around the $Z$-axis. 

By introducing rotation gates, one can better understand the operation of the Pauli gates. 
The gate that rotates the qubit around the $X$-axis by an angle of $\theta$ is defined as 
\begin{equation}
    R_X(\theta) = e^{-i \frac{\theta}{2} X} = \begin{pmatrix} \cos\left(\frac{\theta}{2}\right) & -i\sin\left(\frac{\theta}{2}\right) \\ -i\sin\left(\frac{\theta}{2}\right) & \cos\left(\frac{\theta}{2}\right) \end{pmatrix}        
\end{equation}
Similarly, $R_Y$($\theta$) and $R_Z$($\theta$) gates are defined as follows: 
\begin{equation}
    R_Y(\theta) = e^{-i \frac{\theta}{2} Y} = \begin{pmatrix} \cos\left(\frac{\theta}{2}\right) & -\sin\left(\frac{\theta}{2}\right) \\ \sin\left(\frac{\theta}{2}\right) & \cos\left(\frac{\theta}{2}\right) \end{pmatrix}
\end{equation}
and 
\begin{equation}
    R_Z(\theta) = e^{-i \frac{\theta}{2} Z} = \begin{pmatrix} e^{-i\frac{\theta}{2}} & 0 \\ 0 & e^{i\frac{\theta}{2}} \end{pmatrix}, 
\end{equation}
respectively. 
An example of applying rotation gates is shown in Figure~\ref{fig:sphere2}.
For a rotation angle of $\pi$ radians, the $R_X$, $R_Y$, and $R_Z$ gates can be simplified as follows: 
\begin{equation}
    R_X(\pi) = \begin{pmatrix}  0 & -i \\ -i & 0 \end{pmatrix} = -i X 
\end{equation}
\begin{equation}
    R_Y(\pi) = \begin{pmatrix}  0 & -1 \\ 1 &  0 \end{pmatrix} = -i Y
\end{equation}
\begin{equation}
    R_Z(\pi) = \begin{pmatrix}  -i & 0 \\ 0 &  i \end{pmatrix} = -i Z
\end{equation}
Given that the coefficient $-i$, which can be interpreted as the global phase, does not affect any observable, it can be realized that each Pauli gate is equivalent to the rotation around the specific axis by $\pi$ radians. 

In addition to the fundamental Pauli gates and more flexible rotation gates, it is important to introduce the two gates that we used to construct the QVCs: the Hadamard (H) gate and the Controlled-NOT (CNOT) gate. 
The Hadamard gate effectively places a qubit in a superposition state. 
It maps the state $|0\rangle$ to $\frac{1}{\sqrt{2}}(|0\rangle + |1\rangle)$ and the state $|1\rangle$ to $\frac{1}{\sqrt{2}}(|0\rangle - |1\rangle)$. The matrix notation is 
\begin{equation}
    H = \frac{1}{\sqrt{2}} \begin{pmatrix} 1 & 1 \\ 1 & -1 \end{pmatrix}. 
\end{equation}
Also, we note that if one successively applies the Hadamard gate to a qubit twice, the resultant state is the initial one (i.e., $HH$ is equivalent to the identity matrix).  
The CNOT gate plays a crucial role in inducing the entanglement of two qubits: 
a unique quantum phenomenon, in which the states of two qubits are linked. 
In quantum computing, a controlled gate generally refers to a gate in which the states of qubits (the control qubit) determines whether a single-qubit unitary operation is applied to a target qubit. The CNOT gate is represented by the following matrix:
\begin{equation}
    \text{CNOT} = \begin{pmatrix}
    1 & 0 & 0 & 0 \\
    0 & 1 & 0 & 0 \\
    0 & 0 & 0 & 1 \\
    0 & 0 & 1 & 0
    \end{pmatrix}
\end{equation}
Given that two qubits in the basis state $|00\rangle$ can be expressed as 
\begin{equation}
    |00\rangle = \begin{pmatrix}
    1 \\
    0 \\
    1 \\
    0 
    \end{pmatrix}. 
\end{equation}
This also applies to the other two-qubit states. 
The results of applying a single CNOT gate to the basis states are as follows: 
\begin{equation}
    \text{CNOT}|00\rangle = |00\rangle \nonumber 
\end{equation}    
\begin{equation}
    \text{CNOT}|01\rangle = |01\rangle \nonumber 
\end{equation}    
\begin{equation}    
    \text{CNOT}|10\rangle = |11\rangle \nonumber  
\end{equation}        
\begin{equation}\label{eqn:cnot_trans}
    \text{CNOT}|11\rangle = |10\rangle.
\end{equation}
These transformations can be described as follows:
\begin{itemize}
    \item If the control qubit (left one in each ket vector of \ref{eqn:cnot_trans}) is in the state $|0\rangle$, the target qubit remains unchanged.
    \item If the control qubit is in the state $|1\rangle$, the target qubit is flipped (i.e., a NOT operation is applied to it).
\end{itemize}

\section{Current Situation of Quantum Computing}\label{app:quantum_current} 

The main bottleneck in building quantum computers is noise. Quantum states are highly fragile and susceptible to decoherence, which causes them to lose their quantumness, i.e., quantum coherence. 
The advantages of quantum computing often stem from its massive parallelism and the use of quantum interference. However, when quantum devices are noisy and decoherence occurs, these crucial properties are compromised, significantly diminishing the utility of quantum computations.

To address this challenge, quantum error correction (QEC) and quantum fault tolerance (QFT) are critical for scalable and reliable quantum computing. The QEC employs quantum error-correcting codes, which use multiple physical qubits to encode a single logical qubit. By distributing quantum information across many physical qubits, local errors can be corrected without destroying the logical information.

A foundational result in this area is the threshold theorem, which states that if the noise level of individual qubits is below a certain threshold, logical errors can be exponentially suppressed by increasing the code distance $d$. The code distance represents the minimum number of errors required to cause a logical error. Equivalently, it is the minimum weight of a Pauli operator that can implement a logical operation. The logical error rate $\epsilon_d$ scales as:
\begin{eqnarray}
\left({p\over p_{\text{thr}}}\right)^{{d+1\over 2}}
\end{eqnarray}
where $p$ is the error probability of an individual physical qubit, and $p_{\text{thr}}$ is the threshold error rate. In 2024, Google AI demonstrated error suppression for code distances d = 3, 5, and 7 using the surface code, a well-known quantum error-correcting code. They reported a suppression rate of $\Lambda = \epsilon_d / \epsilon_{d+2}$ greater than two, confirming the exponential scaling predicted by theory.

The fact that a logical qubit consists of many physical qubits adds complexity to logical gate operations. Logical gates must be designed to prevent error propagation within a logical qubit. This is particularly challenging for two-qubit gates, such as the CNOT gate, which can propagate errors between qubits. For instance, an $X$ (bit-flip) error on a control qubit, when followed by a CNOT gate, can lead to $X$ errors on both the control and target qubits.

To address this, transversal gates are employed. A transversal gate implements a logical operation $\bar{A}$ using a tensor product of physical qubit operations, $\bigotimes_{i=1}^{d} A_i$, where $A_i$ acts on physical qubit $i$. Transversal gates inherently prevent error propagation. In December 2023, QuEra et al. achieved simultaneous transversal logical gate operations with (pseudo) threshold fidelities (\cite{2024Natur.626...58B}), representing a significant milestone in scalable quantum computing.

While fault tolerance is critical for the long-term success of quantum computing, active research has been conducted to solve classically intractable problems using noisy quantum computers without full error correction. These approaches operate within the limitations on the number of qubits and gate depths, restricting the complexity of the computations that can be performed reliably.

A prominent technique used for noisy quantum devices is error mitigation (\cite{2017PhRvL.119r0509T, 2016arXiv161109301L}). Unlike QEC, which corrects errors during computation, error mitigation reduces noise effects in post-processing by learning noise properties through experiments. Recently, error mitigation techniques have been applied to the quantum simulation of a two-dimensional transverse field model using a 127-qubit processor (\cite{Kim2023}). Remarkably, this simulation achieved accuracies comparable to or surpassing those achieved by cutting-edge classical algorithms, demonstrating the potential of quantum devices to produce meaningful results even in the NISQ era.

\section{Code for Running Quantum Circuits} \label{app:code}

We introduce a code to simulate quantum circuits using Qiskit, the official 
software for running quantum circuits on IBM quantum computers. 
By slightly modifying the code, one can execute the circuit on real devices. 
First, the setup consists of three parts: defining the quantum circuit executor (backend), 
retrieving a noise model (e.g., "ibm\_brisbane") if noise is to be included in simulations, and 
setting it when defining the executor. 

\begin{lstlisting}[style=python, caption={Setting backend}]
from qiskit_aer import AerSimulator
from qiskit_ibm_runtime import QiskitRuntimeService
from qiskit_aer.noise import NoiseModel 

service     = QiskitRuntimeService()
# set the name of an available quantum computer from which a noise model is retrieved 
noise_model_qc = "ibm_brisbane" 
backend_for_noise_model = service.backend(noise_model_qc)
noise_model = NoiseModel.from_backend(backend_for_noise_model)
backend     = AerSimulator(noise_model=noise_model)
\end{lstlisting}
Next, one constructs a feature map and an ansatz, and
performs qubit measurements. 
The code below uses the feature map and ansatz reported by \citet{Chen2020arXiv200901783Y} and measures each qubit for Pauli-$Z$; i.e., one observes $|0\rangle$ or $|1\rangle$: 
\begin{lstlisting}[style=python, caption={ Constructing layers and defining measurements}]
from qiskit import QuantumCircuit
from qiskit.circuit import ParameterVector
from qiskit.quantum_info import SparsePauliOp
import matplotlib.pyplot as plt 

# set the number of qubits 
num_qubits = 4 

# Feature map 
class Chen20_fm(QuantumCircuit):
    def __init__(self, num_qubits: int, name="encoding-layer"):
        super().__init__(num_qubits, name=name)

        x_input = ParameterVector('x', length=num_qubits)

        for i in range(num_qubits):
            self.h(i)
            self.ry((x_input[i]).arctan(), i)
            self.rz((x_input[i]**2).arctan(), i) 

# Ansatz 
class Chen20_ansatz(QuantumCircuit):
    def __init__(self, num_qubits, ansatz_reps: int, 
                name="ansatz-layer"): 
                # ansatz_reps indicates how many times the ansatz is repeated
        super().__init__(num_qubits, ansatz_reps, name=name)

        alpha = ParameterVector('alpha', 
                                length=num_qubits*ansatz_reps)
        beta  = ParameterVector('beta', 
                                length=num_qubits*ansatz_reps)
        gamma = ParameterVector('gamma', 
                                length=num_qubits*ansatz_reps)

        for rep in range(ansatz_reps):
            for i in range(1, num_qubits-1): 
                for j in range(num_qubits):
                    if j + i < num_qubits:
                        self.cx(j, j+i)
                    else:
                        self.cx(j, j+i-num_qubits)  

            for i in range(num_qubits):
                self.rx(alpha[i+num_qubits*rep], i)
                self.ry(beta[i+num_qubits*rep], i)
                self.rz(gamma[i+num_qubits*rep], i) 

# Define how to measure qubits 
obsv = [SparsePauliOp.from_list(
            [("I"*(num_qubits-i-1) + "Z" + "I"*i,1)]
            ) for i in range(num_qubits)]

# combine the encoding and ansatz layers and the measurement 
ansatz_reps = 3
QVC         = QuantumCircuit(num_qubits) 
feature_map = Chen20_fm(num_qubits)
ansatz      = Chen20_ansatz(num_qubits, ansatz_reps)
QVC.compose(feature_map, inplace=True)
QVC.compose(ansatz, inplace=True)

# save the circuit 
QVC.draw(output="mpl", fold=20)
plt.savefig("circuit.pdf") 
plt.close()
\end{lstlisting}
Finally, using a neural network module of \texttt{EstimatorQNN}, 
which accepts a parametrized quantum circuit and an optional quantum mechanical observable, one can obtain expected values (see the official document for more detail). 
\begin{lstlisting}[style=python, caption={Execution of the quantum circuits}]
# Modify the prepared circuits so that a quantum computer can understand them 
from qiskit.transpiler.preset_passmanagers import generate_preset_pass_manager
from qiskit_machine_learning.neural_networks import EstimatorQNN

pm = generate_preset_pass_manager(backend=backend, optimization_level=1) 
isa_QVC = pm.run(QVC) 
# Note that ISA is Instruction Set Architecture 

# Contruct a neural network 
qnn = EstimatorQNN(
        circuit         = isa_QVC,
        observables     = obsv,
        input_params    = feature_map.parameters,
        weight_params   = ansatz.parameters,
)

from qiskit_machine_learning.connectors import TorchConnector
import torch 

QVC    = TorchConnector(qnn) 
inputs = torch.tensor([0, 1, 2, 3])
expected_values = QVC(inputs)
print(expected_values)
\end{lstlisting}

\section{More Detailed Introduction to Effective Dimension}\label{app:eff}

We here present a more detailed description on the effective dimension than that in the main text. However, please refer to \citet{Abbas2020arXiv201100027A} and \citet{Berezniuk2020arXiv200110872B} for  complete descriptions. 

The complexity of a statistical model $\mathcal{M} = \{P(\cdot|\theta) : \theta \in \Theta\}$, where $\theta$ is a $d$-dimensional parameter ($\theta \in \Theta$ and $\Theta \subset \mathbb{R}^d$), or its ability to fit a large amount of data, is quantified as follows:
\begin{equation}
    \operatorname{COMP}_{n}(\mathcal{M}):= \log \left(\sum_{x^{n} \in \mathcal{X}^{n}} P\left(x^{n} \mid \hat{\theta}\left(x^{n}\right)\right)\right)
\end{equation}
Here, $x^n$ represents a dataset consisting of $n$ data $(x_1, x_2, \ldots, x_n)$ 
and 
$\hat{\theta}$ denotes the set of parameters that maximizes the probability of $x^n$. 
From this definition, it is clear that as the model fits more data, the complexity increases. Under certain conditions, the above expression can be rewritten as
\begin{equation}
    \operatorname{COMP}_{n}(\mathcal{M})=\frac{d}{2} \log \frac{n}{2 \pi}+\log \left(\int_{\Theta} \sqrt{\operatorname{det} \boldsymbol{F}(\theta)} d \theta\right)+o(1)
\end{equation}
Here, $\boldsymbol{F}(\theta)$ is the Fisher information matrix, which is used as the metric for the manifold of ($\mathcal{M}$, $\Theta$), and is defined as
\begin{equation}
    F_{i j}(\theta):=\mathbb{E}\left[-\frac{\partial^{2}}{\partial \theta_{i} \partial \theta_{j}} \log P(X \mid \theta)\right], \quad i, j \in\{1, \ldots, d\}
\end{equation}
By transforming this into $g_{ij} = (n/2\pi) F_{ij}$, the complexity can be expressed as
\begin{equation}
    \operatorname{COMP}_{n}(\mathcal{M}) = \log \left(\int_{\Theta} \sqrt{\det g(\theta)} \, d\theta \right)
\end{equation}
This indicates that the complexity corresponds to the volume of the manifold mapped by $(g_{ij}, \Theta)$, or $(F_{ij}, \sqrt{n/2\pi} \Theta)$. 
The parameters that contribute to this volume are crucial and define the effective dimension. For a simpler understanding, consider a diagonalized metric, where the lengths in the redefined parameter directions, or the eigenvalues of the eigenvectors, are obtained. The larger a length, or an eigenvalue ($s^2_i$), the more largely the parameter contributes to the volume, or the complexity (e.g., $dl_i = s_i d\theta_i$).
Such a largely contributing parameter can be considered an effective parameter.

The effective dimension introduced above is determined using the box-counting method for the space, or the manifold, created by the Fisher information matrix and the expanded parameter space: i.e., $F_{ij}$ and $\sqrt{n/2\pi} \Theta$. This box-counting method calculates the dimension by covering an object with boxes of small edge length ($\epsilon$) in the same dimension, and then determines the dimension as 
\begin{equation}
    \lim_{\epsilon \to 0} \frac{\log N(\epsilon)}{|\log \epsilon|}.
\end{equation}
This matches our empirical understanding of these dimensions. For example, by considering a three-dimensional cube and gradually reducing the size of the boxes, the number of boxes required increases along all axes. From this definition, the dimension of 3 is obtained as $N \propto \epsilon^{-3} \to$ dim $ = \lim_{\epsilon \to 0} \log(\epsilon^{-3}) / \log \epsilon^{-1} = 3$. In addition, if one considers a thin sheet with no thickness, the number of boxes required does not increase in the thickness direction, leading to a two-dimensional result: i.e., $N \propto \epsilon^{-2}$, dim $= \lim_{\epsilon \to 0} \log(\epsilon^{-2}) / \log \epsilon^{-1} = 2$.

Using the box-counting method described above, we calculate the effective dimension of the manifold. According to the method in \citet{Abbas2020arXiv201100027A}, for a fixed hypercube size, as the number of data samples increases, the parameter space $\sqrt{n/2\pi} \Theta$ expands, thus improving the corresponding resolution in proportion to $1/\sqrt{n}$. That is, when the size of each hypercube side is $\epsilon$, and the number of hypercubes needed to cover the parameter space $\sqrt{n/2\pi} \Theta$ is $N(\epsilon)$, the specific calculation is written as follows:
\begin{equation}
    \dim = \lim_{\epsilon \to 0}  \frac{\log N(\epsilon)}{|\log(\epsilon)|} = \lim_{\epsilon \to 0} \frac{\log N(\epsilon)}{|\log(\sqrt{1/n})|}
\end{equation}
Considering a simple case in which the Fisher matrix is diagonalized as $\operatorname{diag}(s^2_1, s^2_2, \ldots, s^2_d)$, the size 1 of the hypercube side corresponds to $1/s_i$ in the parameter space through the metric (i.e., $ds^2 = g_{ij}d\theta_i d\theta_j$, and therefore $d\theta_i = 1/s_i$). The hypercube can be expressed as
\begin{equation}
    \left[0, \frac{1}{s_{1}}\right] \times \ldots \times \left[0, \frac{1}{s_{d}}\right], \quad s_{i}>0.
\end{equation}
The parameter space $\sqrt{n/2\pi} \Theta$ is $[0, \sqrt{n/2\pi}]^d$ by assuming $\Theta = [0, 1]^d$, and according to the definition of the box-counting method, the number of hypercubes needed to cover the parameter space is expressed as
\begin{equation}
    \prod_{i=1}^{d}
    \lceil\sqrt{\frac{n}{2 \pi}} s_{i}\rceil \approx \sqrt{\prod_{i=1}^{d}\left(1+\frac{n}{2 \pi} s_{i}^{2}\right)}=\sqrt{\operatorname{det}\left(\operatorname{Id}_{d}+\frac{n}{2 \pi} \boldsymbol{F}\right)}
\end{equation}
This equation indicates that the parameters contributing to the dimensionality increase are those with eigenvalues ($s_i^2$) such that $s^2_i > 2\pi/n$. Conversely, if the eigenvalue is nearly zero, the contribution remains small, even with a large $n$, indicating that the parameter does not contribute to the dimension. Based on the above discussion, the effective dimension heuristically introduced by \citet{Abbas2020arXiv201100027A} is as follows:
\begin{equation}
    \operatorname{dim}_{\mathrm{eff}, n}(\mathcal{M}):=2 \frac{\log \left(\frac{1}{V_{\Theta}} \int_{\Theta} \sqrt{\operatorname{det}\left(\operatorname{Id}_{d}+\frac{n}{2 \pi} \hat{\boldsymbol{F}}(\theta)\right)} d \theta\right)}{\log \frac{n}{2 \pi}} 
\end{equation}
where
\begin{equation}
    \hat{F}_{i j}:=d \frac{V_{\Theta}}{\int_{\Theta} \operatorname{tr} \boldsymbol{F}(\theta) d \theta} F_{i j},
\end{equation}
and 
\begin{equation}
    V_{\Theta} := \int_{\Theta} d\theta.
\end{equation}

\section{Predictions by LSTM Models}\label{app:demos}

\begin{figure*}
 \begin{center}
  \includegraphics[width=17.5cm]{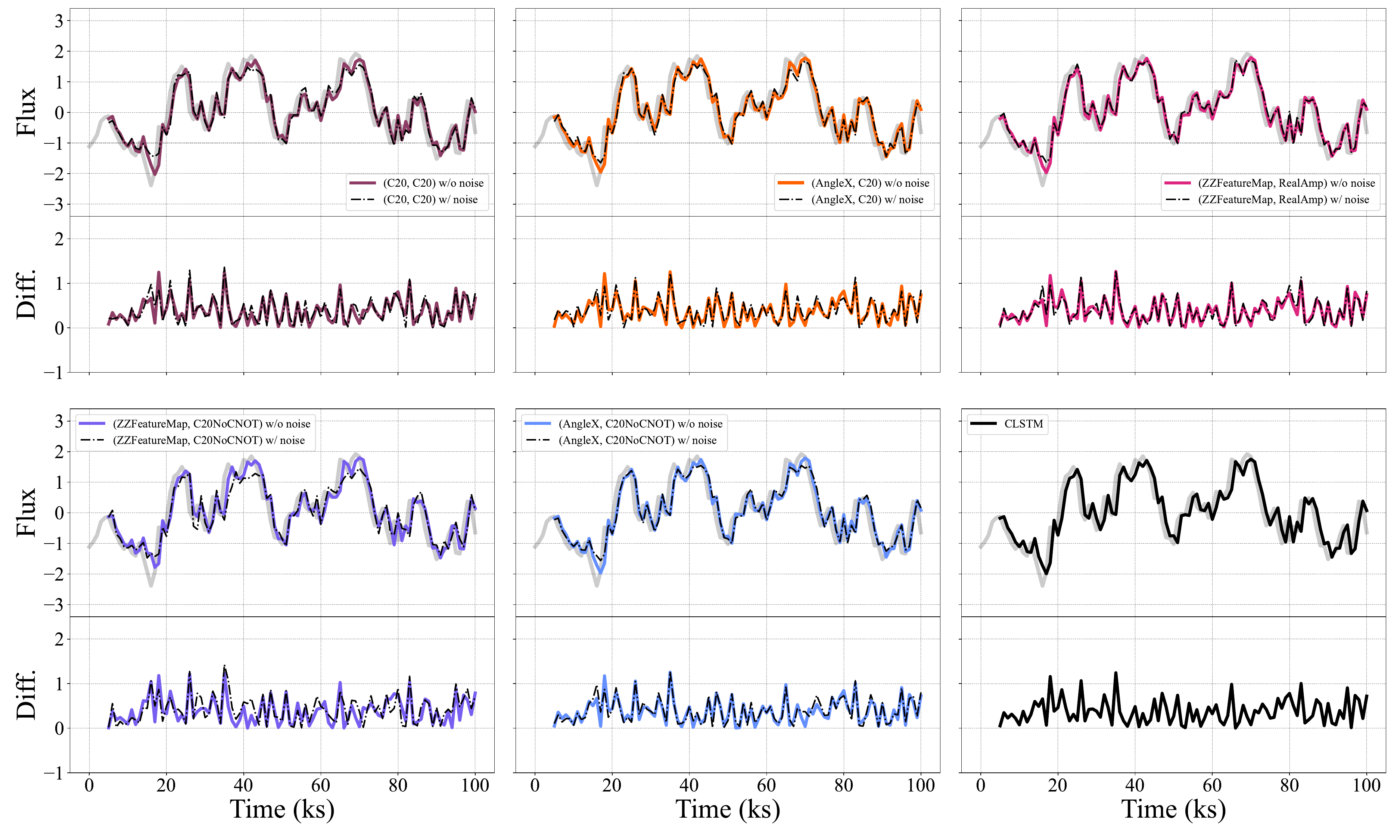}   
  \includegraphics[width=17.5cm]{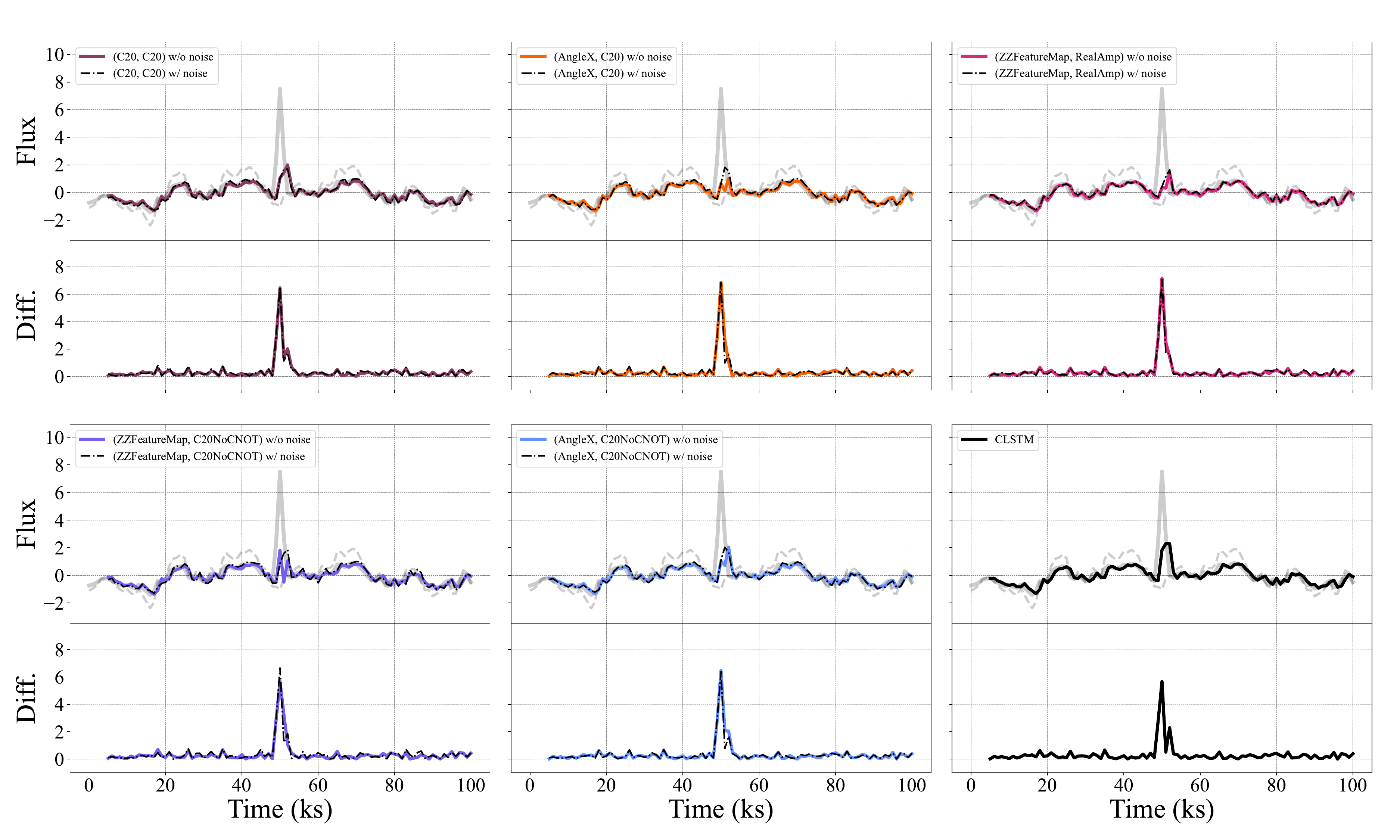}     
 \end{center}
 \caption{
    Top two rows: the prediction results for the light curve simulated with $\sigma_{\rm NXS}^{2} = 10^{-3}$ and the PSD index of $-2.0$ using 
    the CLSTM and five QLSTM models excluding the fiducial QLSTM model. 
    The fluxes are represented in arbitrary units. 
    Bottom two rows: the results for the anomaly light curve simulated with the flare amplitude of 1.5. 
 }
 \label{fig:demo_var_models} 
\end{figure*}

In Figure~\ref{fig:demo_var_models}, 
we present the predictions for the unseen normal data in Set C (i.e., nine representative AGN-like light curves and those with added anomaly flares) made using the CLSTM and 
five QLSTM models excluding the fiducial QLSTM model using (ZZFeatureMap, C20). 
To avoid redundant figures, only the results for the light curve generated with $\sigma_{\rm NXS}^{2} = 10^{-3}$ and a PSD index of $-2.0$ are displayed.
The results for both the noise-free and noisy conditions are shown.

\section{Catalog of XMM-Newton detections with anomalies}\label{app:catalog}

\begin{figure*}
 \begin{center}
  \includegraphics[width=3.9cm]{anomaly_fig_dir/J001427p0-303546_P0743850101PNS003SRCTSR802A_lc_1000_stand_wpred.pdf}
  \includegraphics[width=3.9cm]{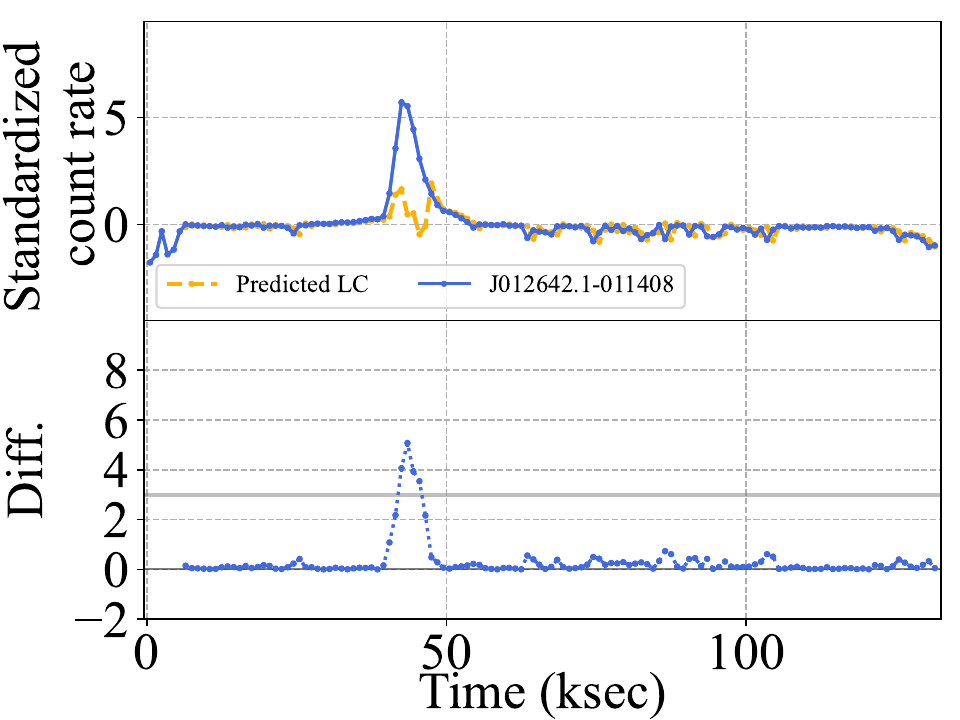}
  \includegraphics[width=3.9cm]{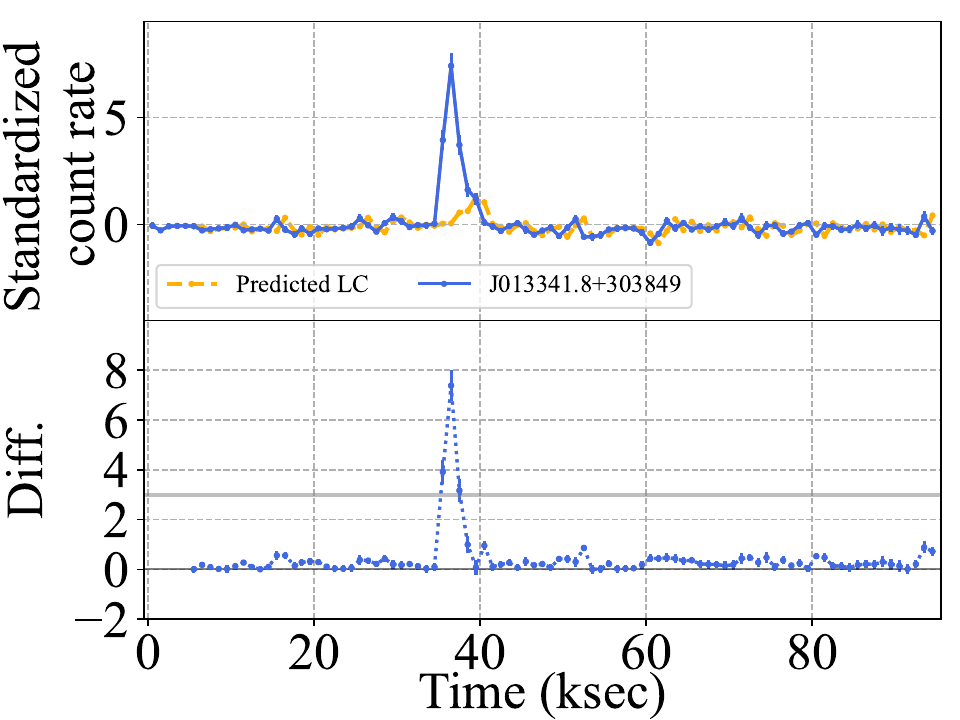}
  \includegraphics[width=3.9cm]{anomaly_fig_dir/J015709p1+373739_P0149780101PNS003SRCTSR8015_lc_1000_stand_wpred.pdf}
  \includegraphics[width=3.9cm]{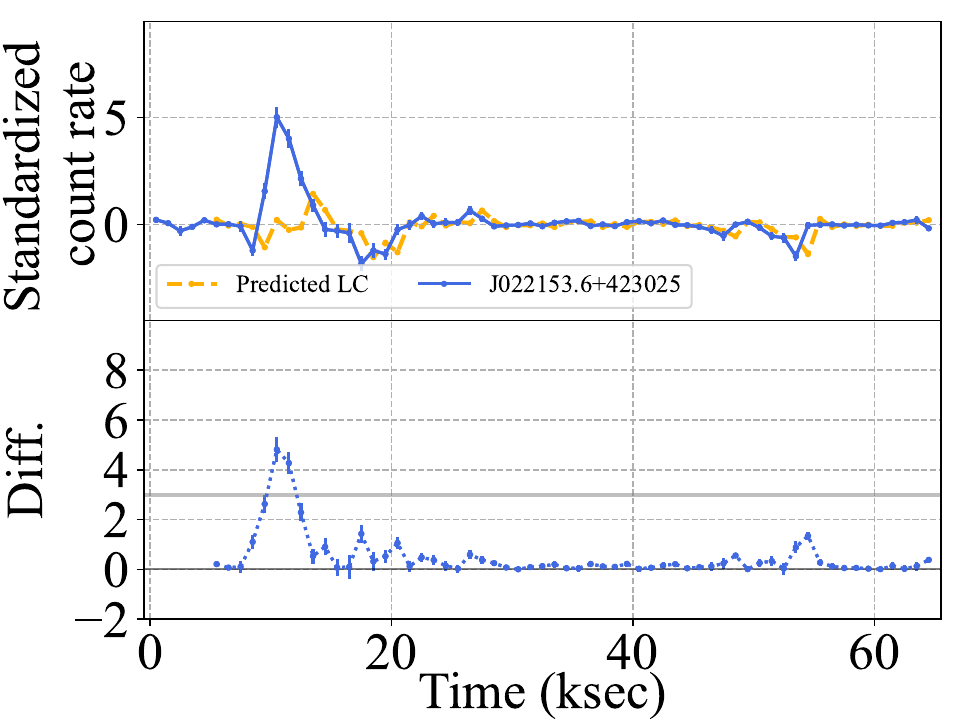}
  \includegraphics[width=3.9cm]{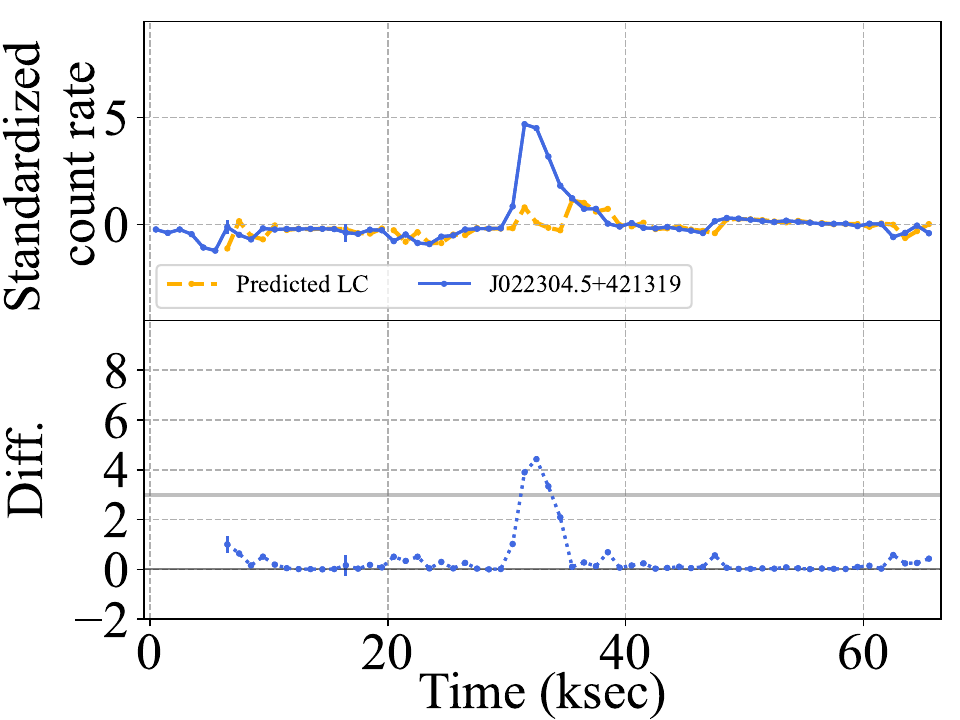}
  \includegraphics[width=3.9cm]{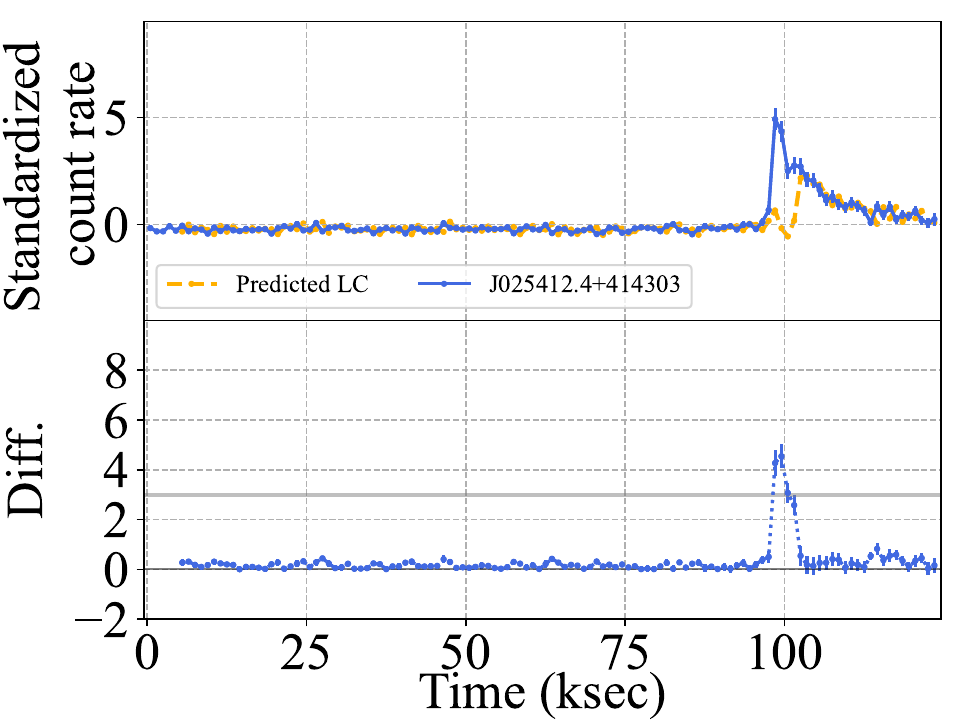}
  \includegraphics[width=3.9cm]{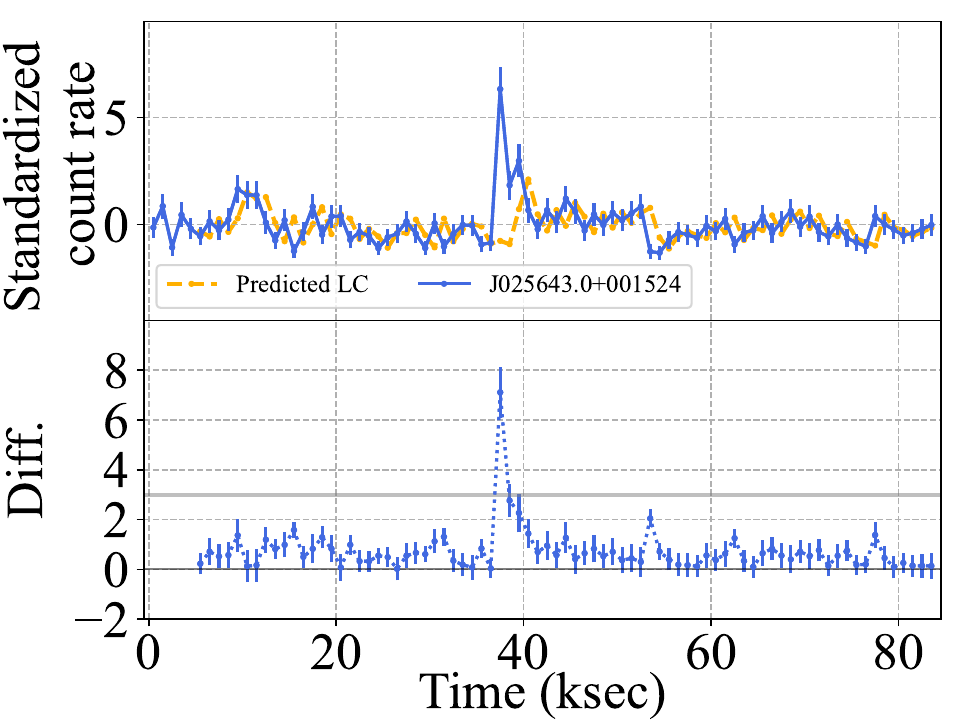}
  \includegraphics[width=3.9cm]{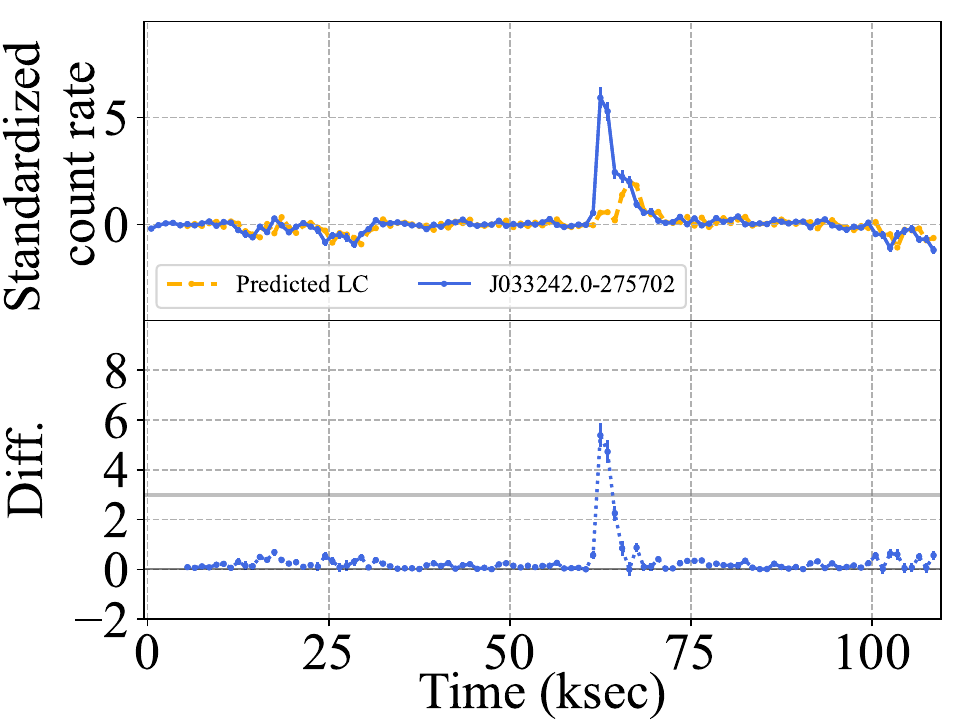}
  \includegraphics[width=3.9cm]{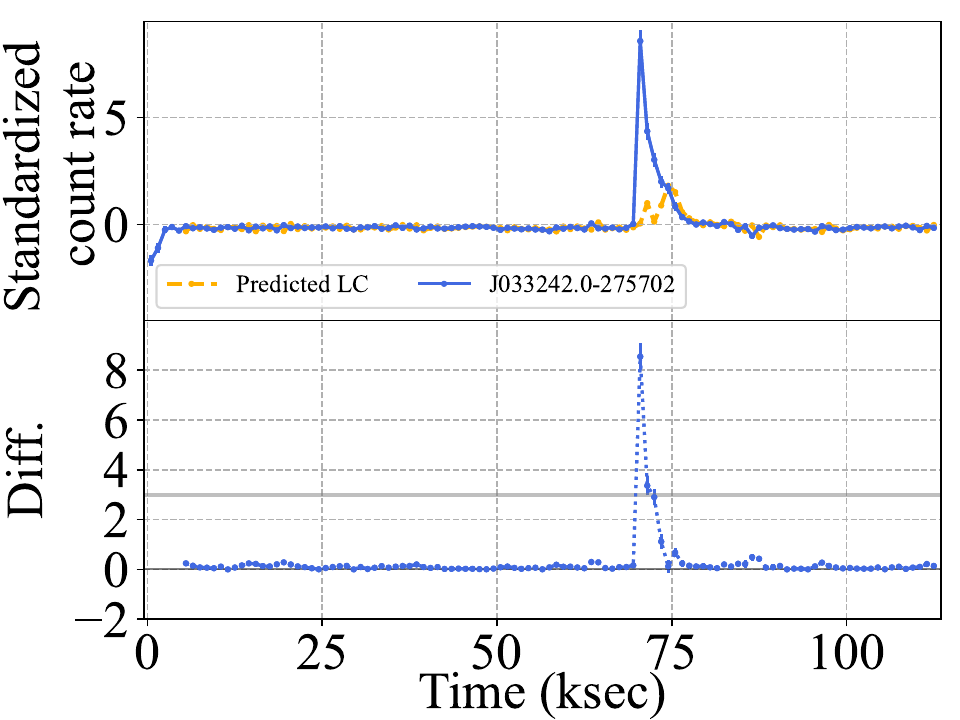}
  \includegraphics[width=3.9cm]{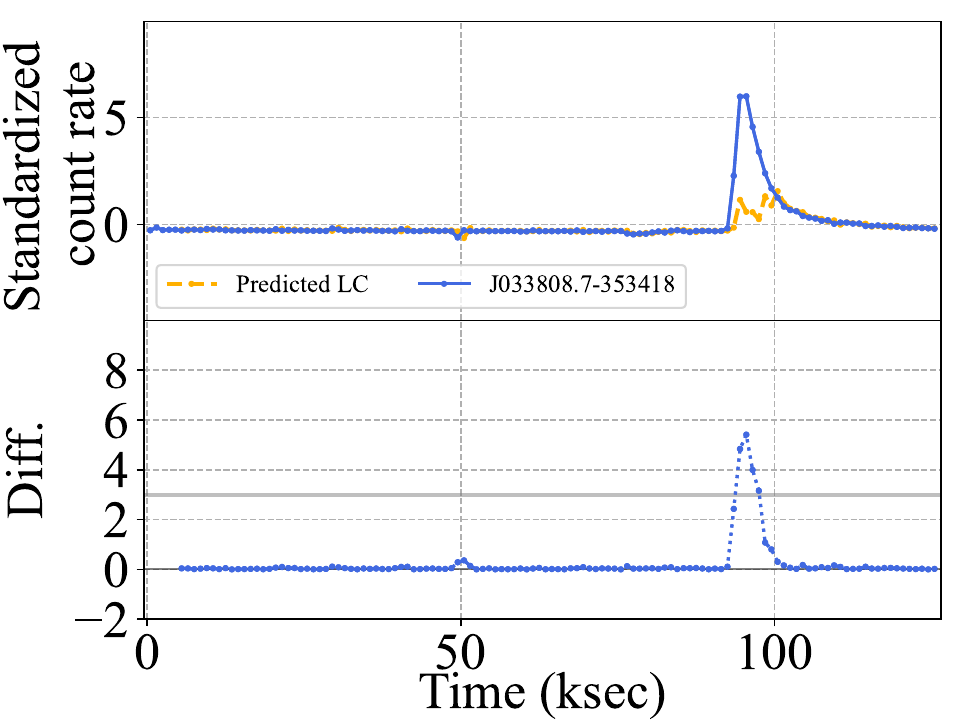}
  \includegraphics[width=3.9cm]{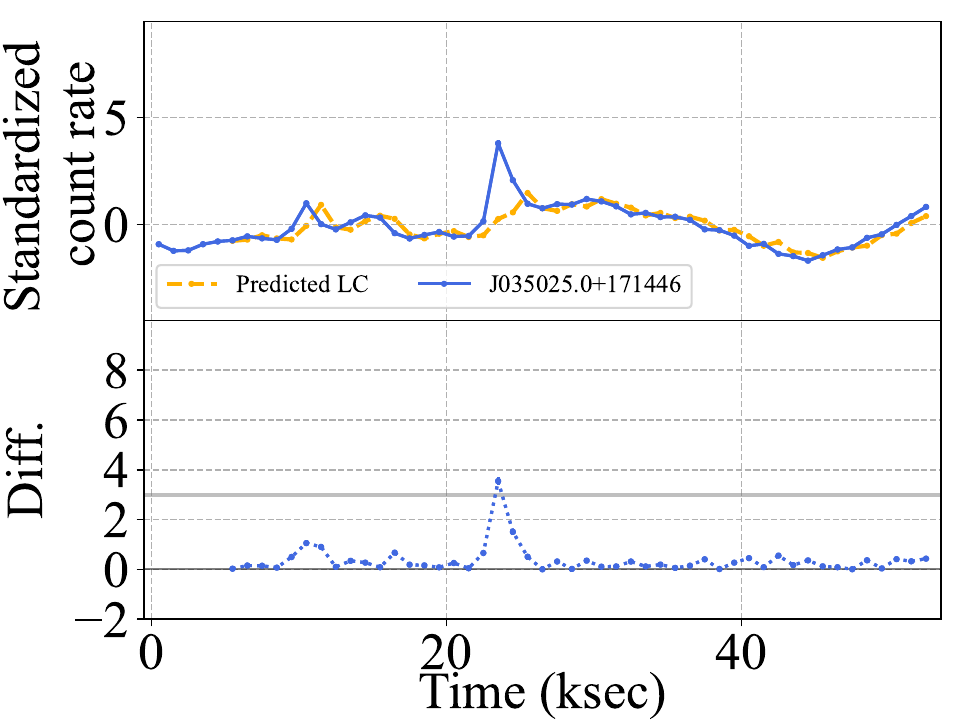}
  \includegraphics[width=3.9cm]{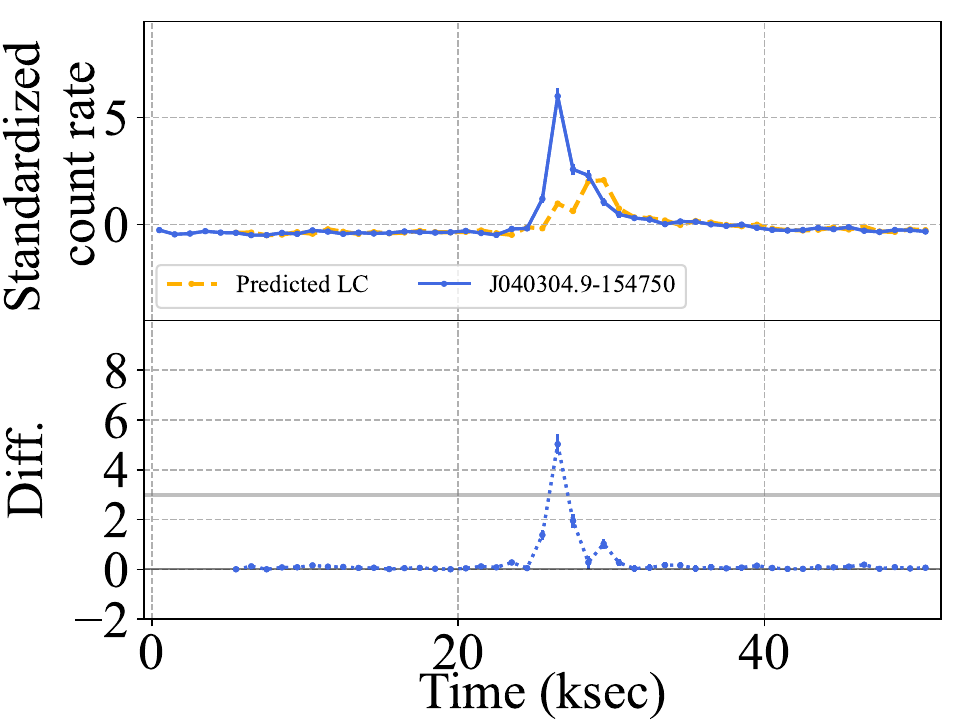}
  \includegraphics[width=3.9cm]{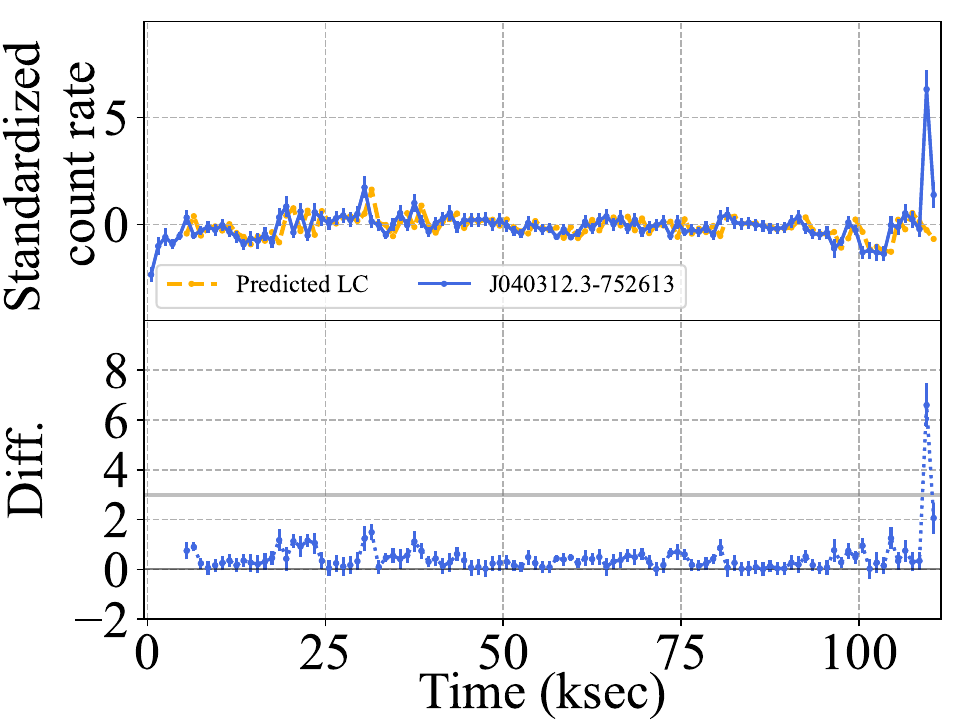}
  \includegraphics[width=3.9cm]{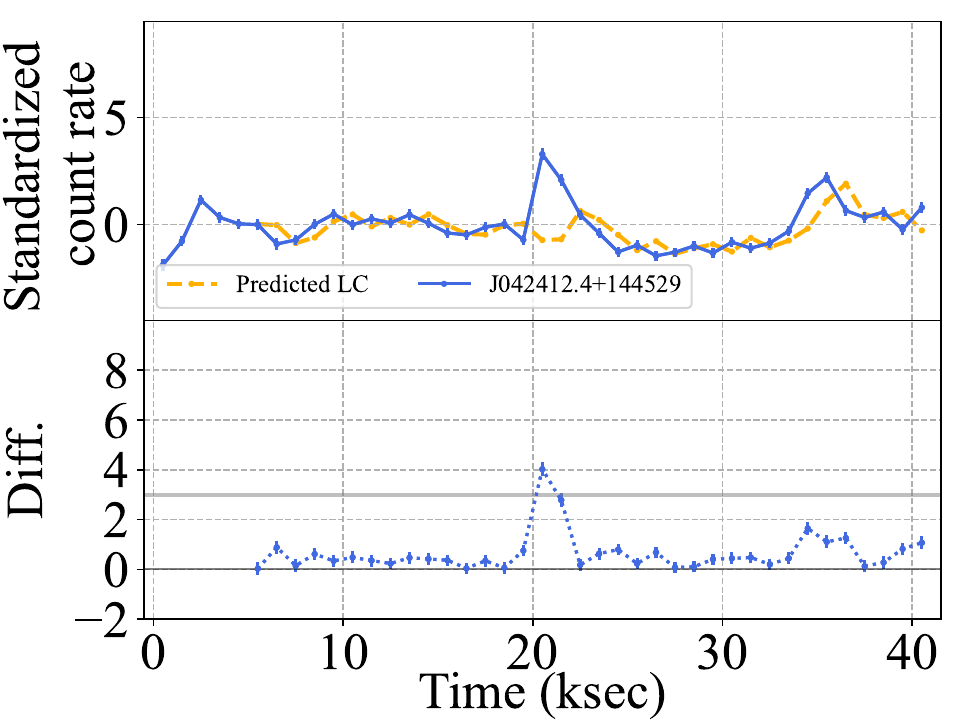}
  \includegraphics[width=3.9cm]{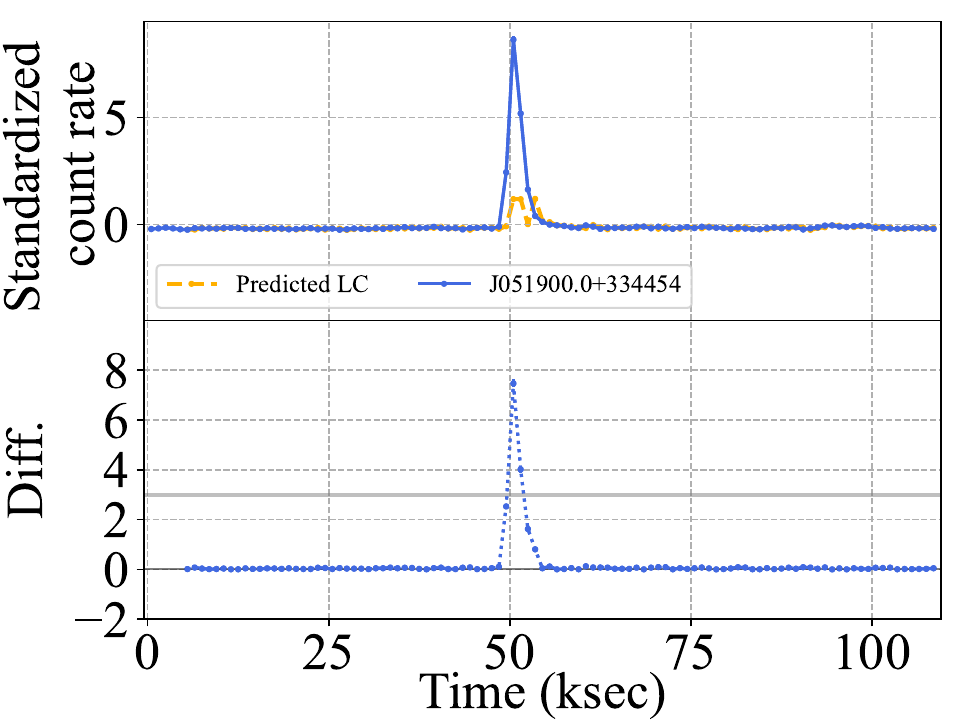}
  \includegraphics[width=3.9cm]{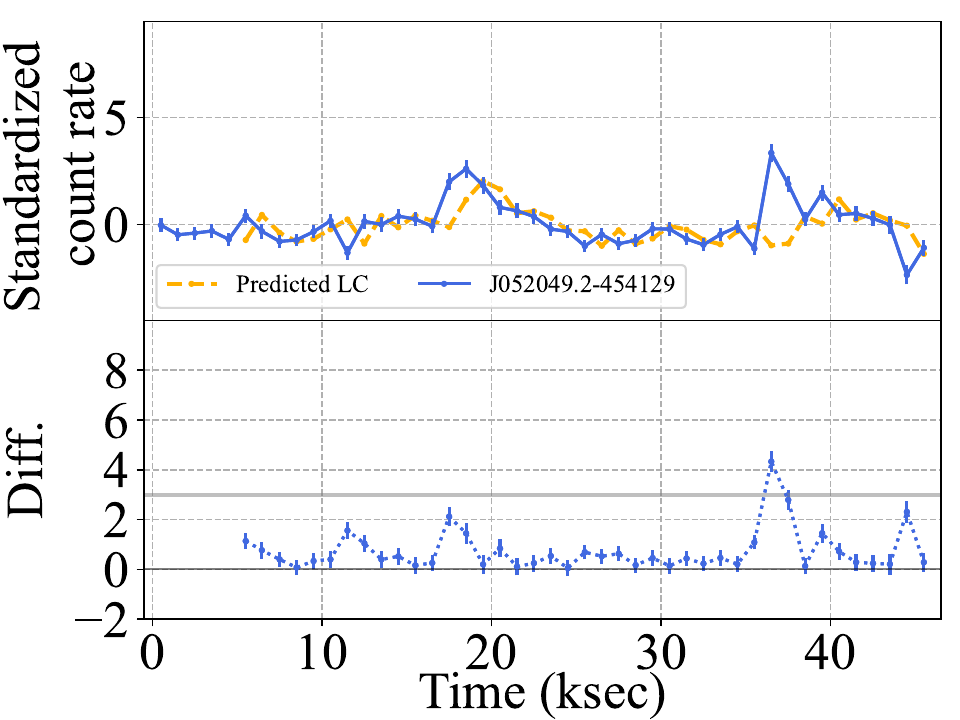}
  \includegraphics[width=3.9cm]{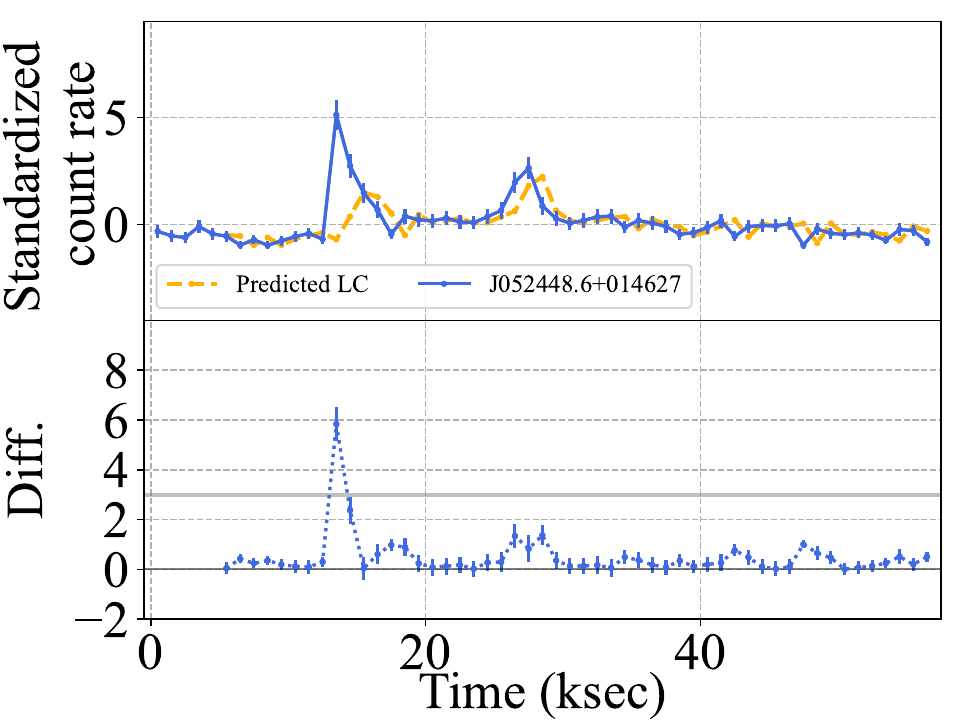}
  \includegraphics[width=3.9cm]{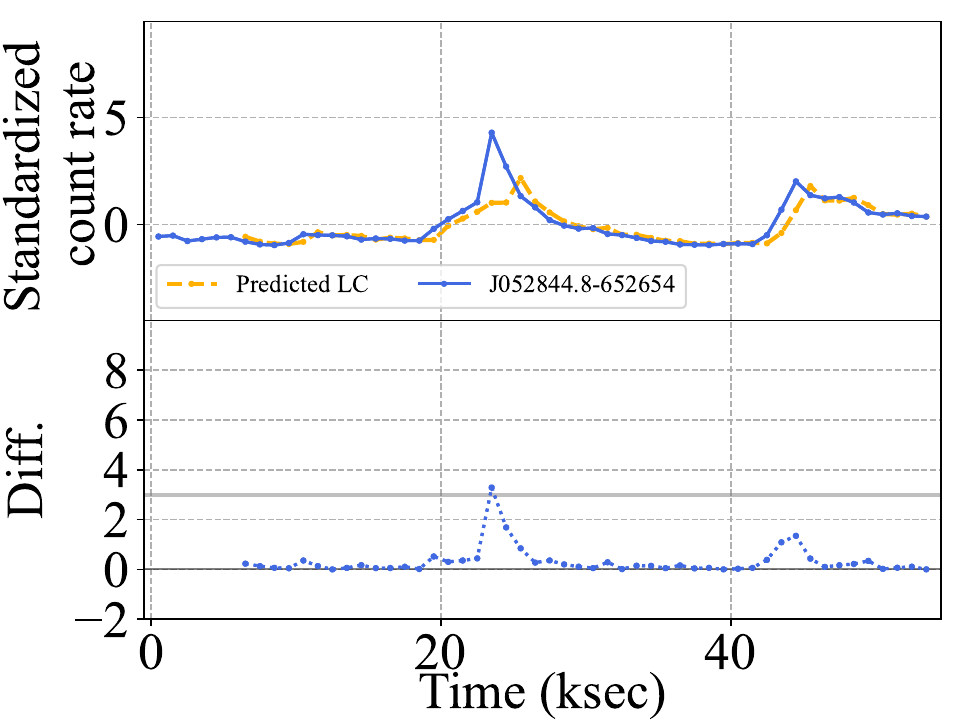}
  \includegraphics[width=3.9cm]{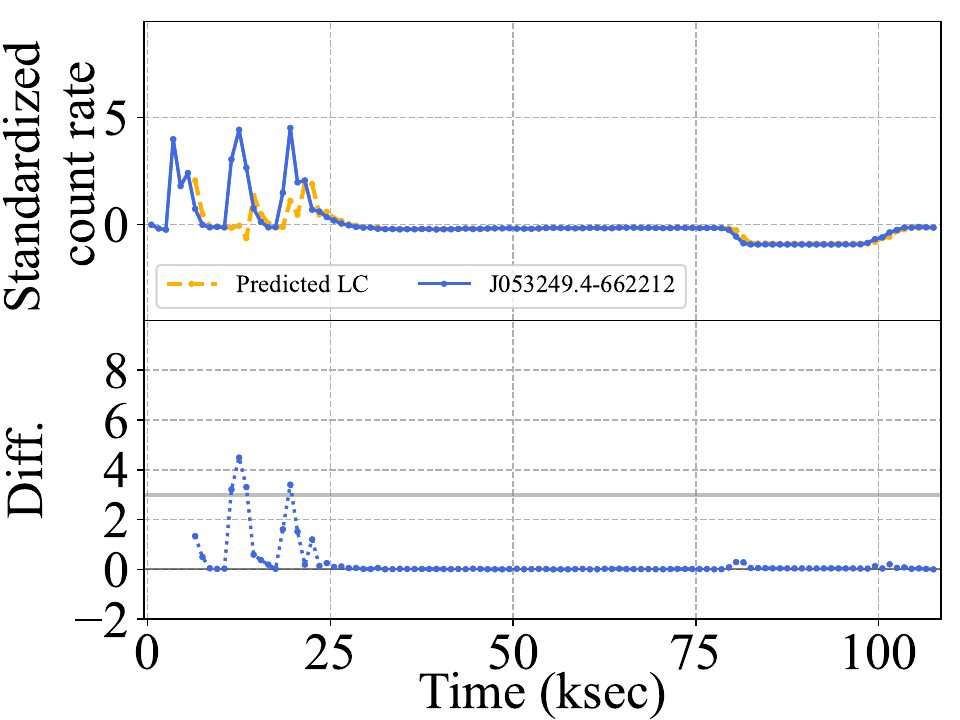}
  \includegraphics[width=3.9cm]{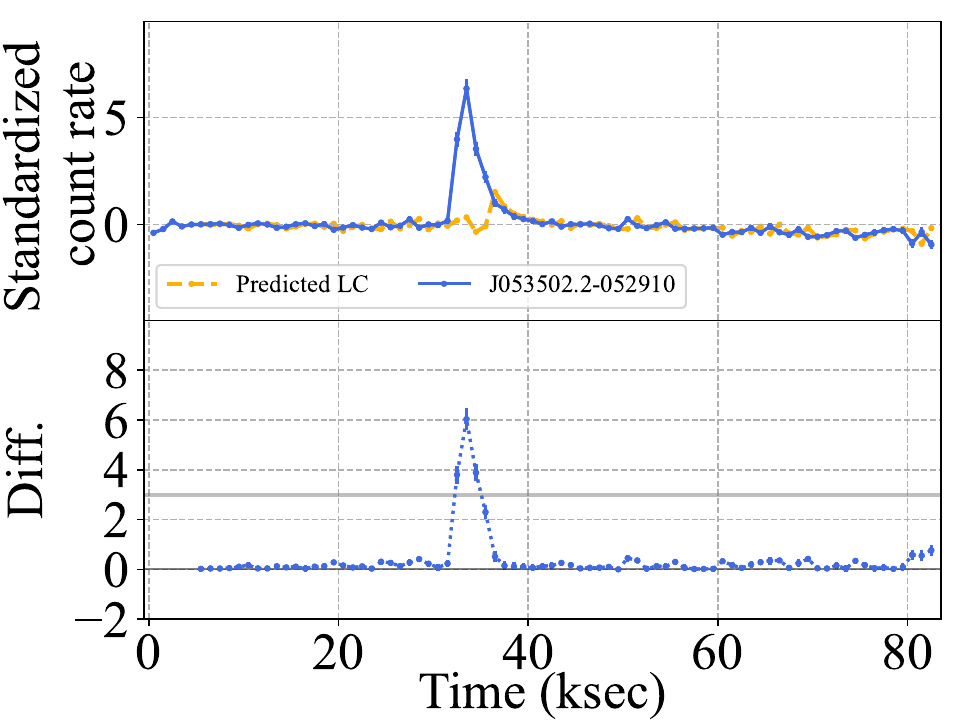}
  \includegraphics[width=3.9cm]{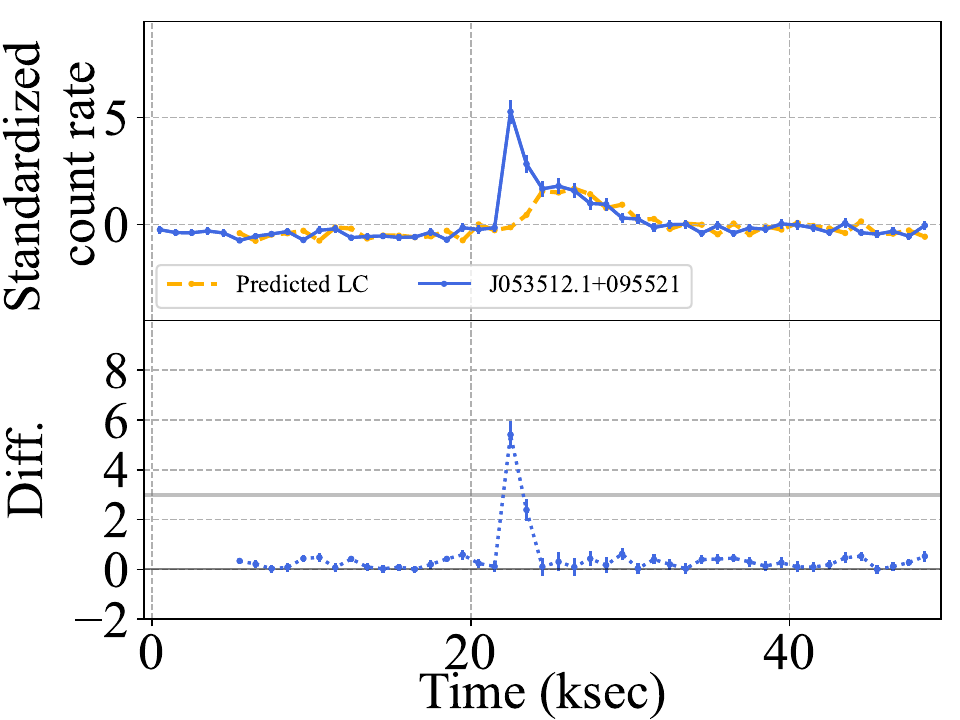}
  \includegraphics[width=3.9cm]{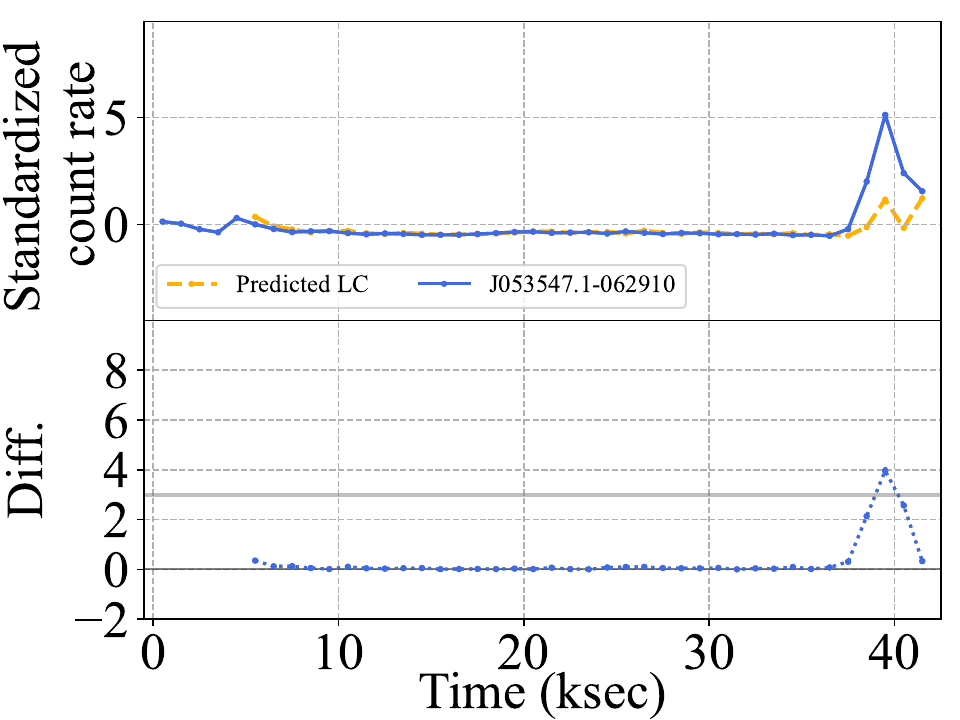}
  \includegraphics[width=3.9cm]{anomaly_fig_dir/J054837p0-510320_P0044740601PNS002SRCTSR801B_lc_1000_stand_wpred.pdf}
  \includegraphics[width=3.9cm]{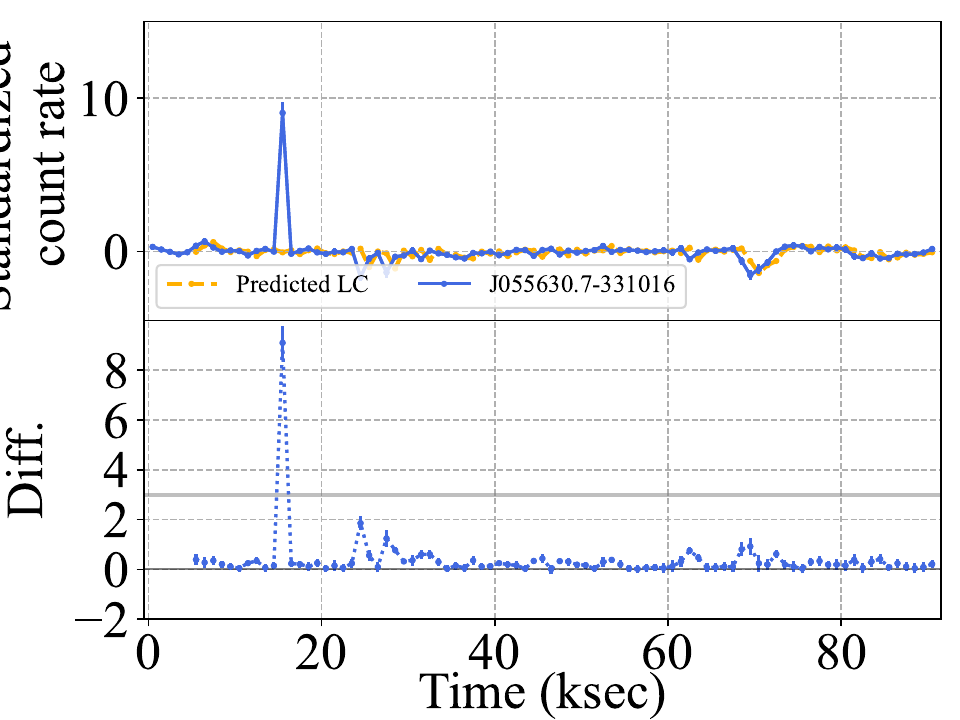}
  \includegraphics[width=3.9cm]{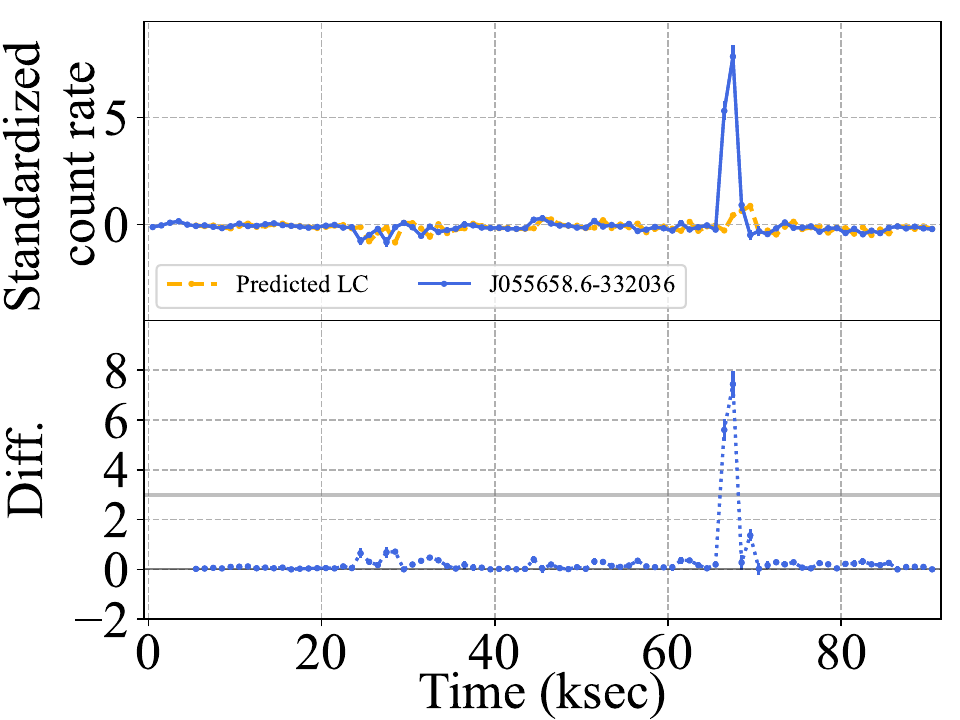}
  \includegraphics[width=3.9cm]{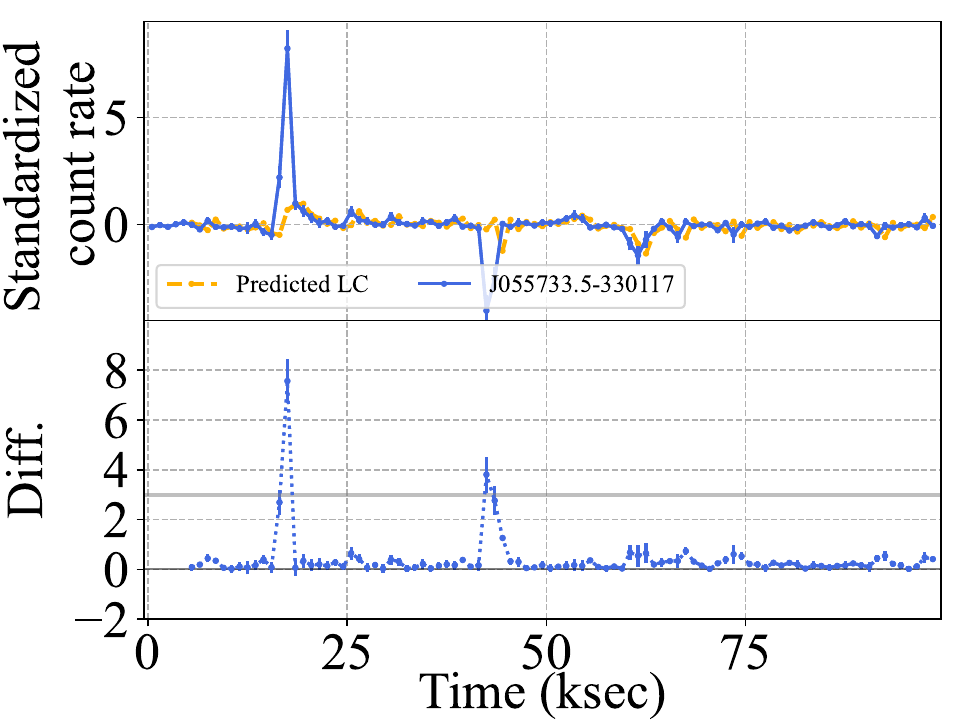}
  \includegraphics[width=3.9cm]{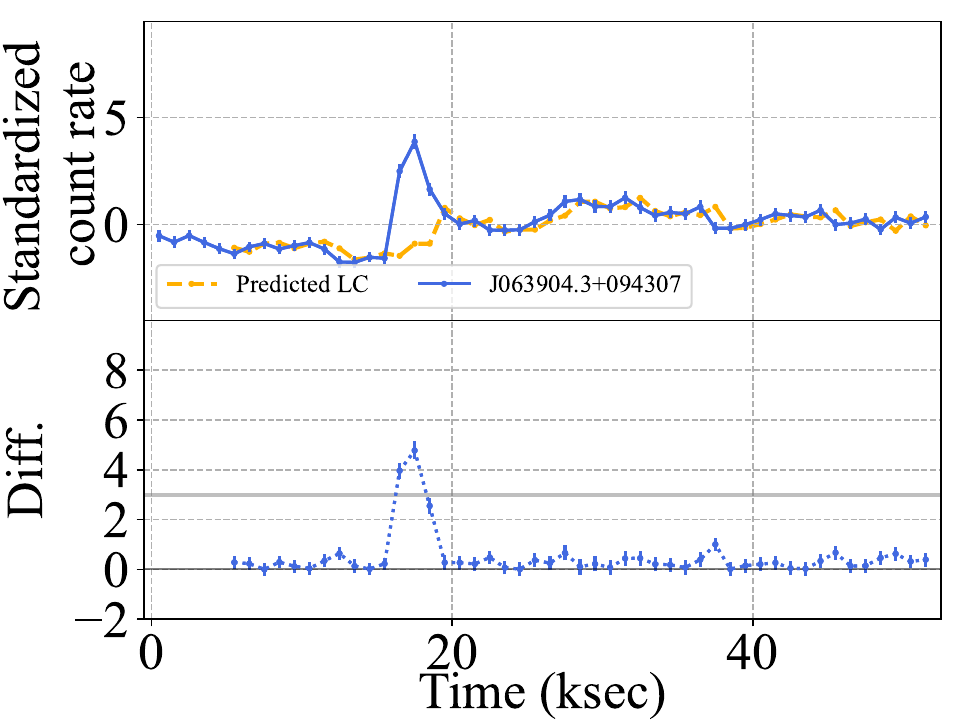}
 \end{center}
 \caption{
        Standardized XMM-Newton light curves in the 0.2--12 keV band (blue) and predicted ones (orange) based on the QLSTM model constructed with ZZFeatureMap and C20. The lower panels show 
        the absolute values of the differences between the real and predicted values. In each figure, the XMM-Newton name is denoted in the legend.
 }
 \label{fig:xmm_lc_supp0}
\end{figure*}
\begin{figure*}
 \begin{center}
  \includegraphics[width=3.9cm]{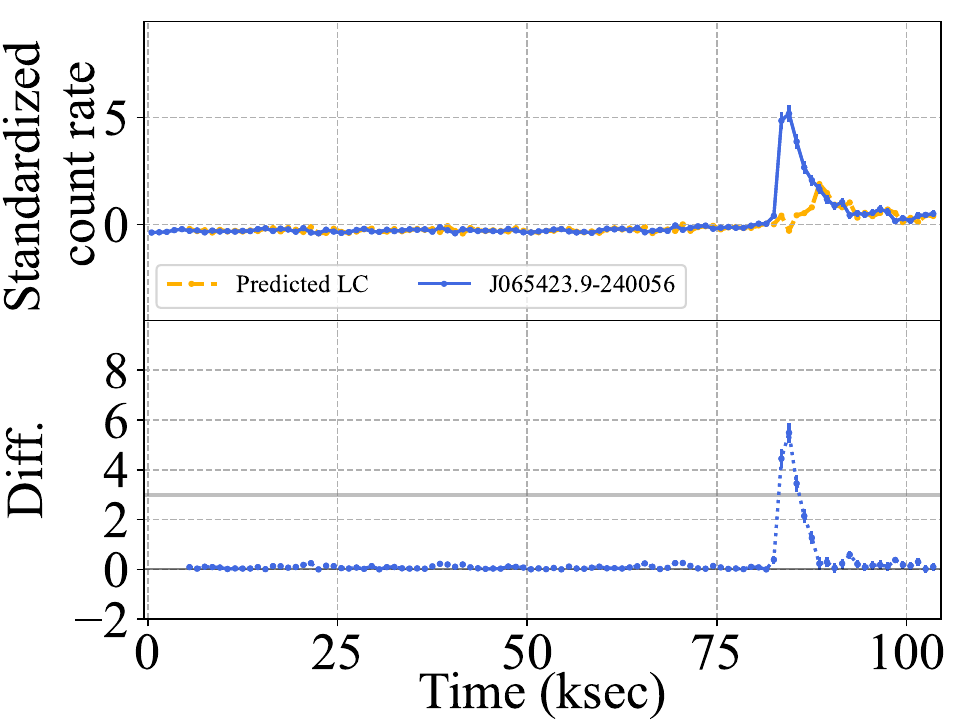}
  \includegraphics[width=3.9cm]{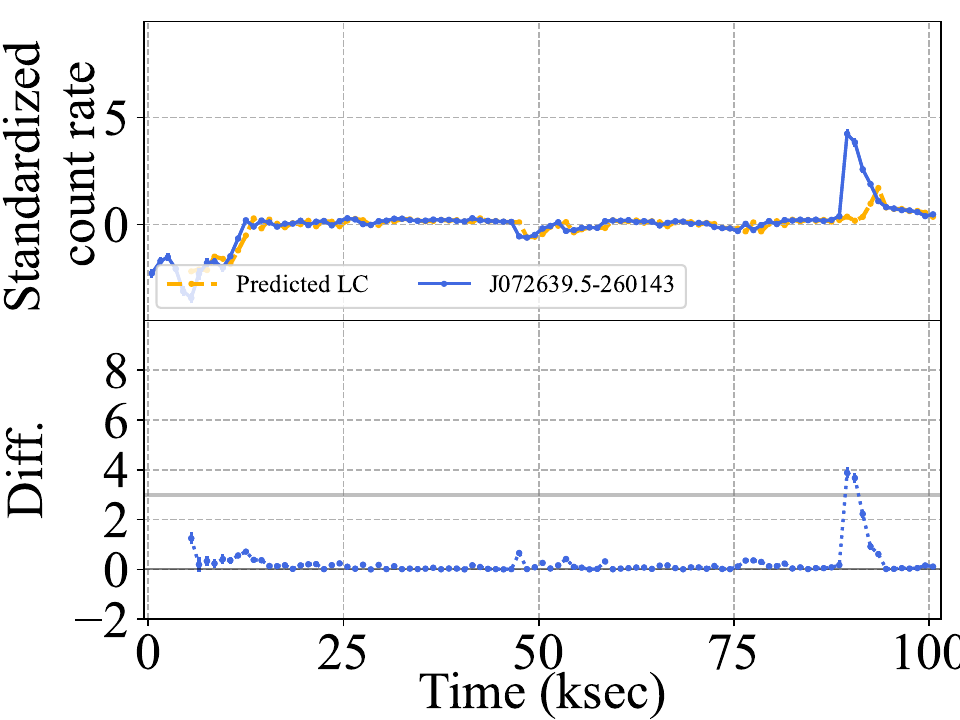}
  \includegraphics[width=3.9cm]{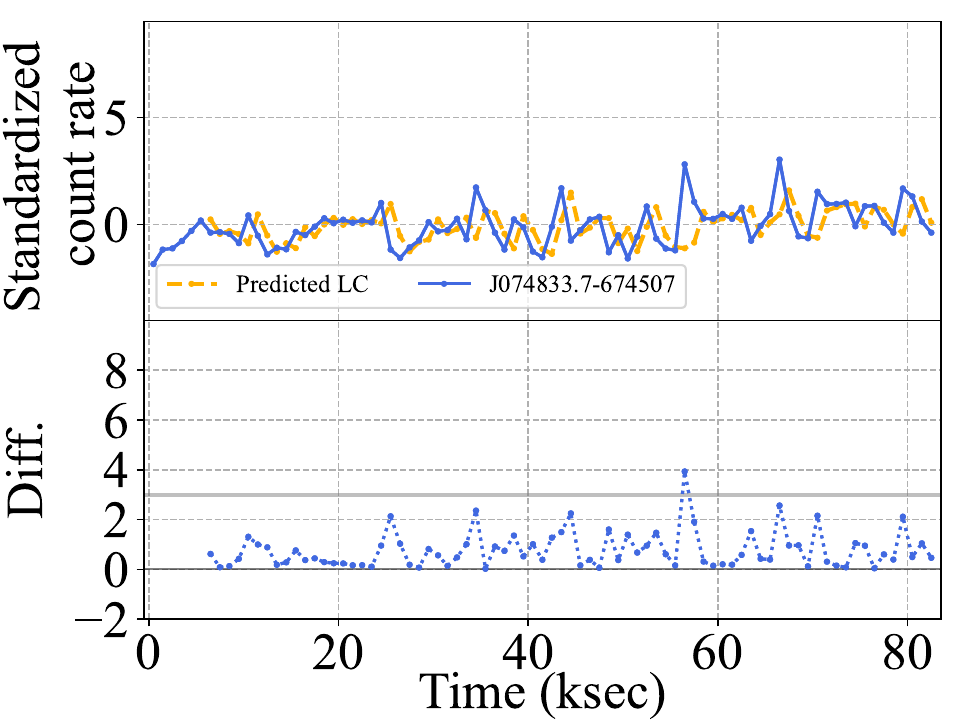}
  \includegraphics[width=3.9cm]{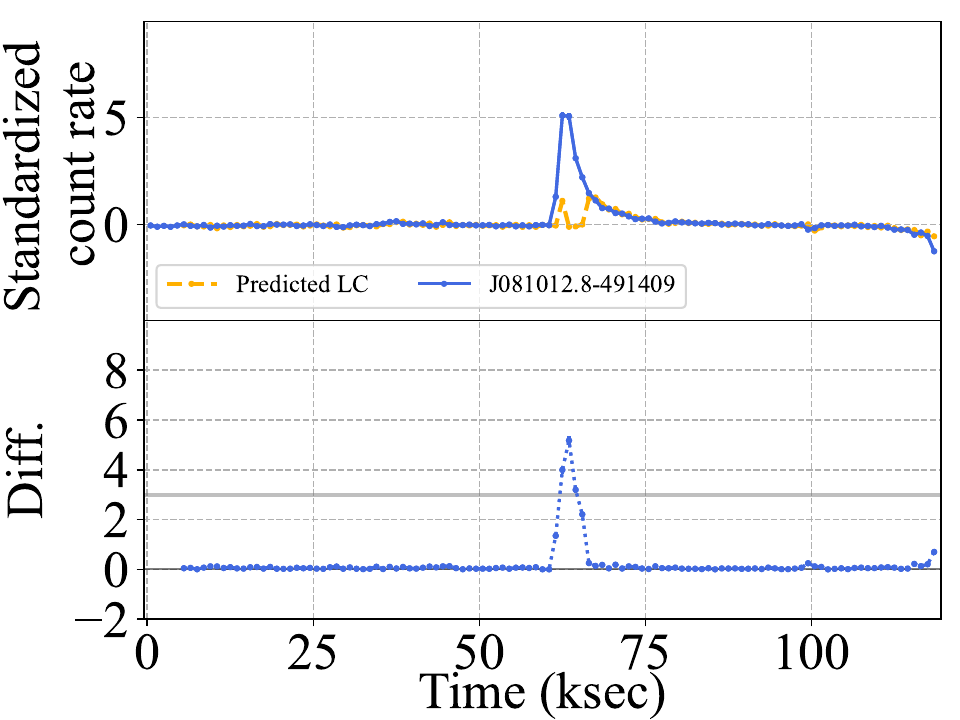}
  \includegraphics[width=3.9cm]{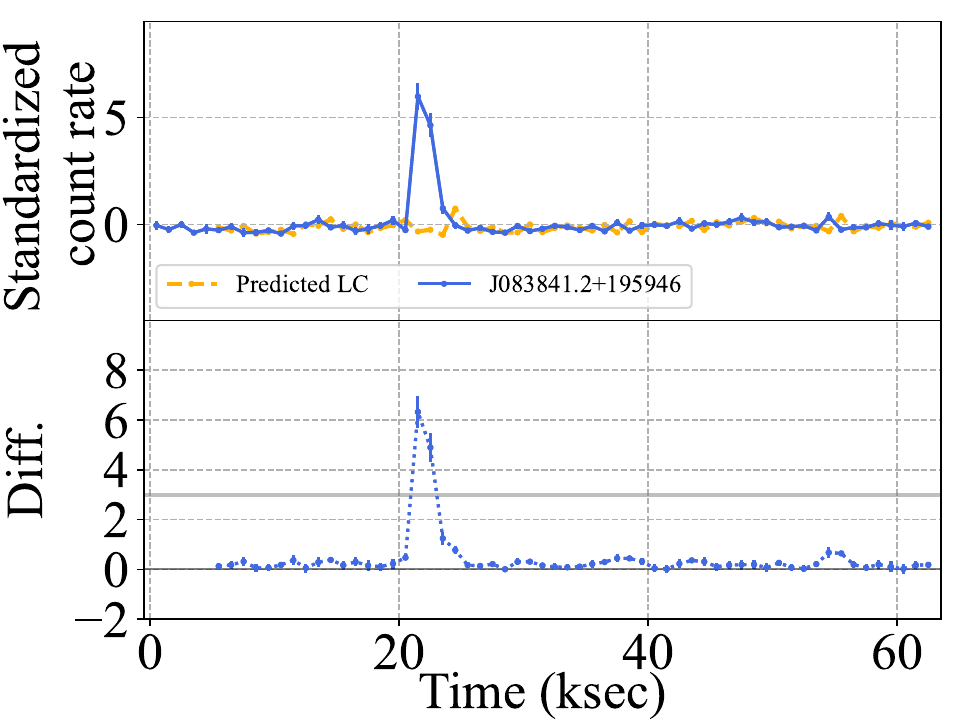}
  \includegraphics[width=3.9cm]{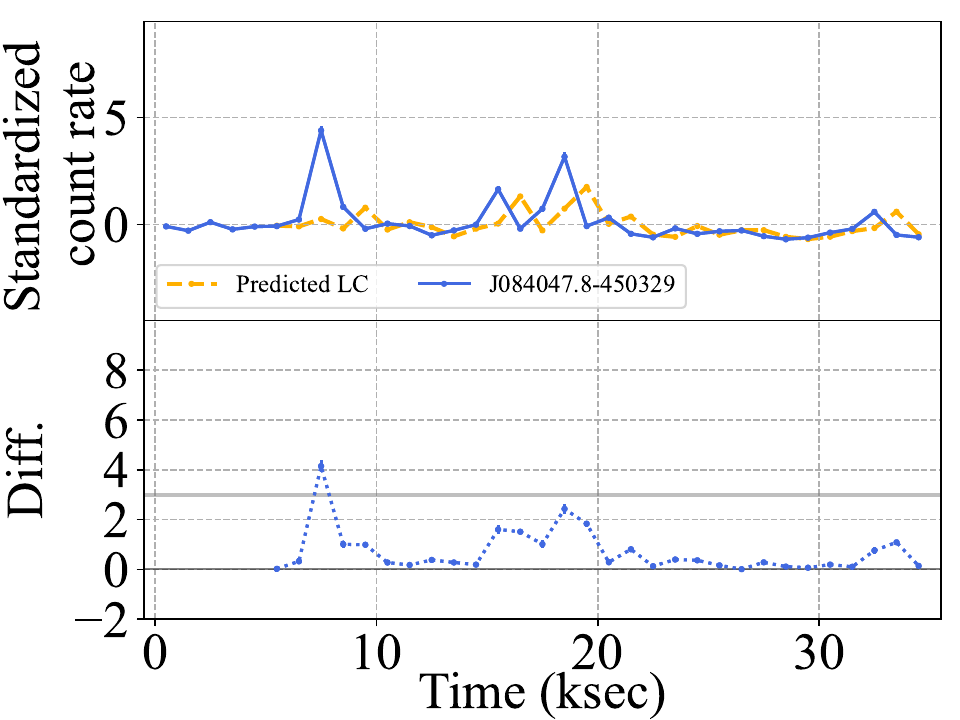}
  \includegraphics[width=3.9cm]{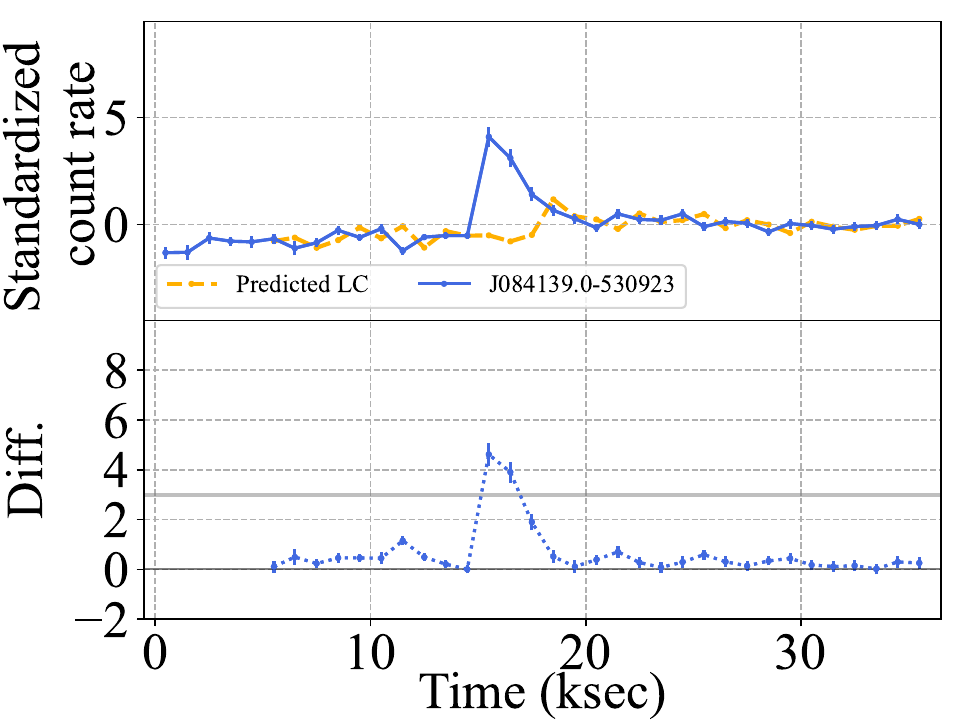}
  \includegraphics[width=3.9cm]{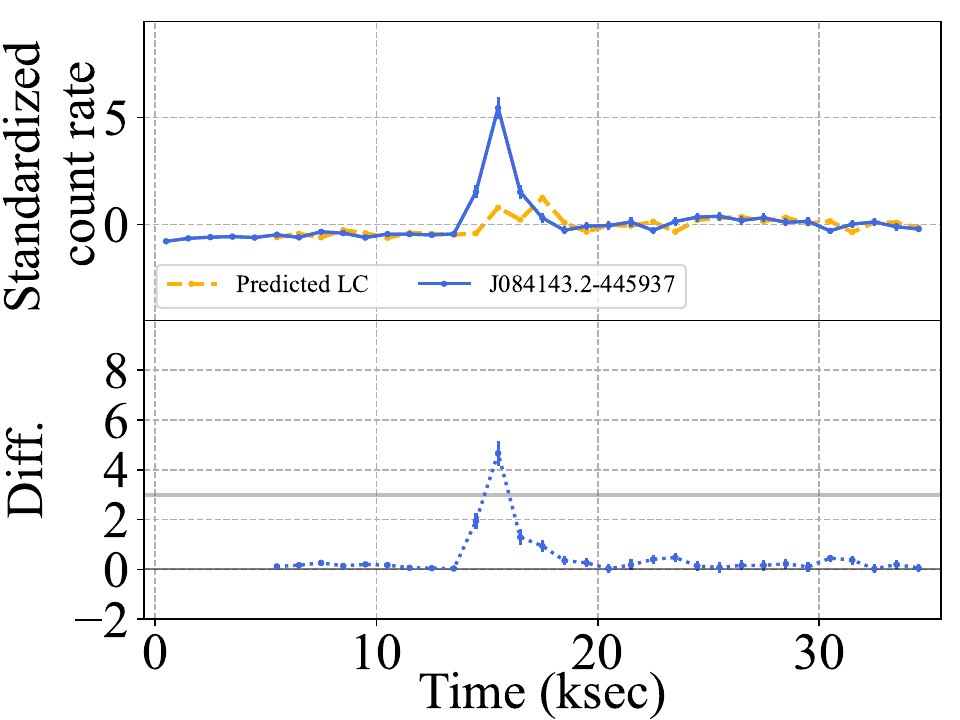}
  \includegraphics[width=3.9cm]{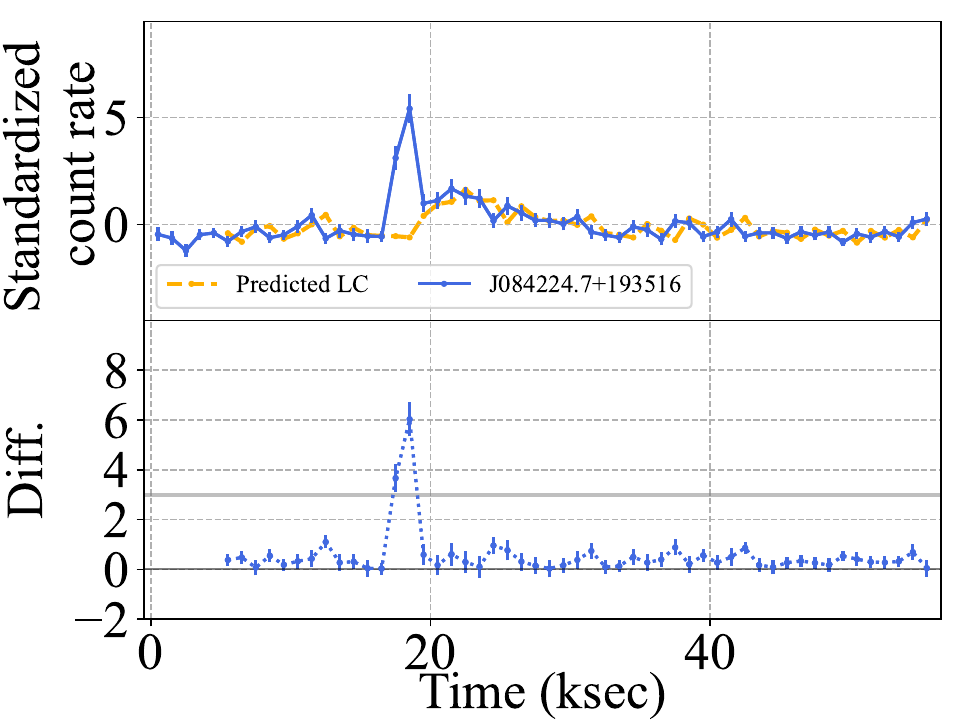}
  \includegraphics[width=3.9cm]{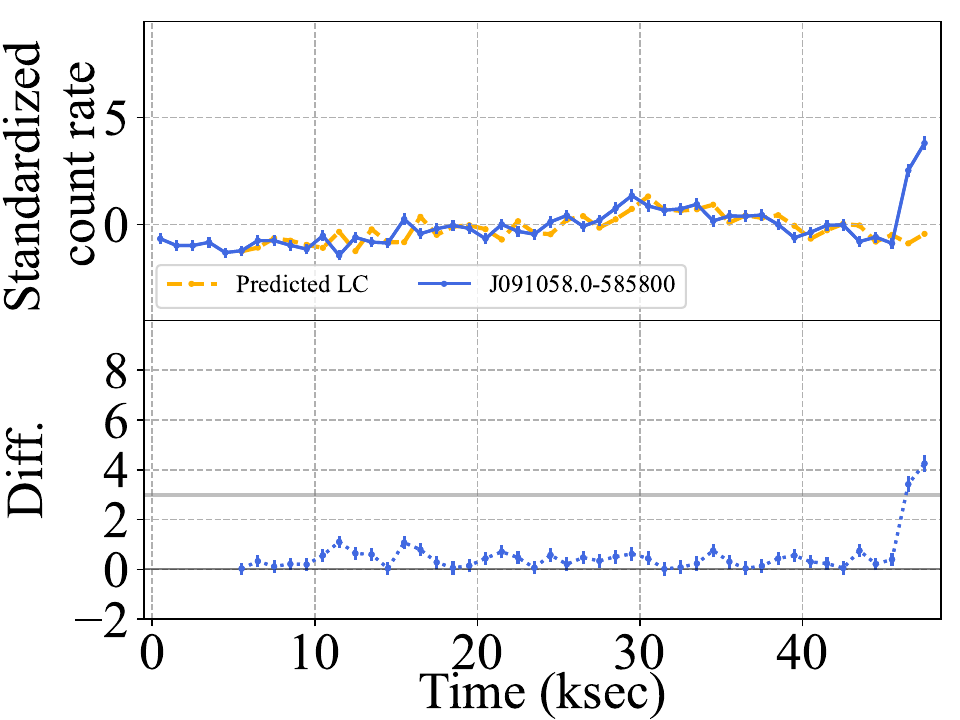}
  \includegraphics[width=3.9cm]{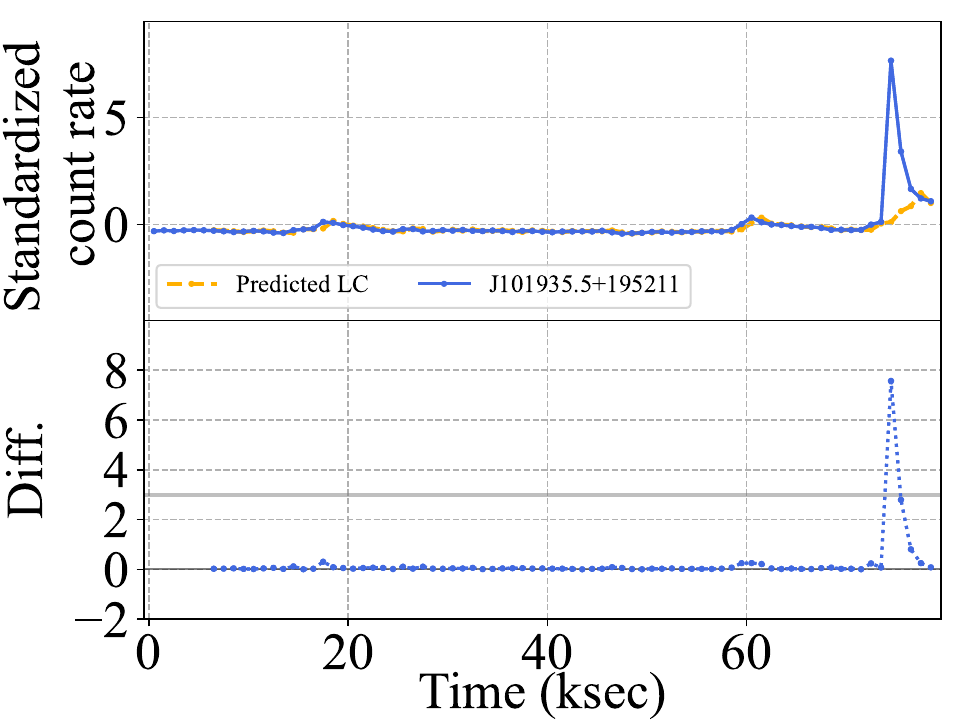}
  \includegraphics[width=3.9cm]{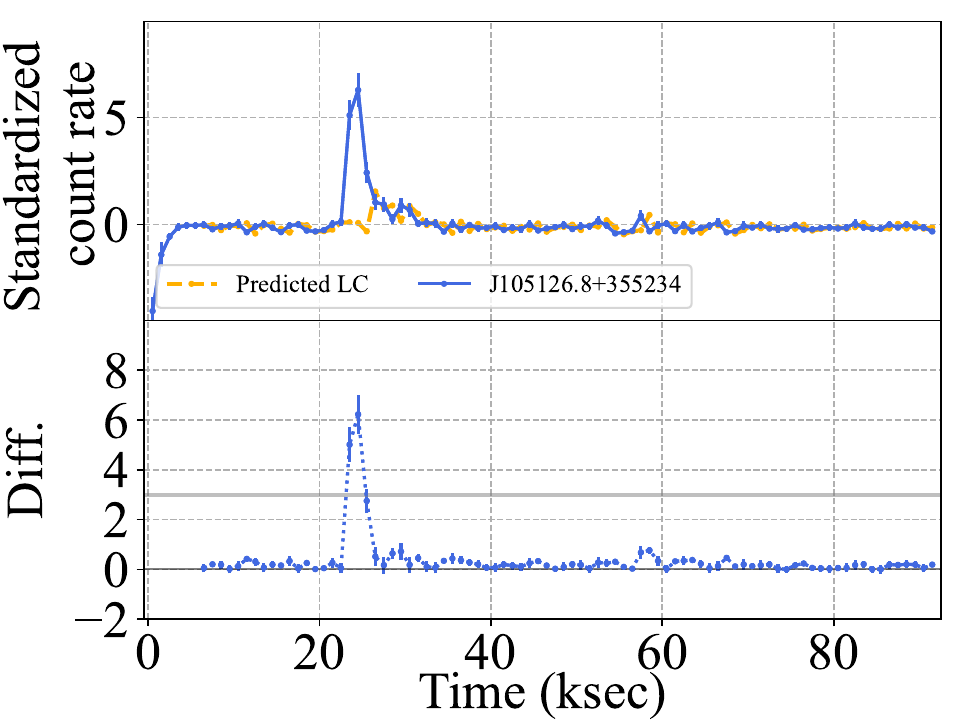}
  \includegraphics[width=3.9cm]{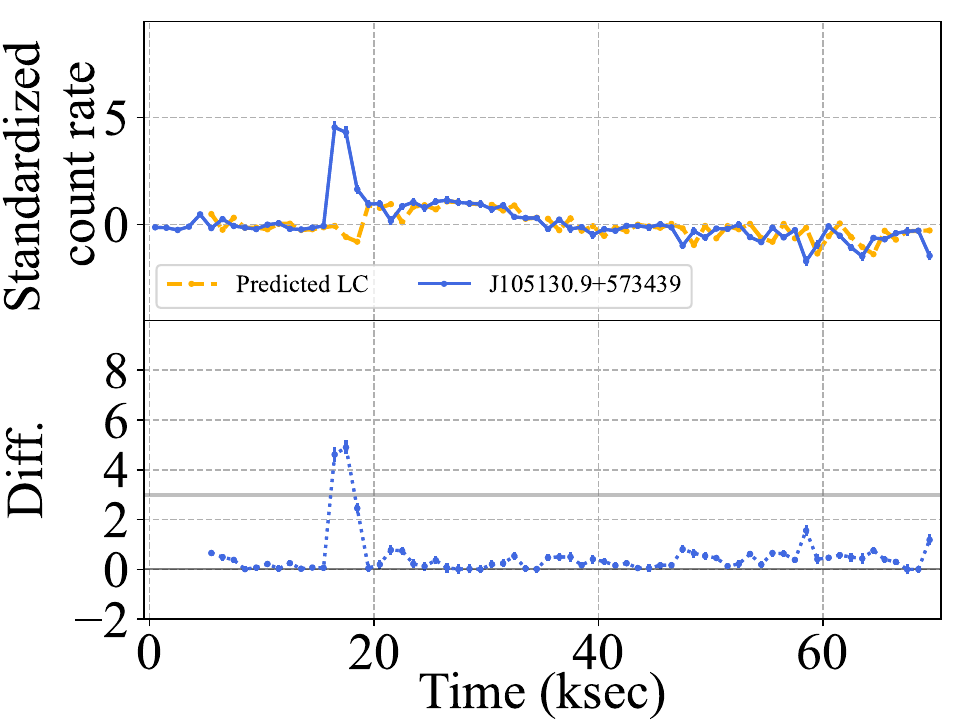}
  \includegraphics[width=3.9cm]{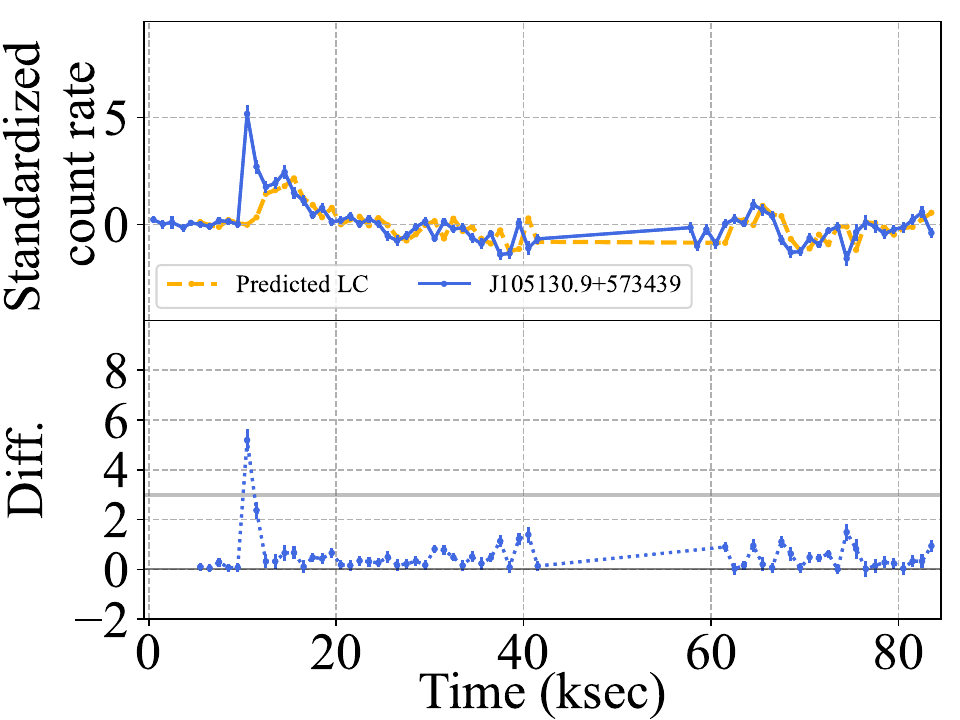}
  \includegraphics[width=3.9cm]{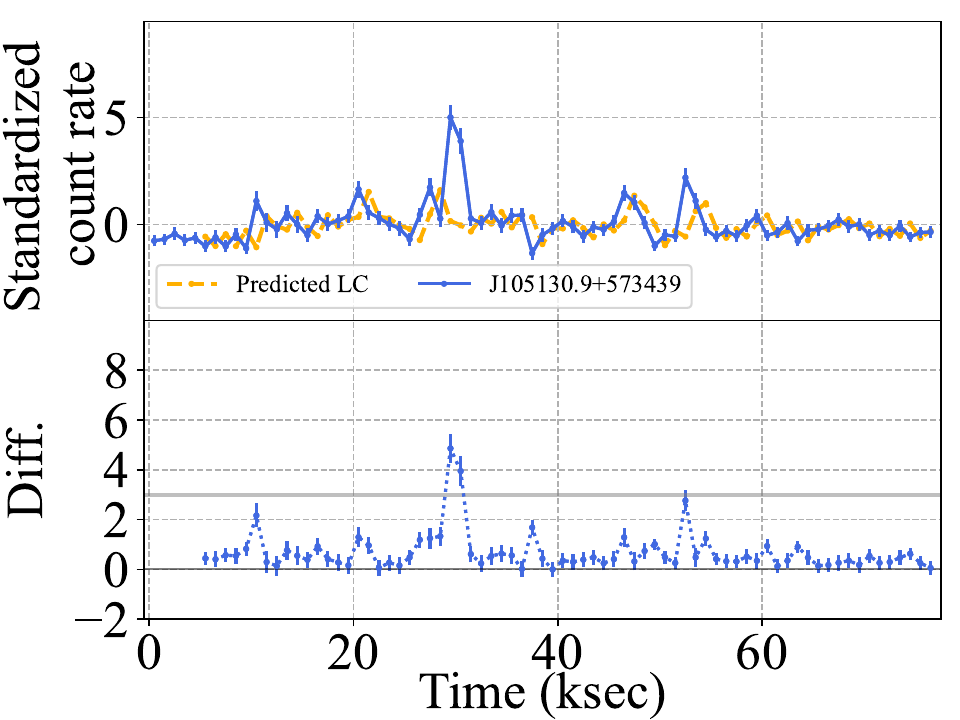}
  \includegraphics[width=3.9cm]{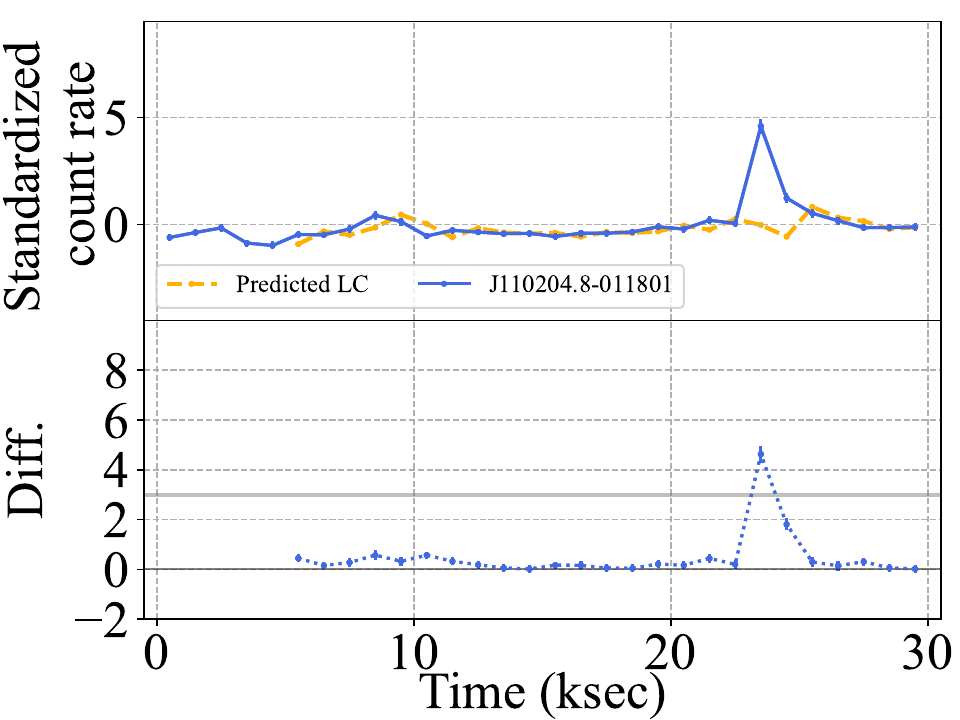}
  \includegraphics[width=3.9cm]{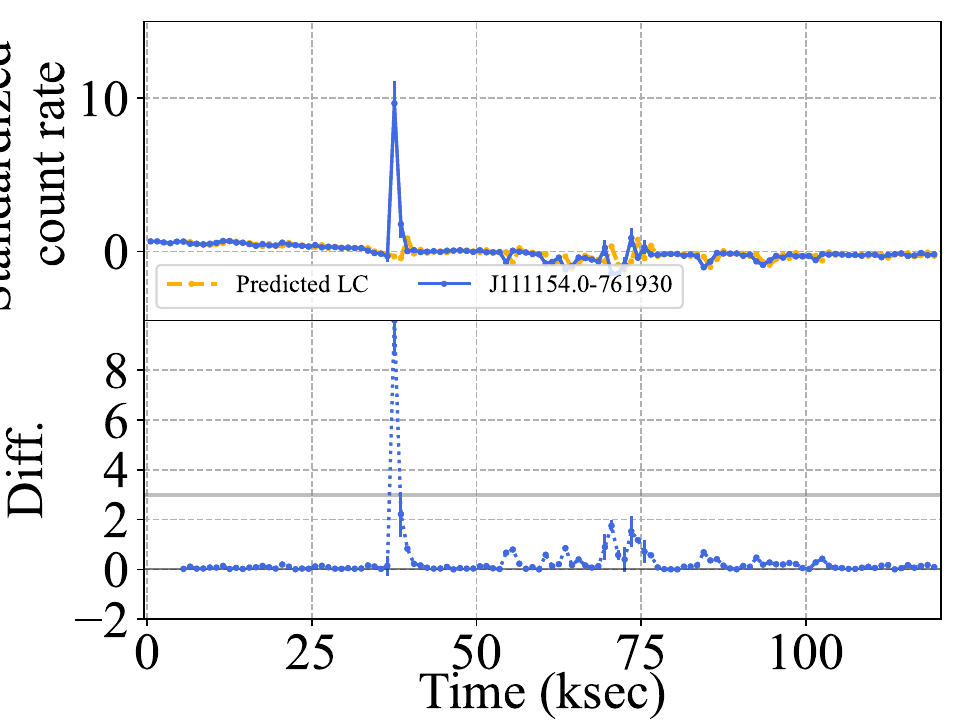}
  \includegraphics[width=3.9cm]{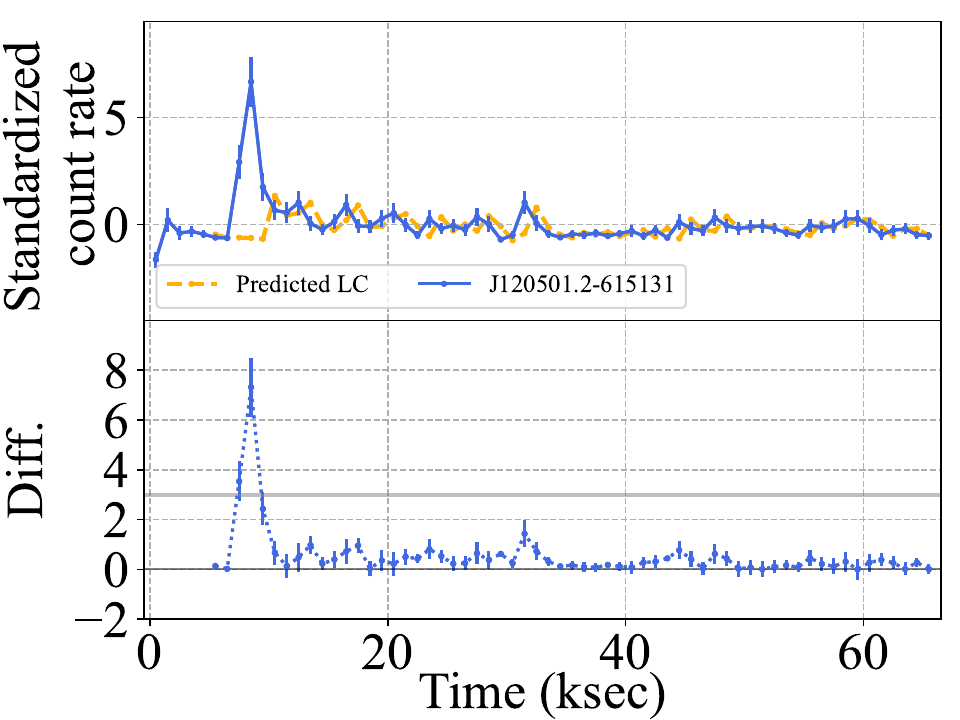}
  \includegraphics[width=3.9cm]{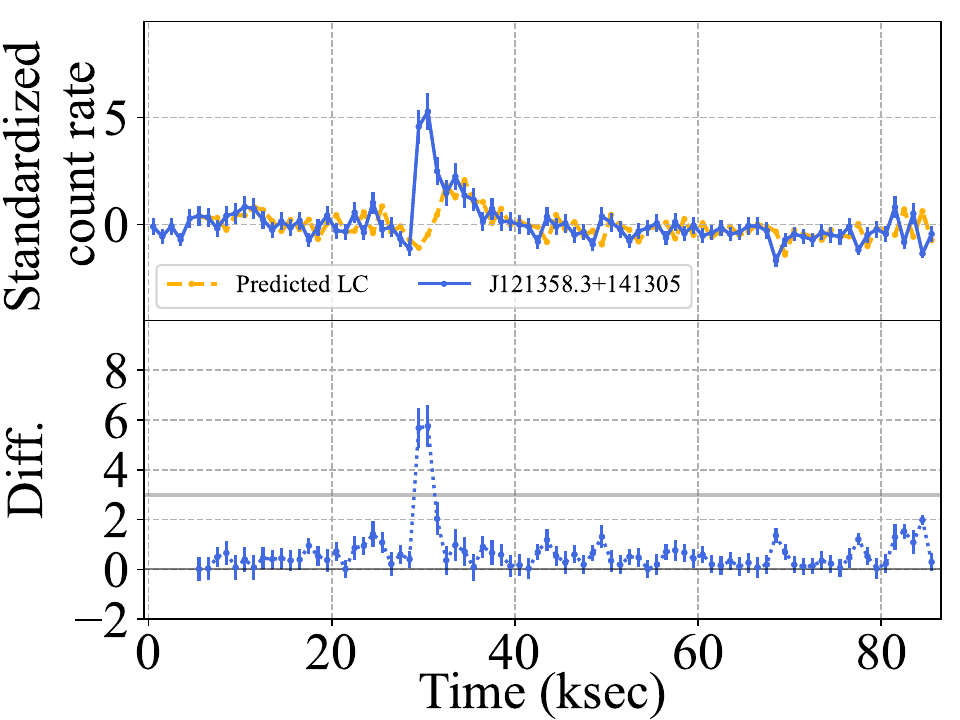}
  \includegraphics[width=3.9cm]{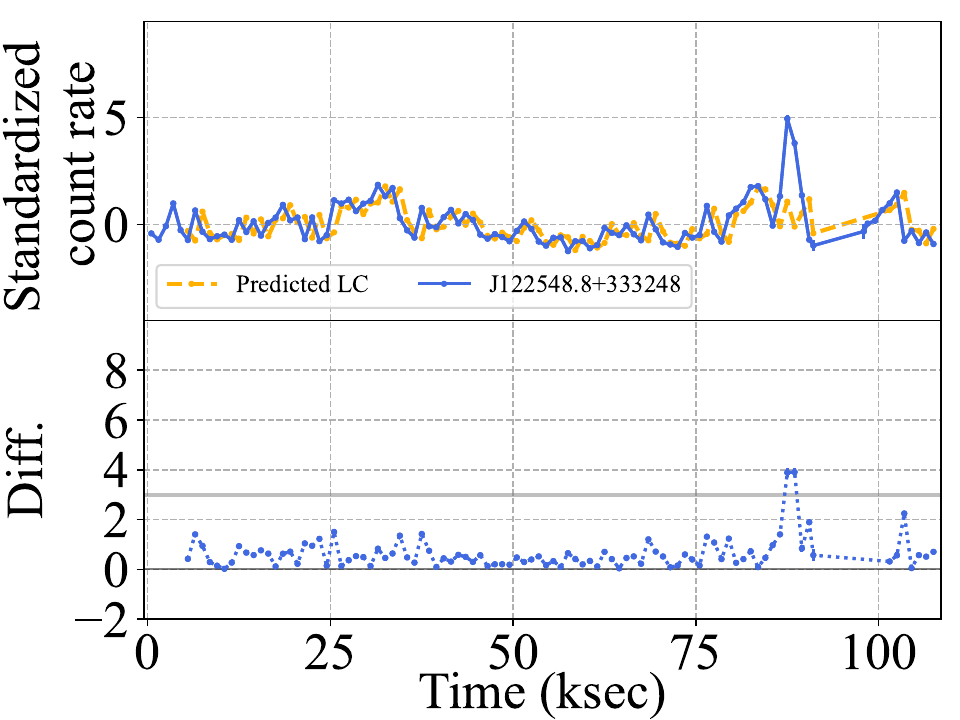}
  \includegraphics[width=3.9cm]{anomaly_fig_dir/J122944p8+075238_P0761630201PNS003SRCTSR8007_lc_1000_stand_wpred.pdf}
  \includegraphics[width=3.9cm]{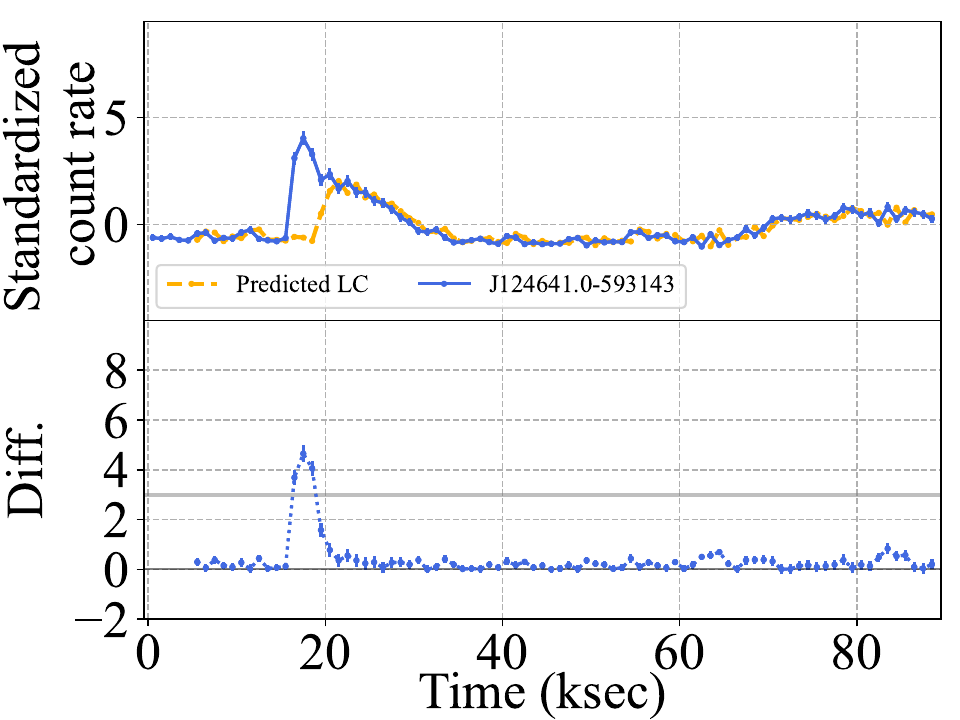}
  \includegraphics[width=3.9cm]{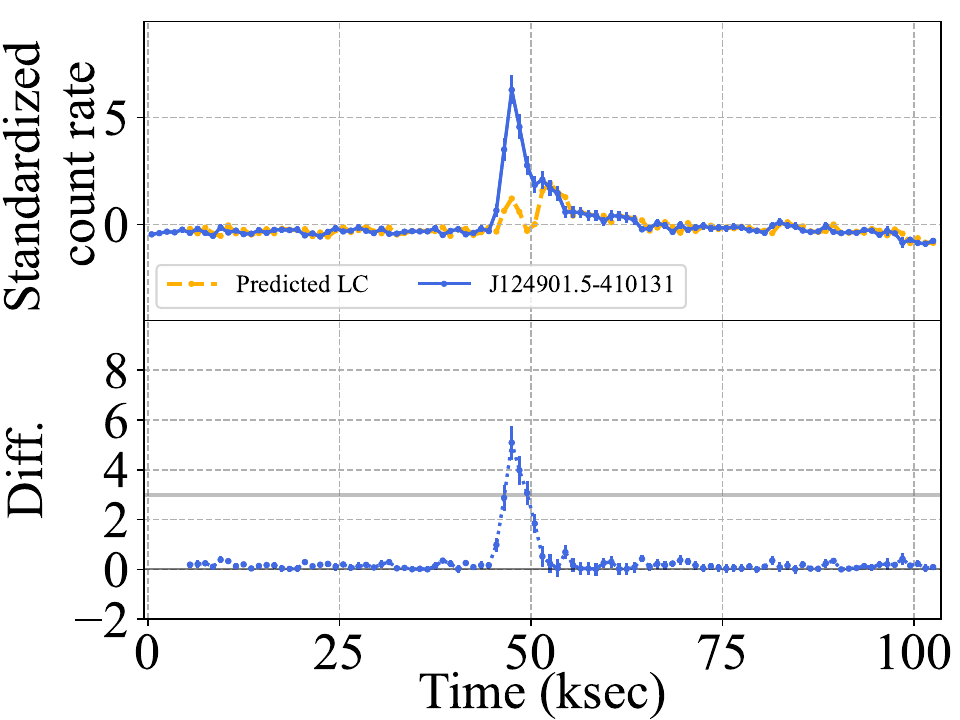}
  \includegraphics[width=3.9cm]{anomaly_fig_dir/J125517p5-043210_P0801930801PNS003SRCTSR8011_lc_1000_stand_wpred.pdf}
  \includegraphics[width=3.9cm]{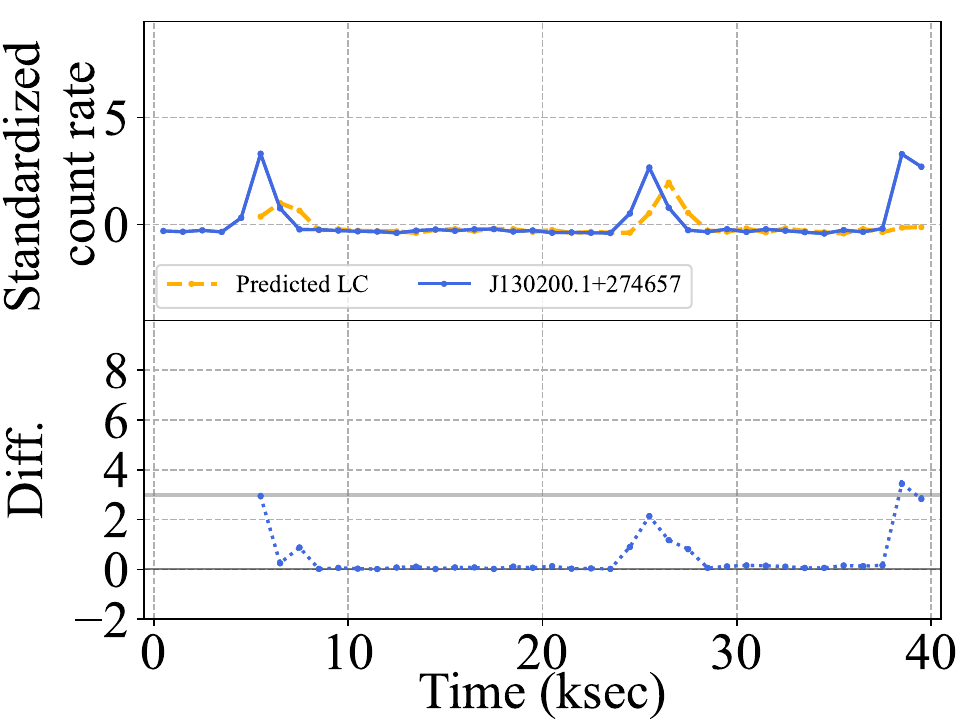}
  \includegraphics[width=3.9cm]{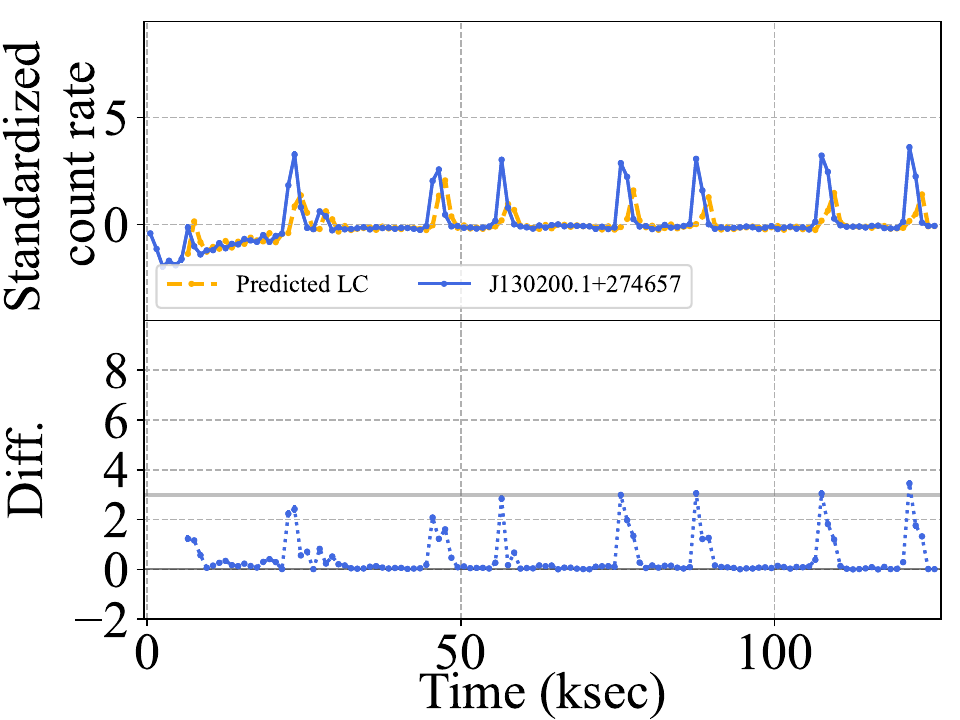}
  \includegraphics[width=3.9cm]{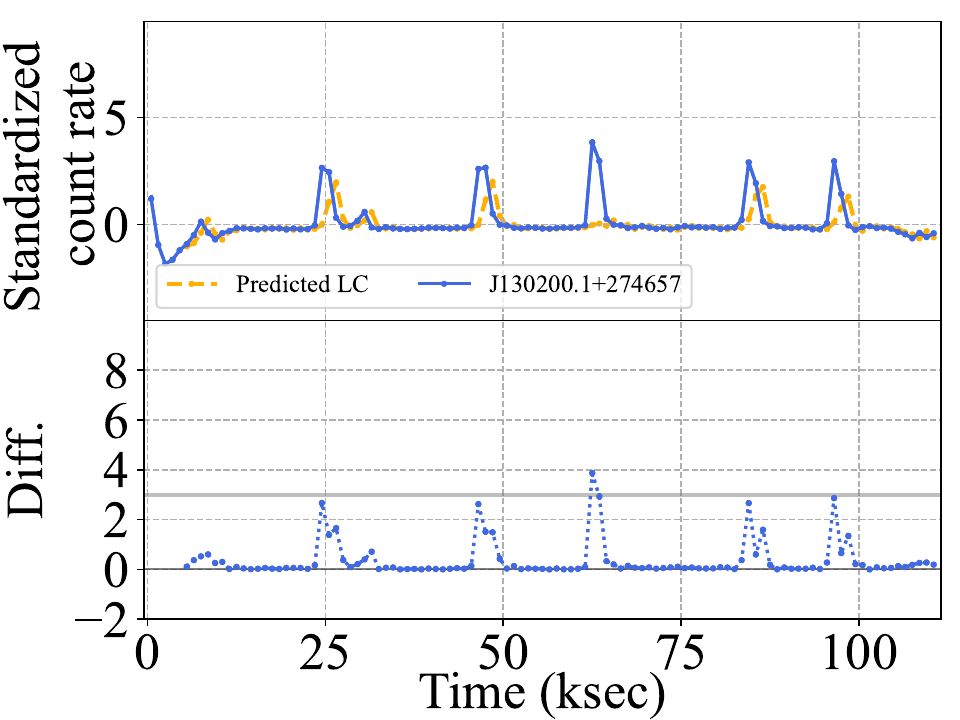}
  \includegraphics[width=3.9cm]{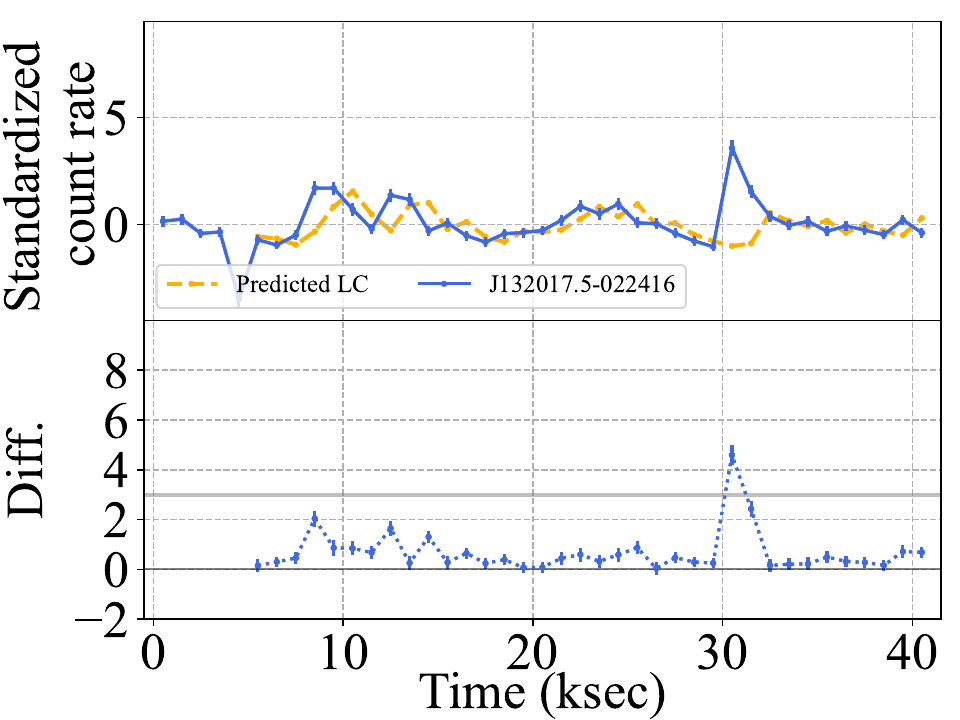}
 \end{center}
 \caption{
        Standardized XMM-Newton light curves in the 0.2--12 keV band (blue) and predicted ones (orange) based on the QLSTM model constructed with ZZFeatureMap and C20. The lower panels show 
        the absolute values of the differences between the real and predicted values. In each figure, the XMM-Newton name is denoted in the legend.
 }
 \label{fig:xmm_lc_supp1}
\end{figure*}
\begin{figure*}
 \begin{center}
  \includegraphics[width=3.9cm]{anomaly_fig_dir/J132456p1-430258_P0863890201PNS003SRCTSR800D_lc_1000_stand_wpred.pdf}
  \includegraphics[width=3.9cm]{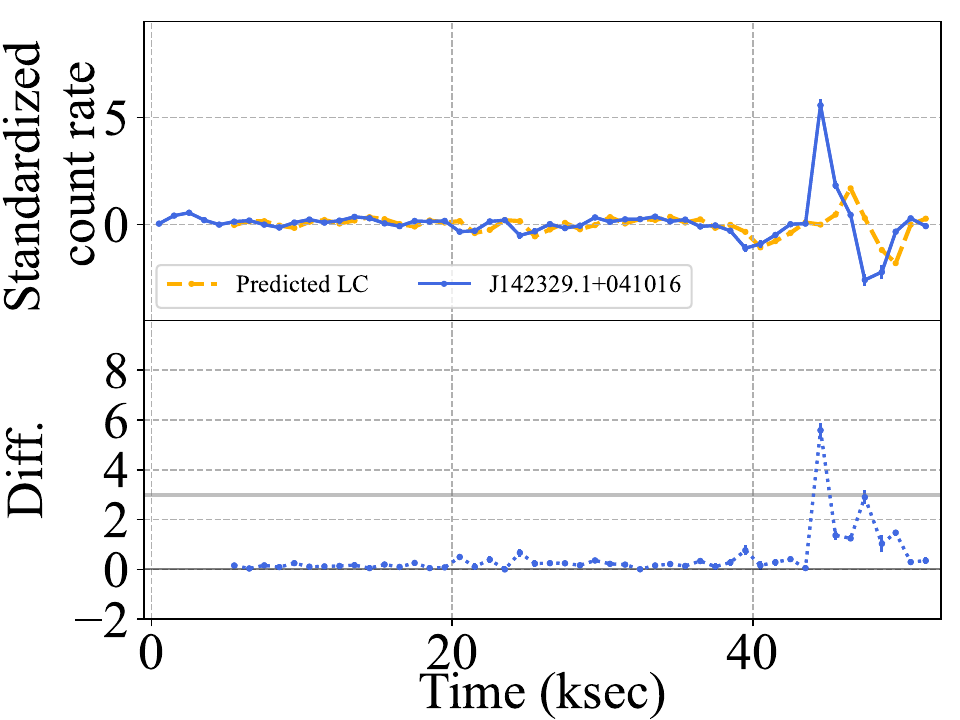}
  \includegraphics[width=3.9cm]{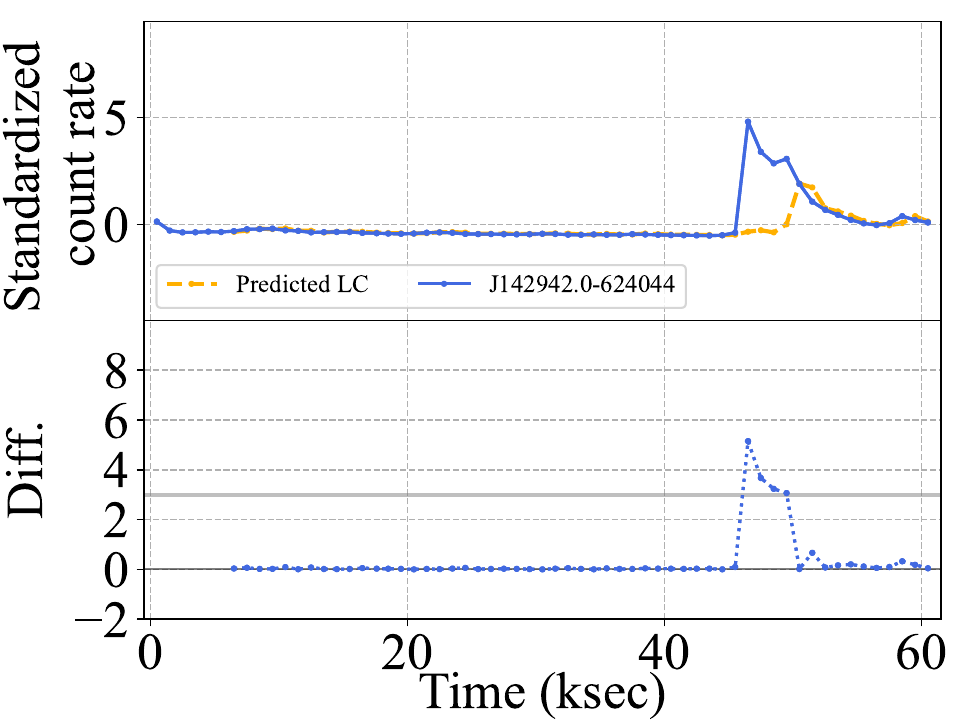}
  \includegraphics[width=3.9cm]{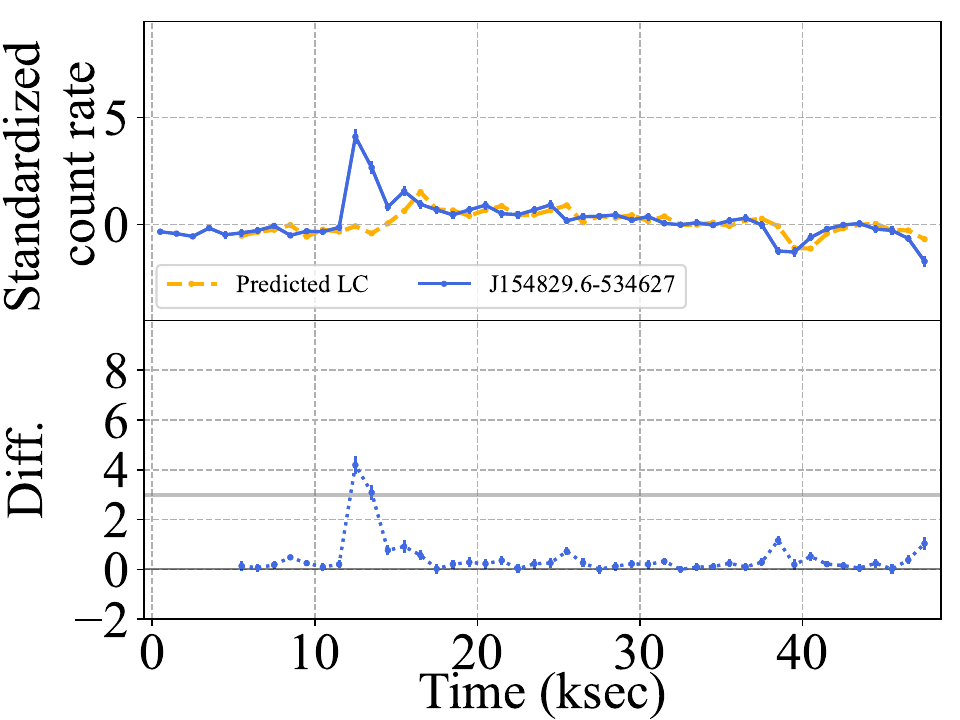}
  \includegraphics[width=3.9cm]{anomaly_fig_dir/J154910p4-534225_P0740640101PNU002SRCTSR8008_lc_1000_stand_wpred.pdf}
  \includegraphics[width=3.9cm]{anomaly_fig_dir/J155717p3-373907_P0882060601PNS003SRCTSR8002_lc_1000_stand_wpred.pdf}
  \includegraphics[width=3.9cm]{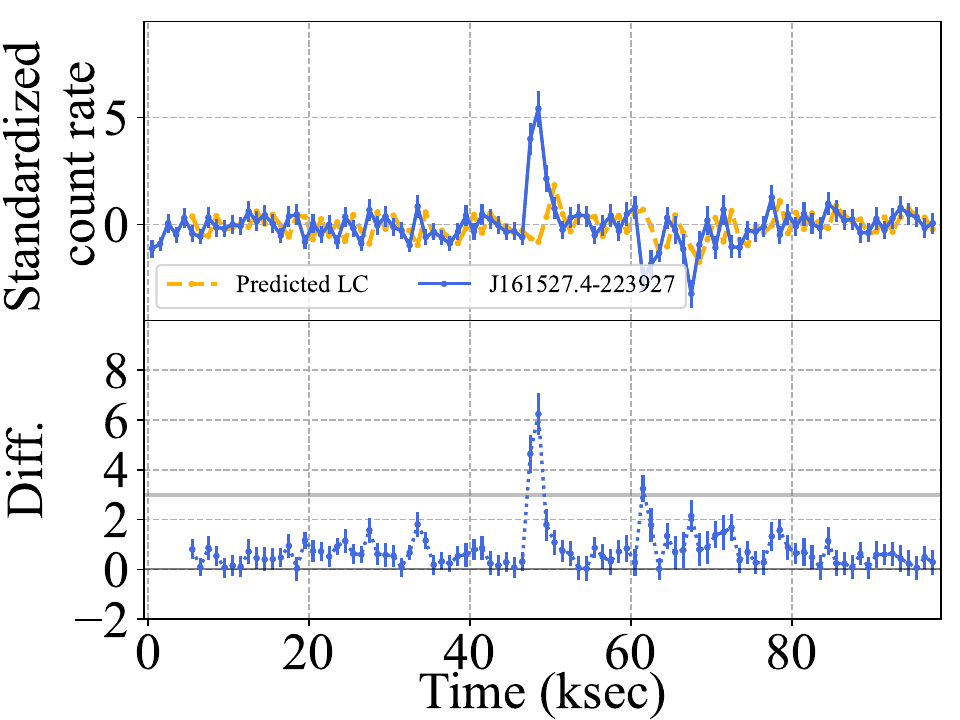}
  \includegraphics[width=3.9cm]{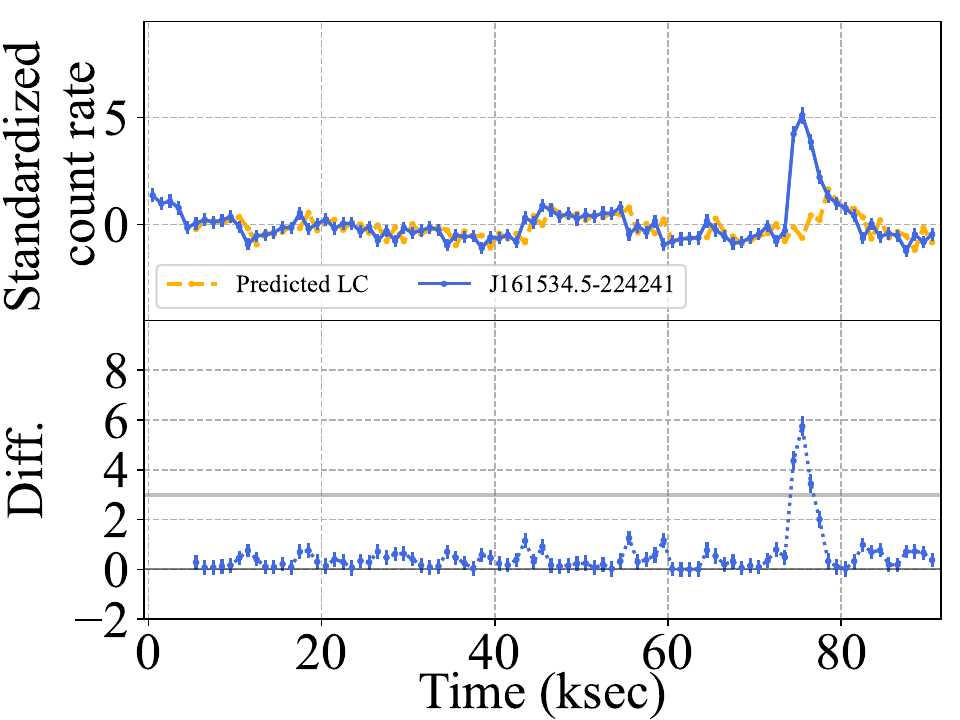}
  \includegraphics[width=3.9cm]{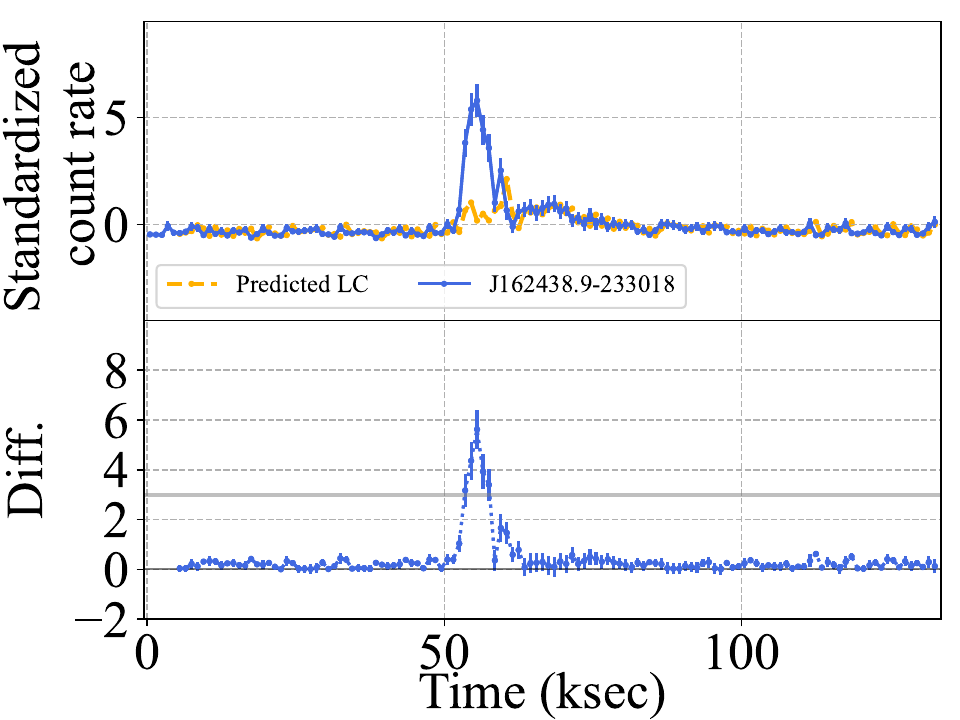}
  \includegraphics[width=3.9cm]{anomaly_fig_dir/J162522p1-233502_P0760900101PNS003SRCTSR800E_lc_1000_stand_wpred.pdf}
  \includegraphics[width=3.9cm]{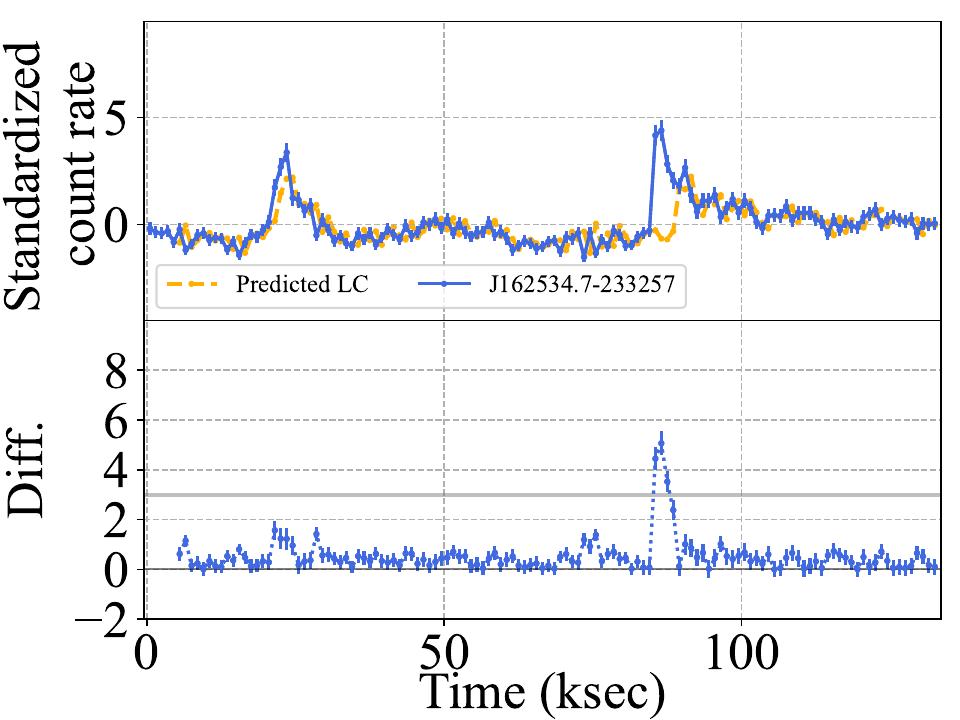}
  \includegraphics[width=3.9cm]{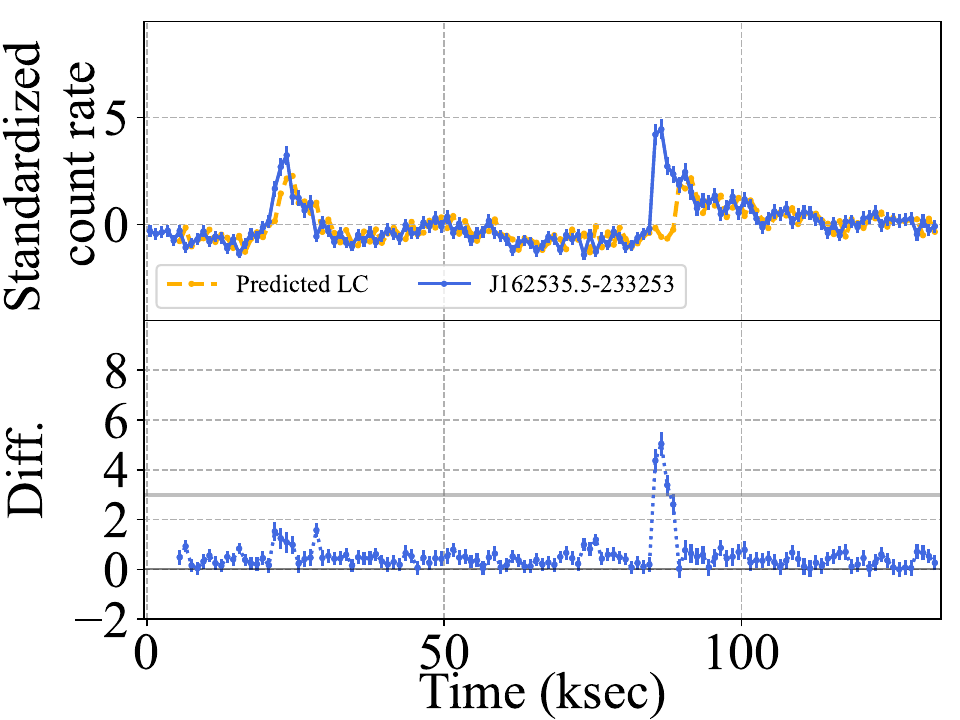}
  \includegraphics[width=3.9cm]{anomaly_fig_dir/J162544p8-232905_P0760900101PNS003SRCTSR8004_lc_1000_stand_wpred.pdf}
  \includegraphics[width=3.9cm]{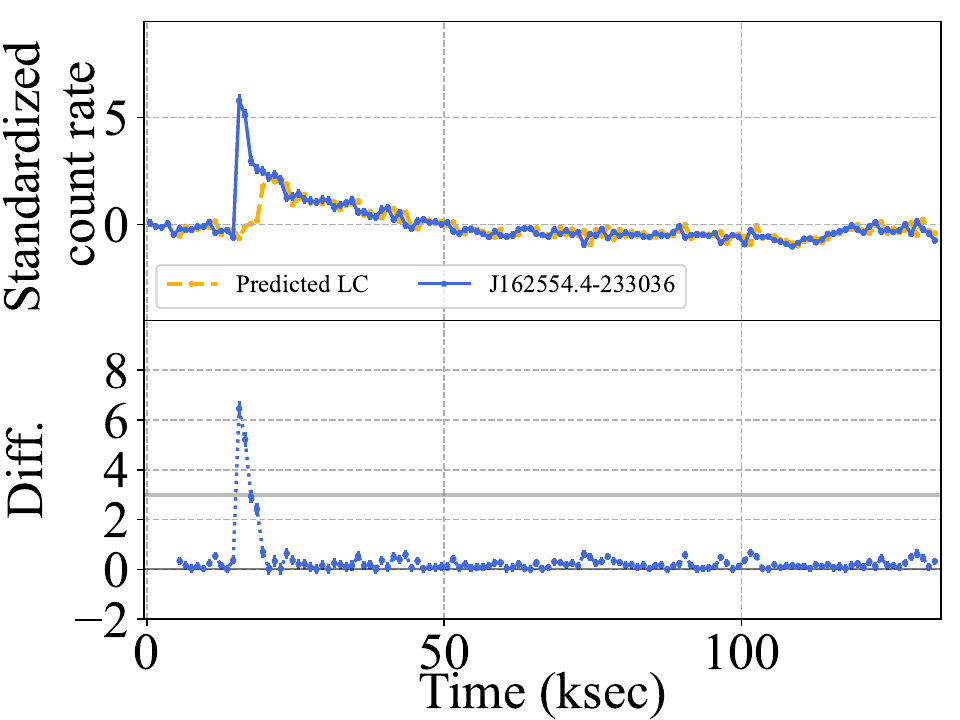}
  \includegraphics[width=3.9cm]{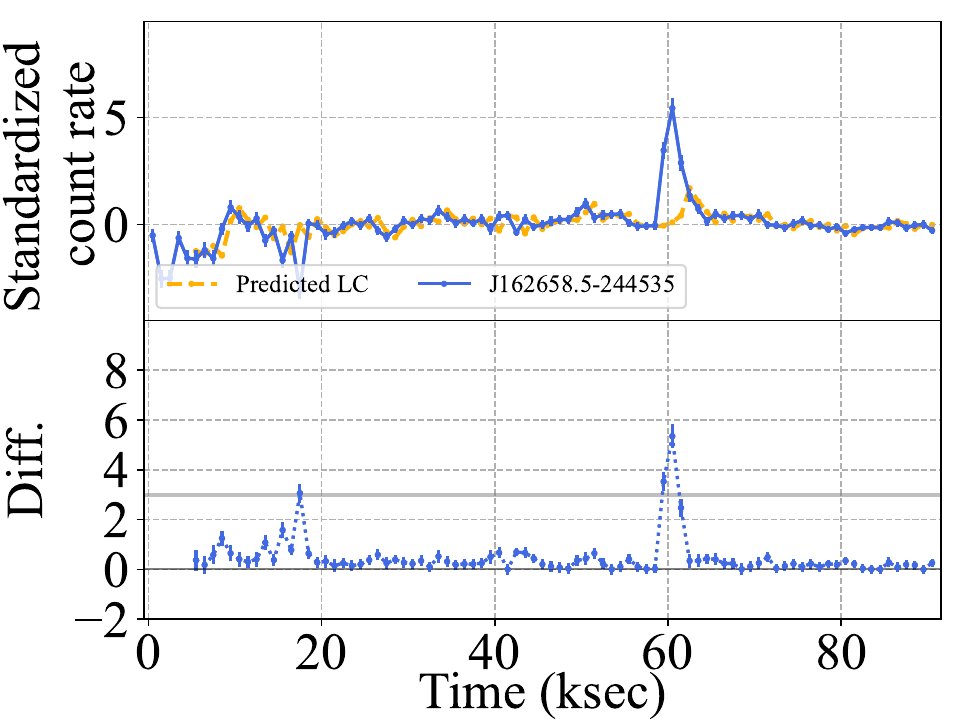}
  \includegraphics[width=3.9cm]{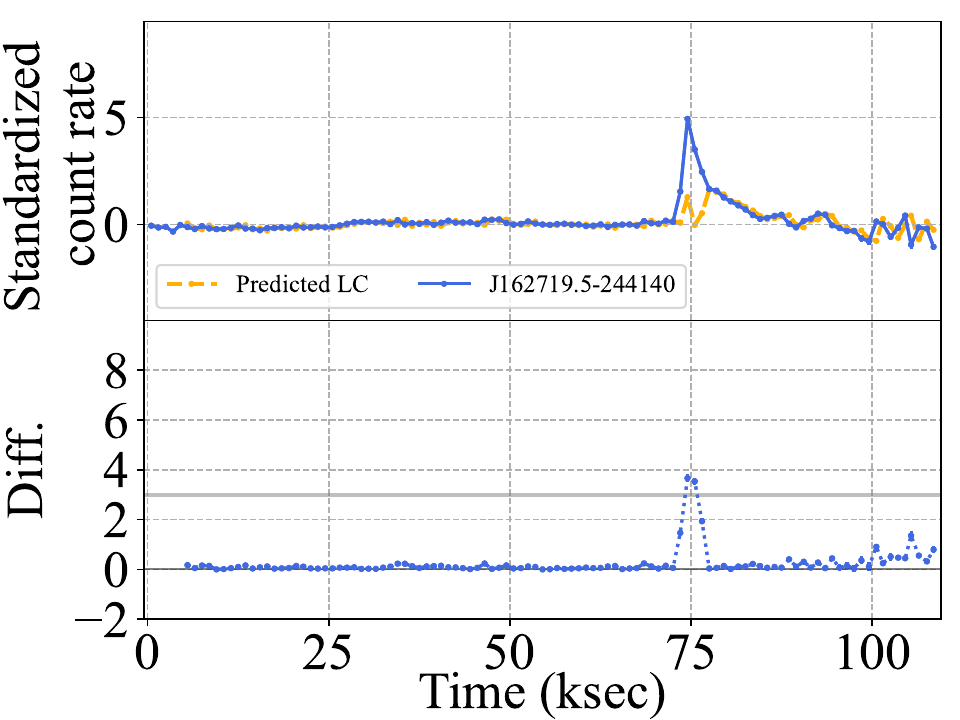}
  \includegraphics[width=3.9cm]{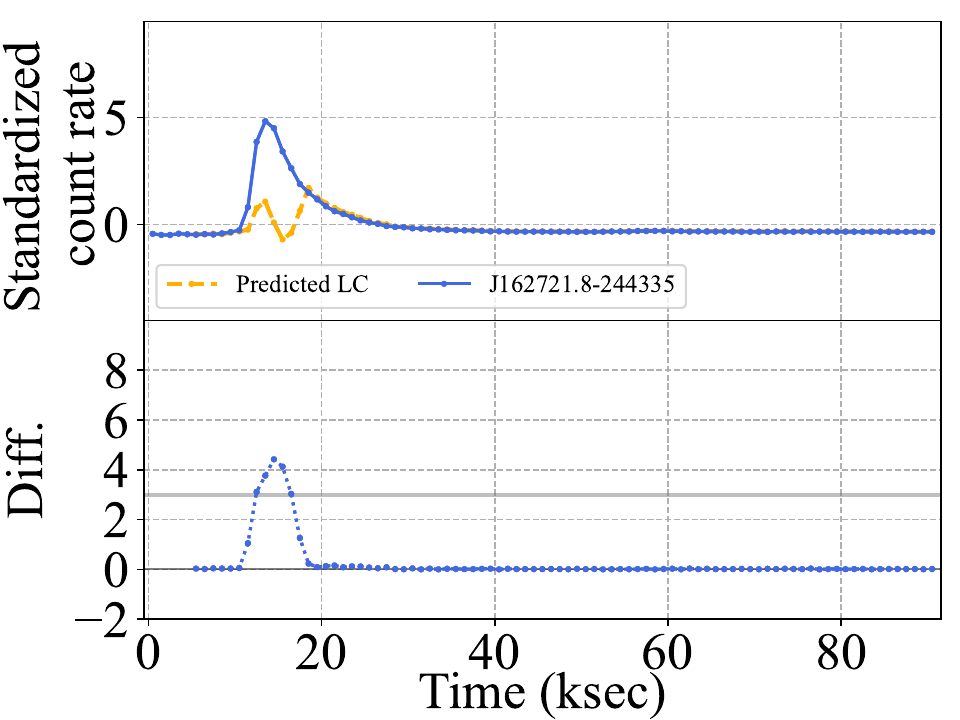}
  \includegraphics[width=3.9cm]{anomaly_fig_dir/J162722p2-244325_P0800031001PNU002SRCTSR801D_lc_1000_stand_wpred.pdf}
  \includegraphics[width=3.9cm]{anomaly_fig_dir/J162724p5-242934_P0800031001PNU002SRCTSR8003_lc_1000_stand_wpred.pdf}
  \includegraphics[width=3.9cm]{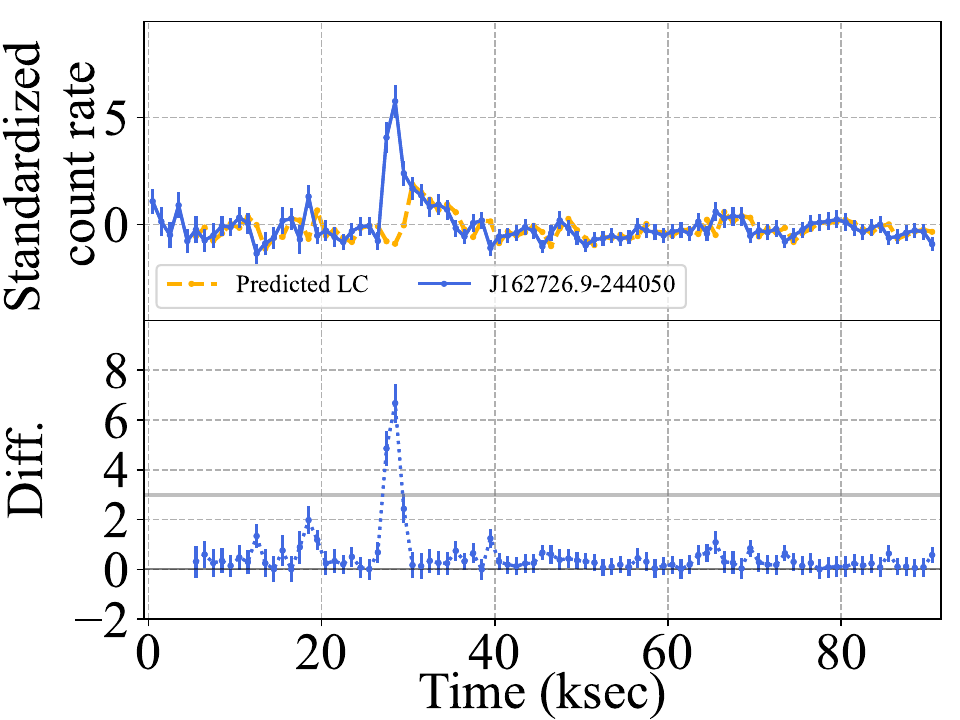}
  \includegraphics[width=3.9cm]{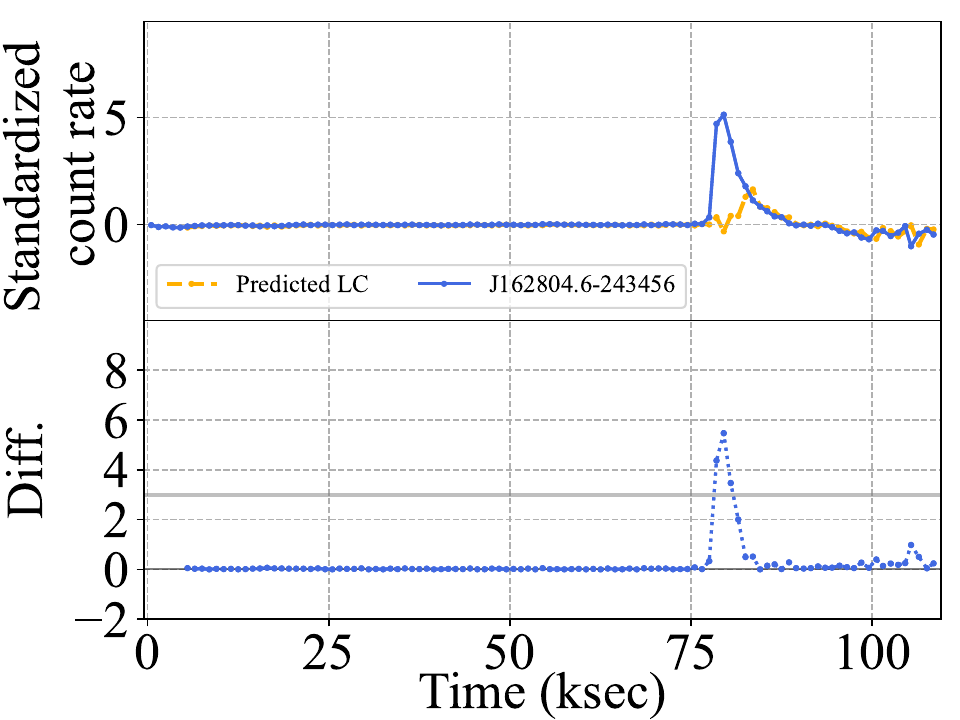}
  \includegraphics[width=3.9cm]{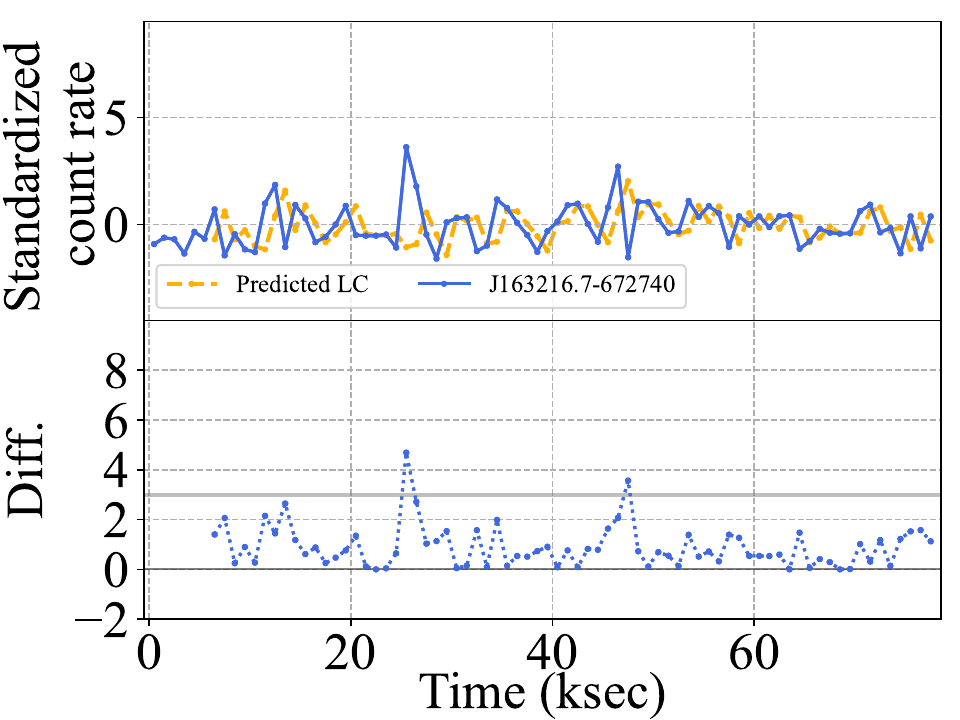}
  \includegraphics[width=3.9cm]{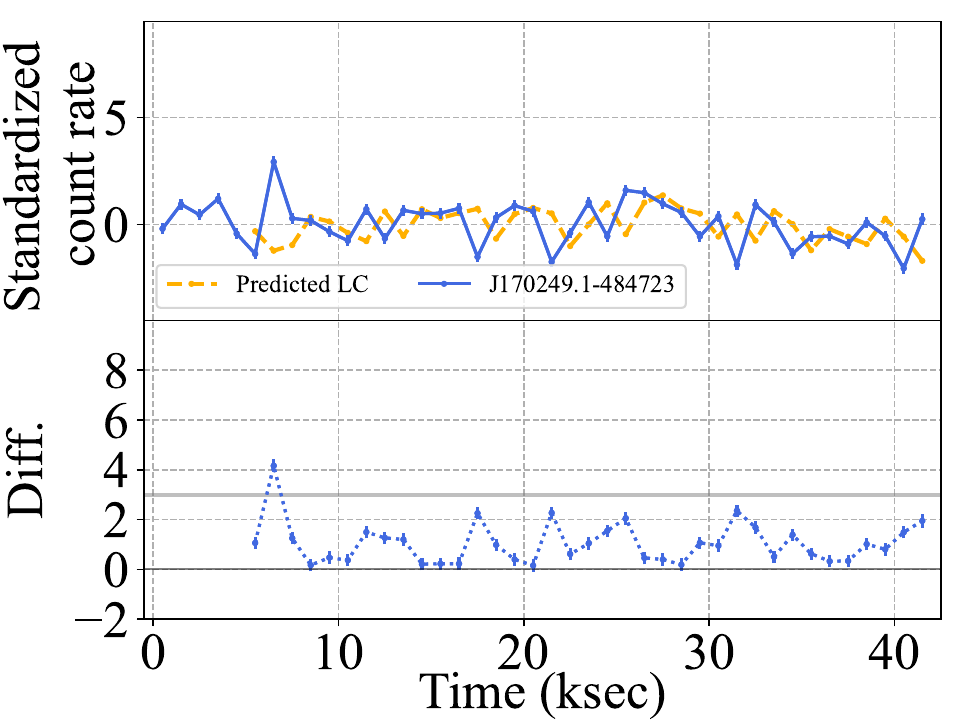}
  \includegraphics[width=3.9cm]{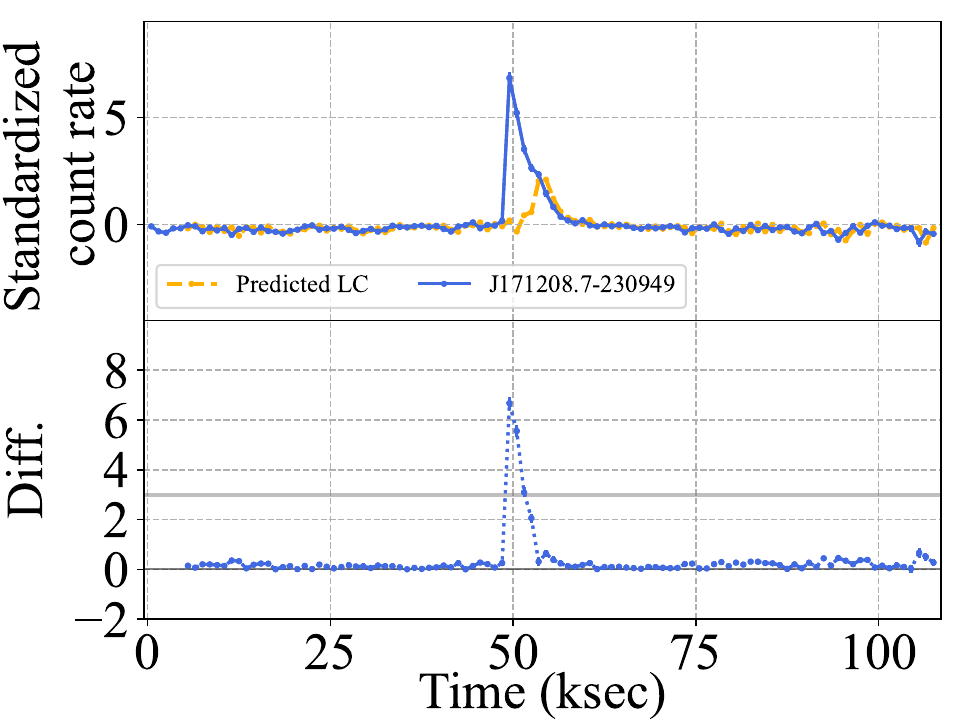}
  \includegraphics[width=3.9cm]{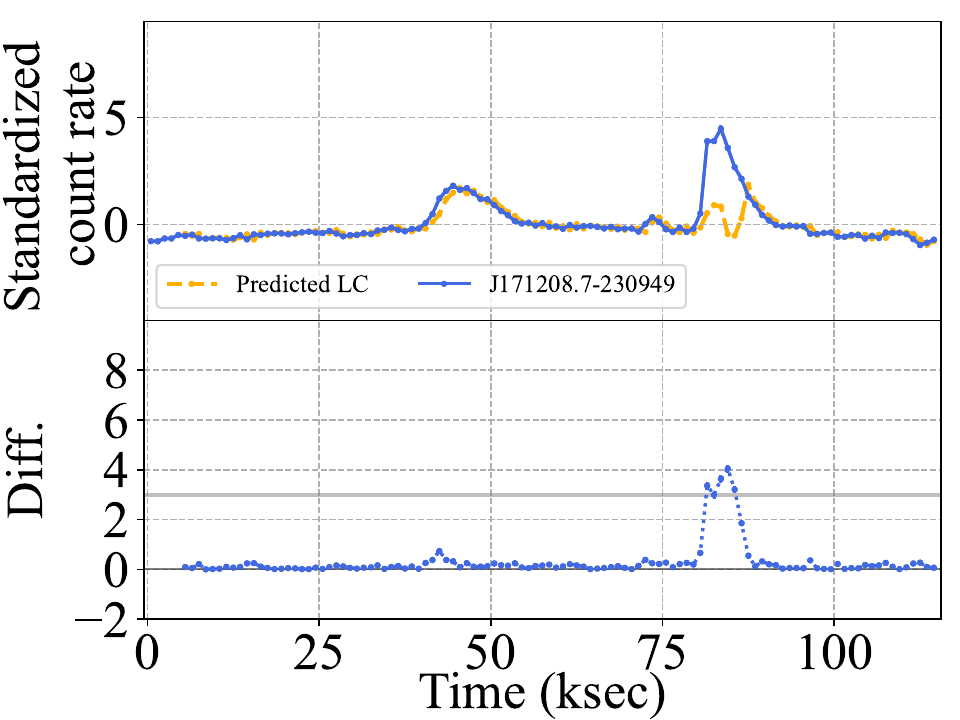}
  \includegraphics[width=3.9cm]{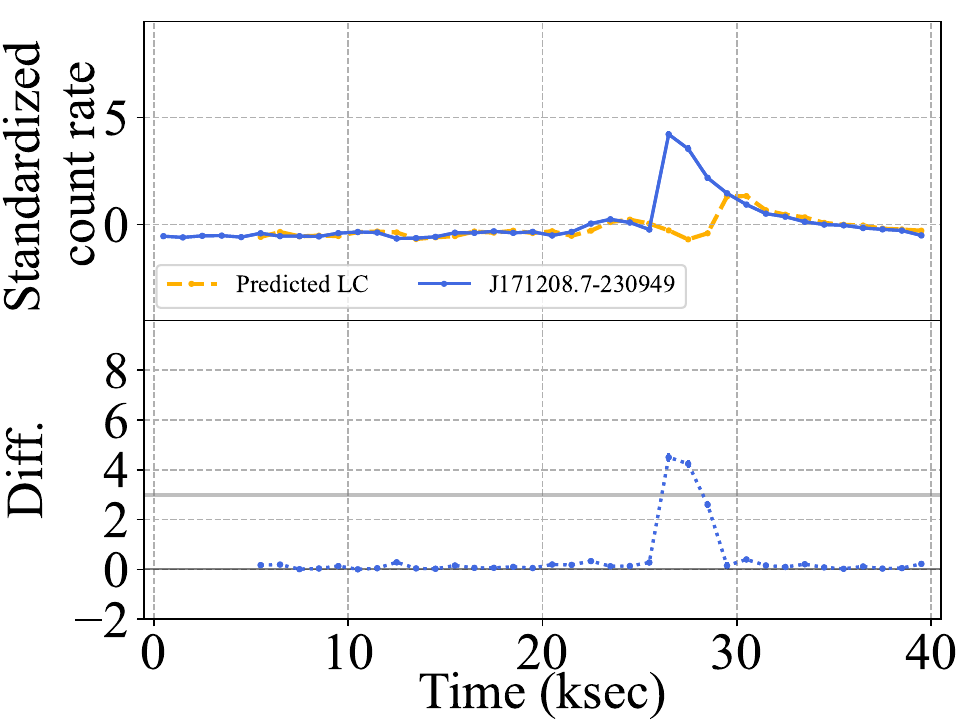}
  \includegraphics[width=3.9cm]{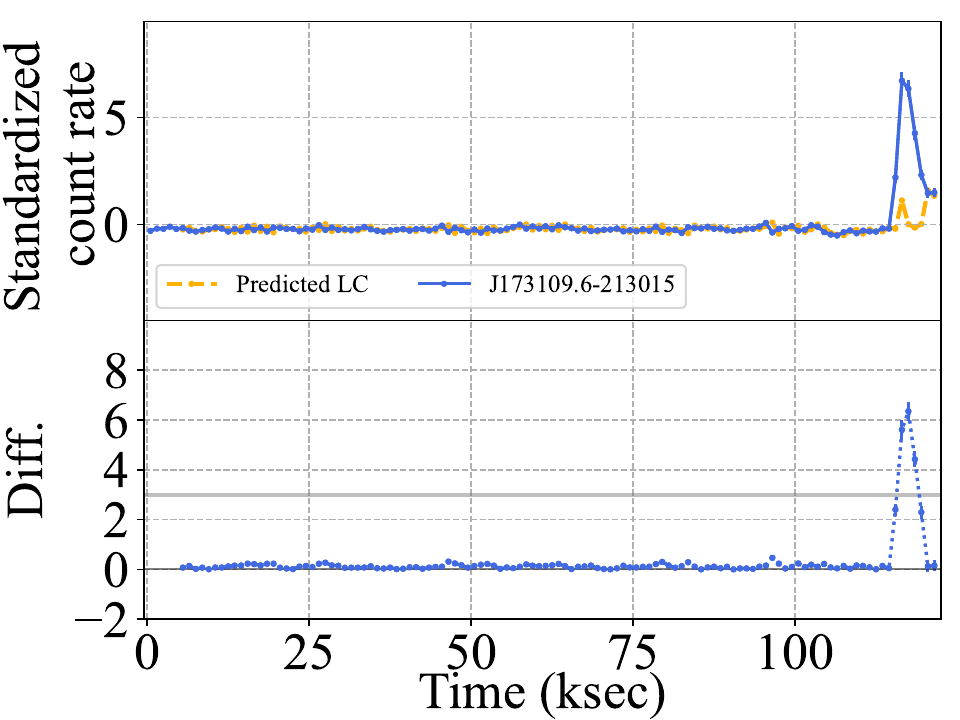}
  \includegraphics[width=3.9cm]{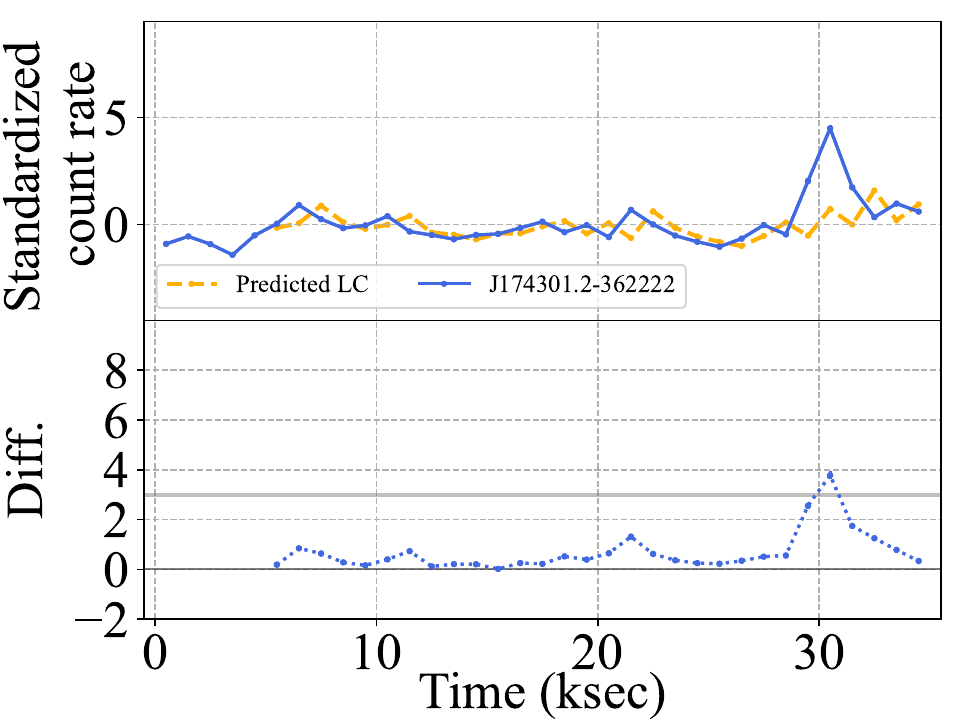}
 \end{center}
 \caption{
        Standardized XMM-Newton light curves in the 0.2--12 keV band (blue) and predicted ones (orange) based on the QLSTM model constructed with ZZFeatureMap and C20. The lower panels show 
        the absolute values of the differences between the real and predicted values. In each figure, the XMM-Newton name is denoted in the legend.
 }
 \label{fig:xmm_lc_supp2}
\end{figure*}
\begin{figure*}
 \begin{center}
  \includegraphics[width=3.9cm]{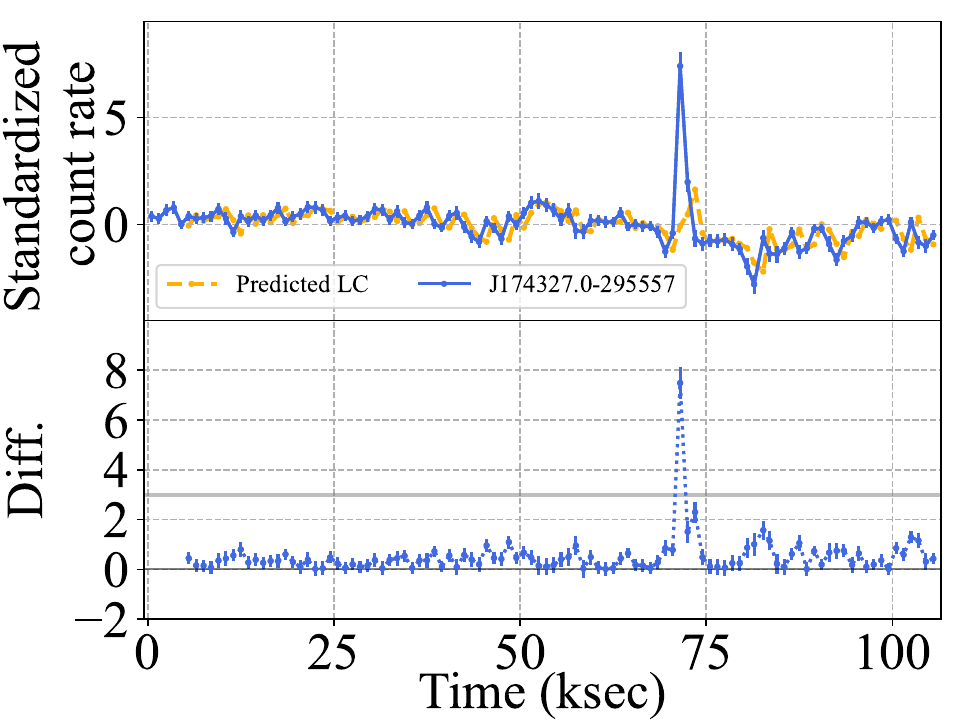}
  \includegraphics[width=3.9cm]{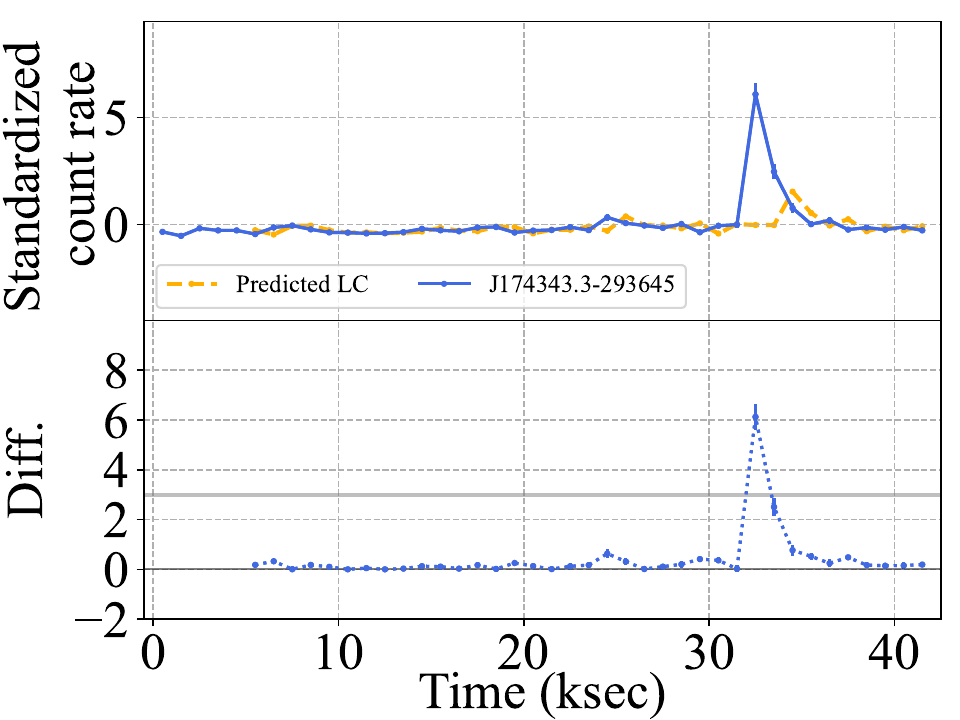}
  \includegraphics[width=3.9cm]{anomaly_fig_dir/J174502p3-285449_P0402430301PNS001SRCTSR8002_lc_1000_stand_wpred.pdf}
  \includegraphics[width=3.9cm]{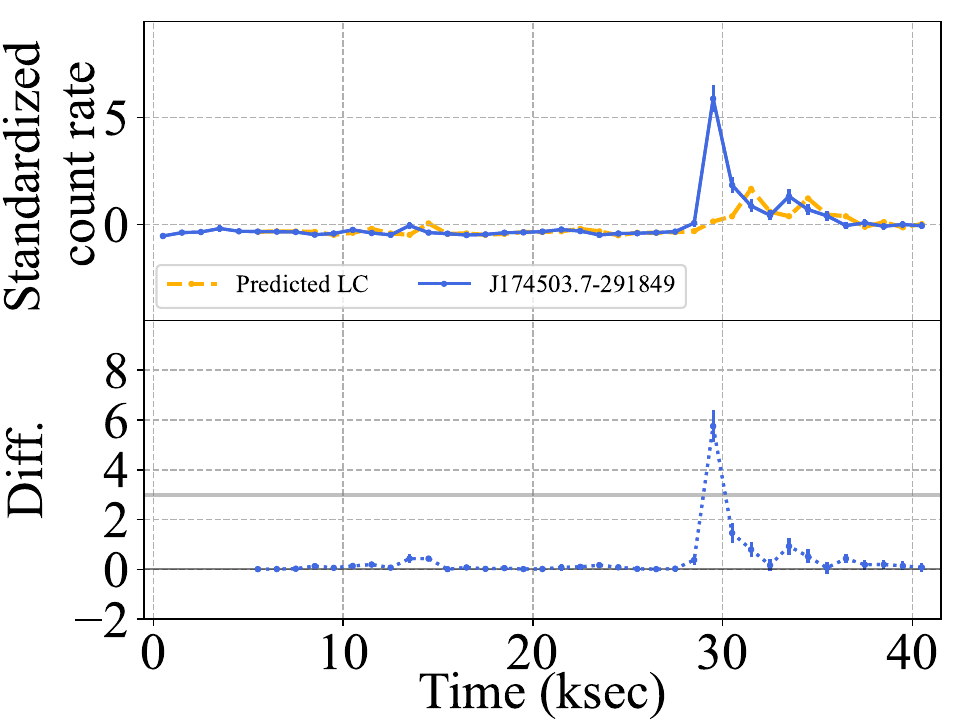}
  \includegraphics[width=3.9cm]{anomaly_fig_dir/J174505p3-291445_P0694640201PNS003SRCTSR8001_lc_1000_stand_wpred.pdf}
  \includegraphics[width=3.9cm]{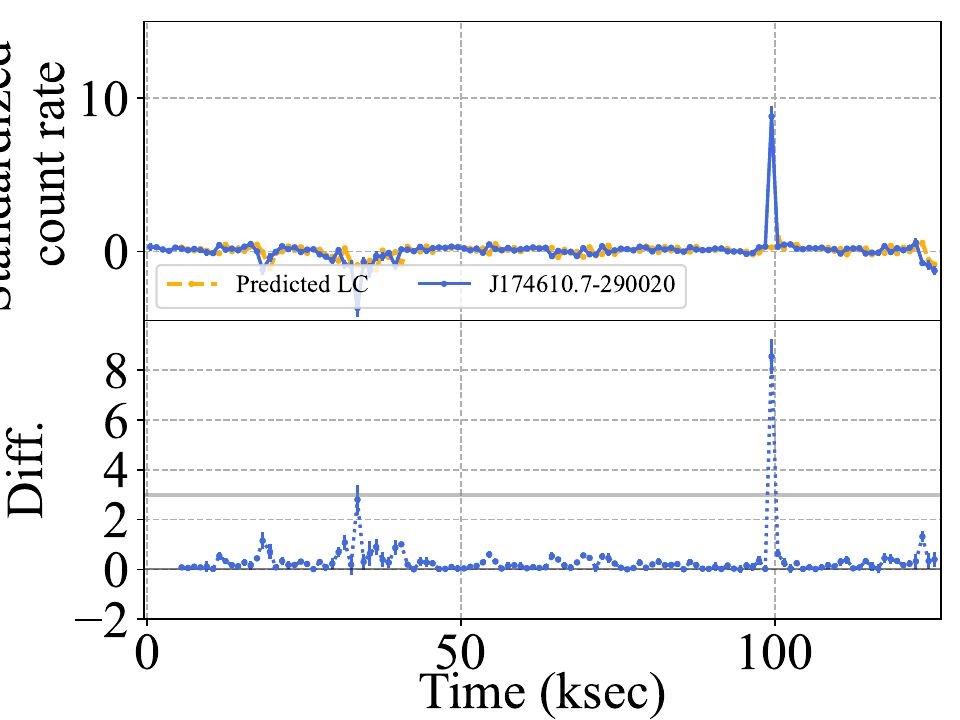}
  \includegraphics[width=3.9cm]{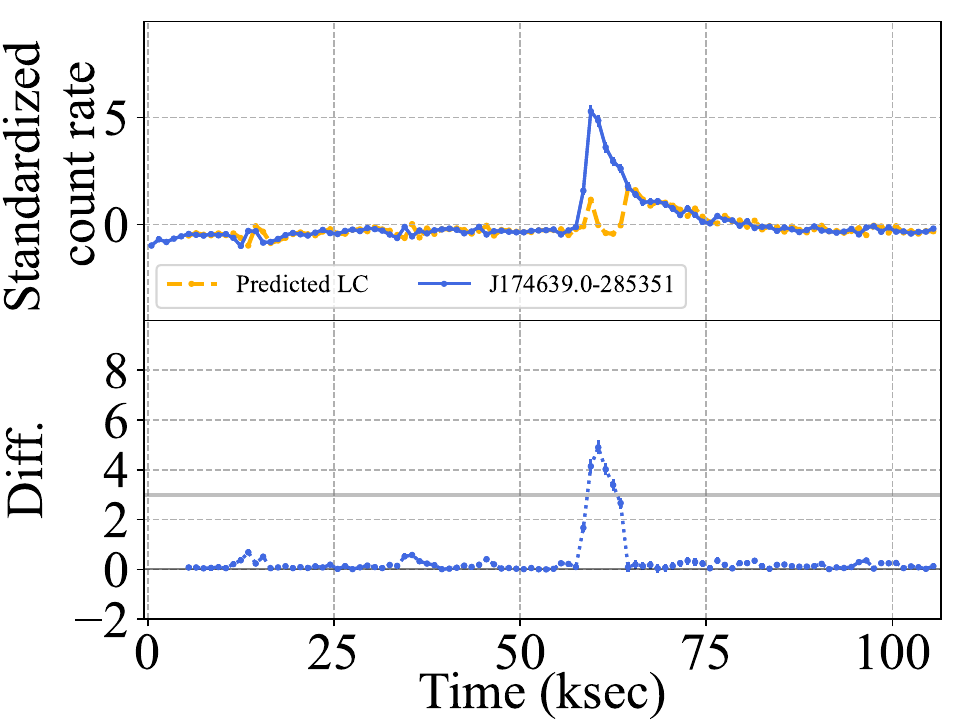}
  \includegraphics[width=3.9cm]{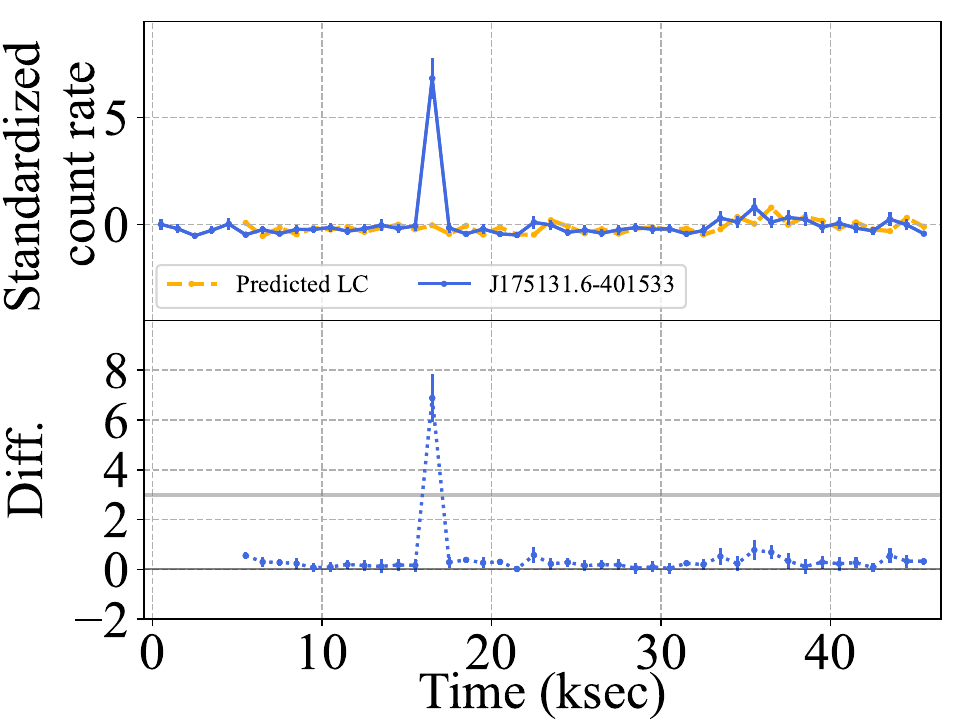}
  \includegraphics[width=3.9cm]{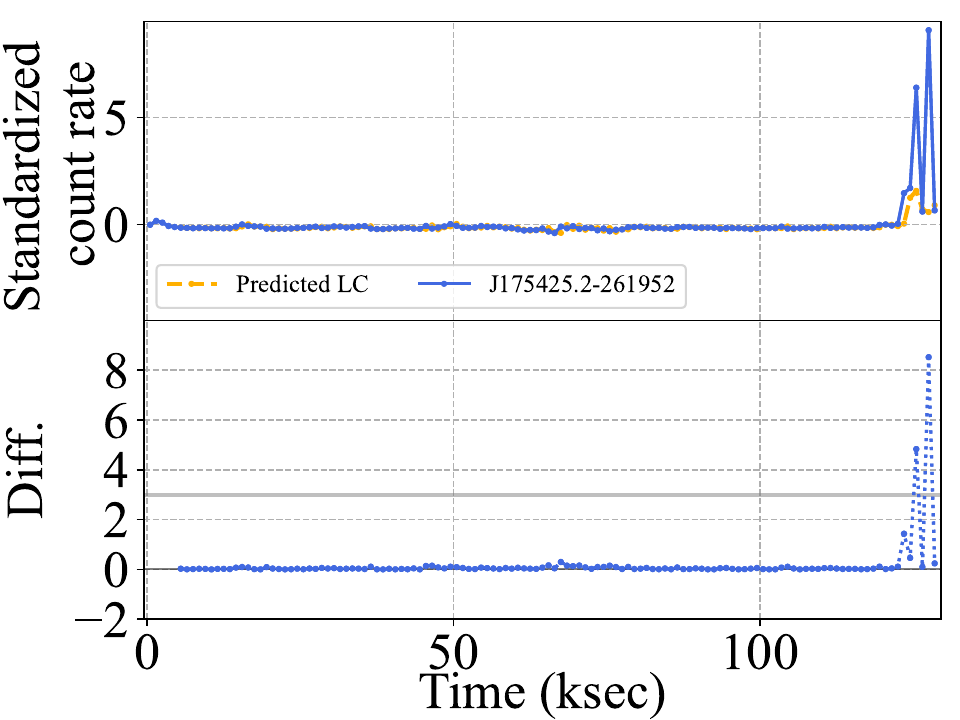}
  \includegraphics[width=3.9cm]{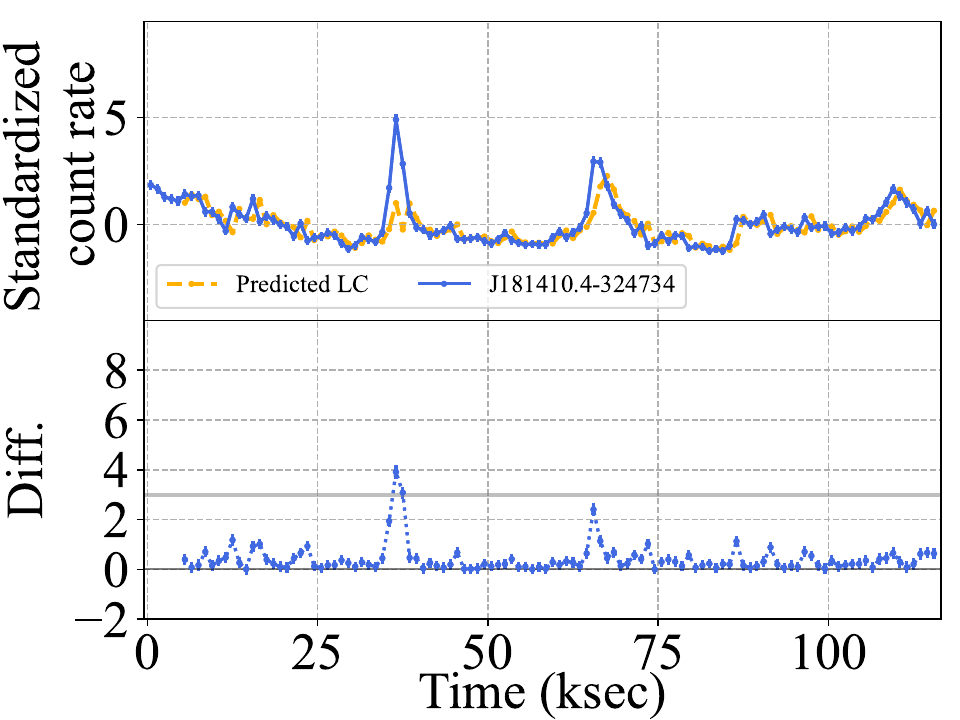}
  \includegraphics[width=3.9cm]{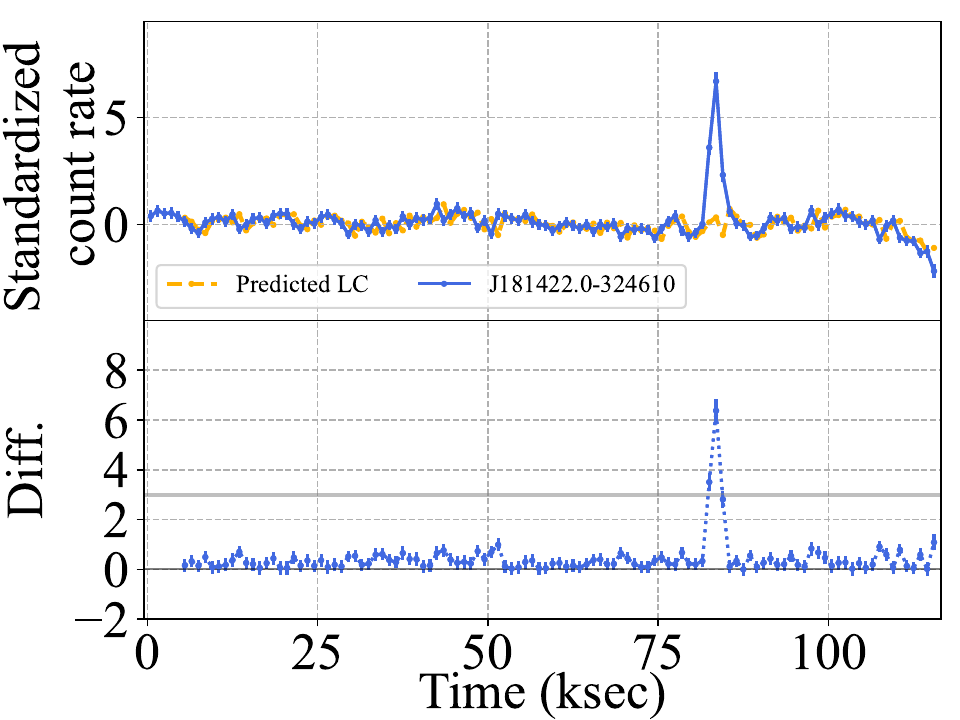}
  \includegraphics[width=3.9cm]{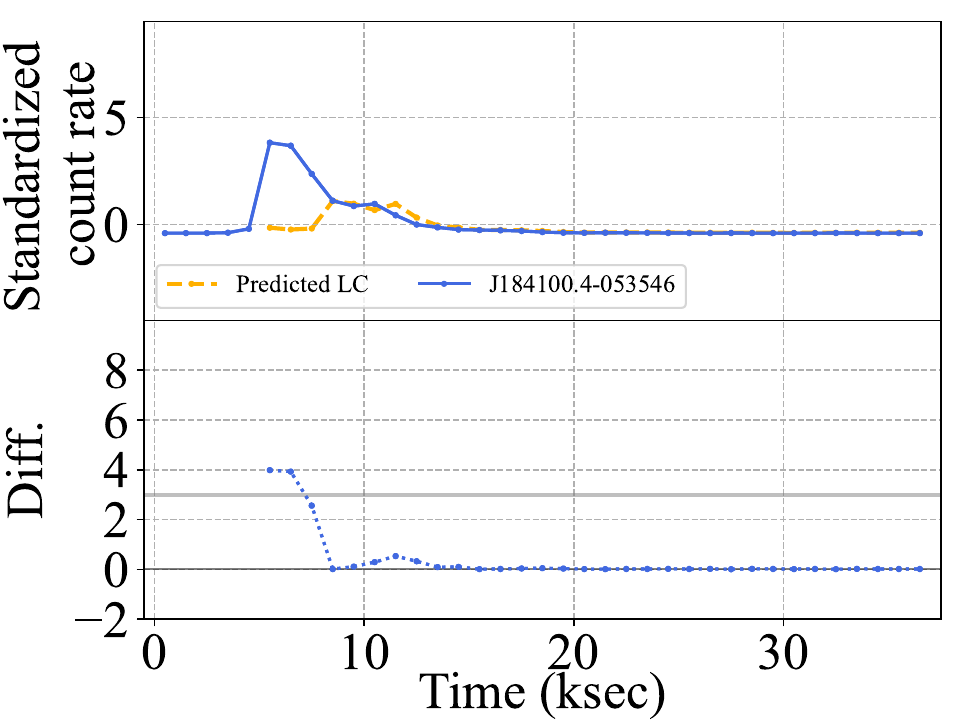}
  \includegraphics[width=3.9cm]{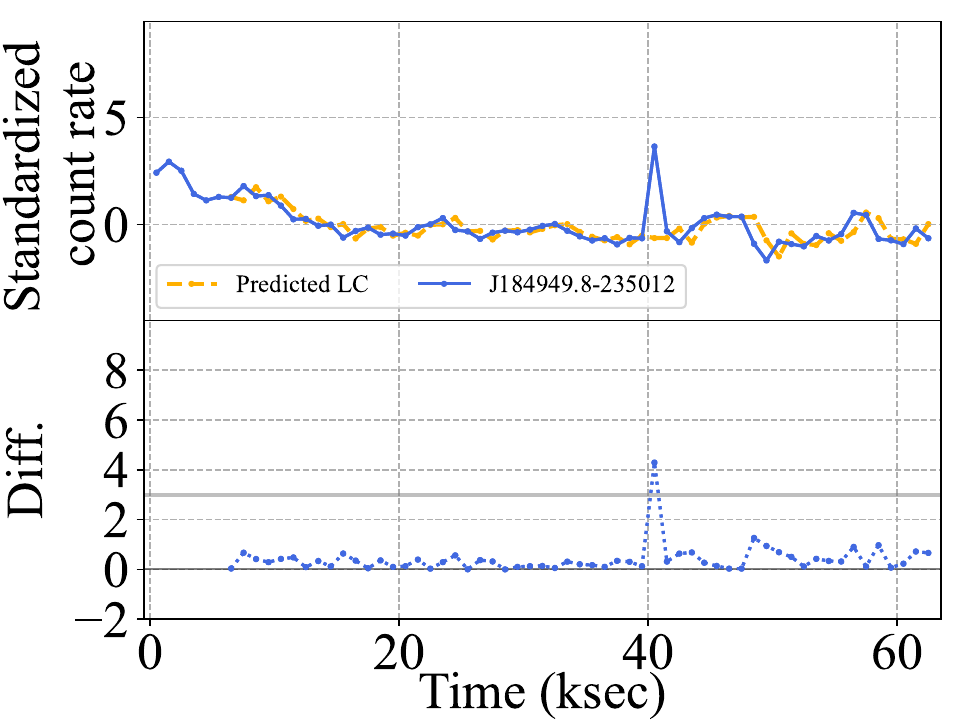}
  \includegraphics[width=3.9cm]{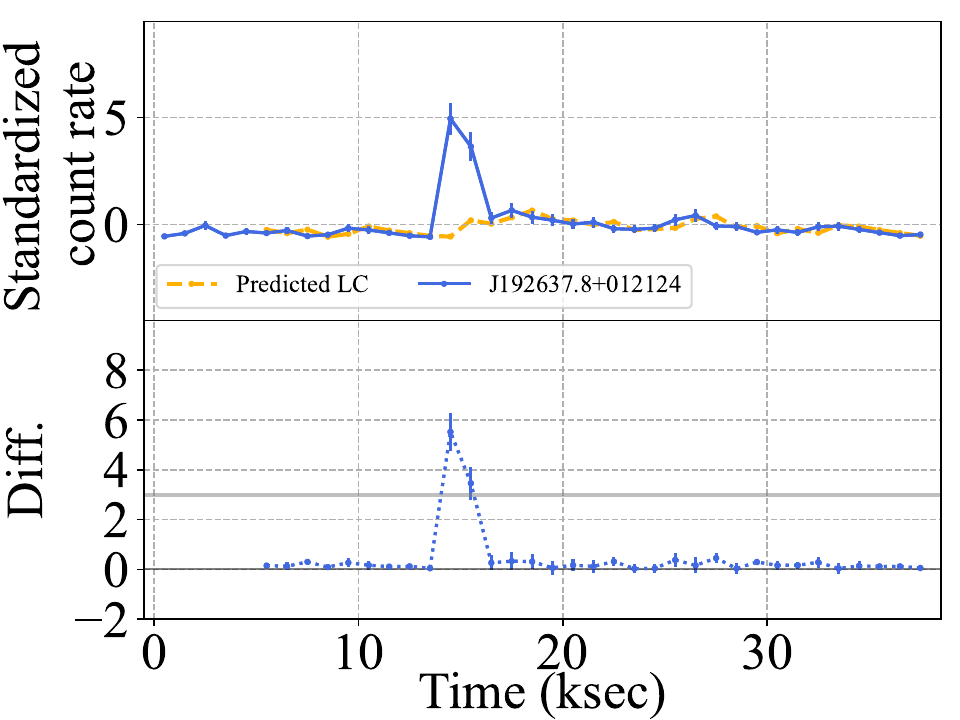}
  \includegraphics[width=3.9cm]{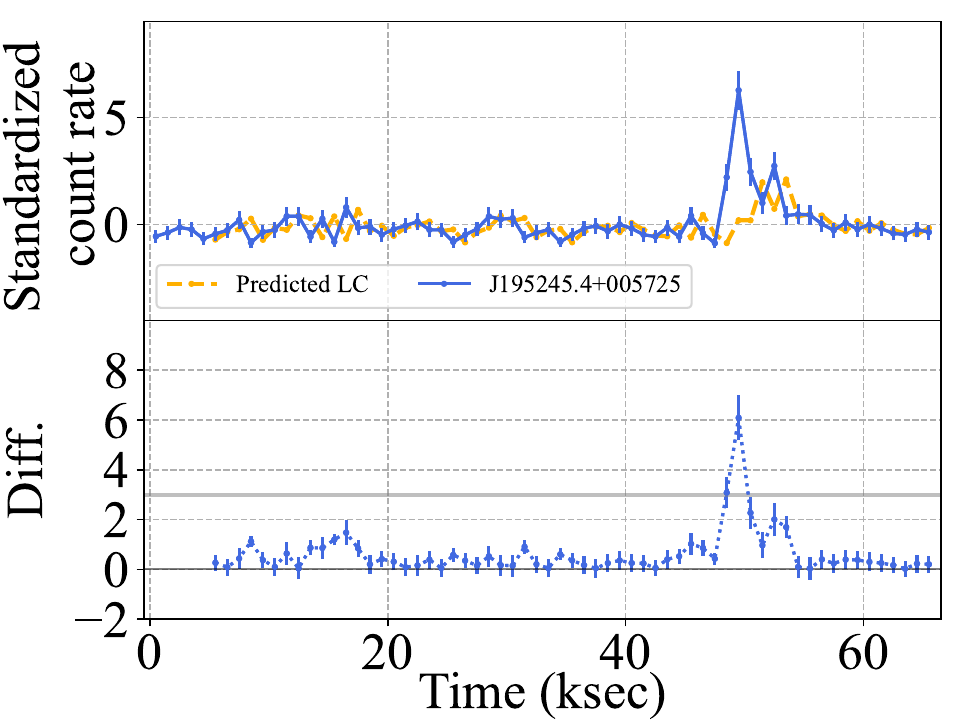}
  \includegraphics[width=3.9cm]{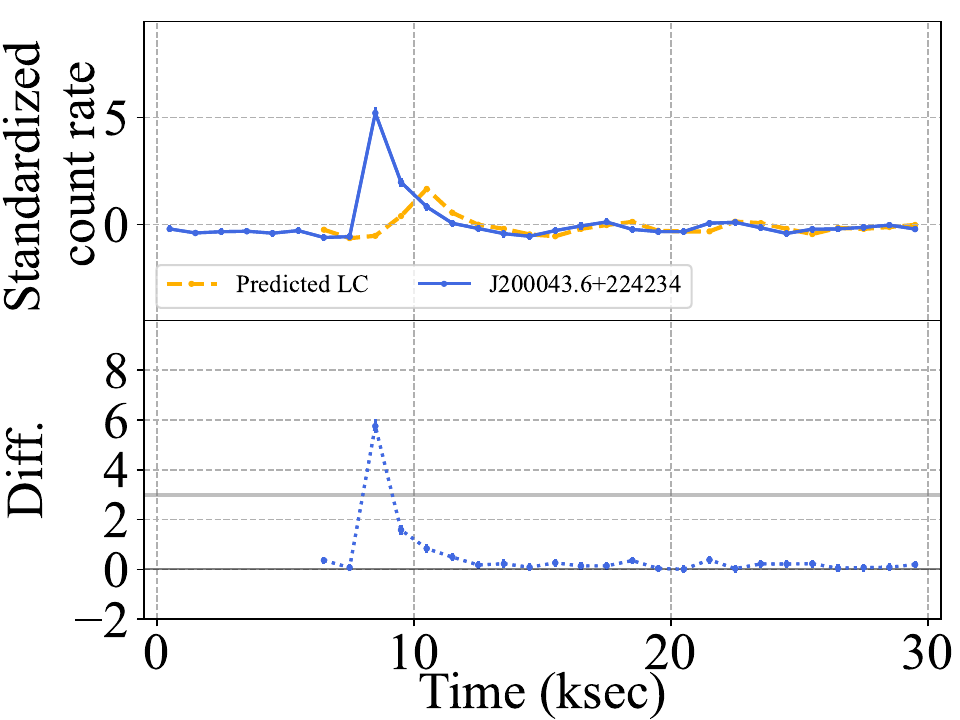}
  \includegraphics[width=3.9cm]{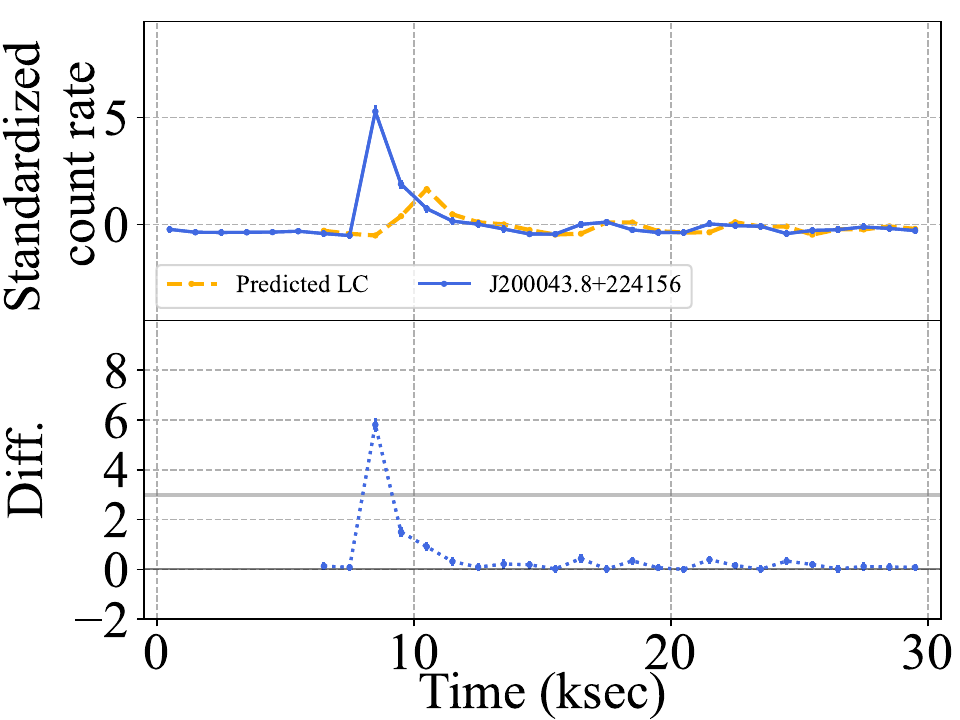}
  \includegraphics[width=3.9cm]{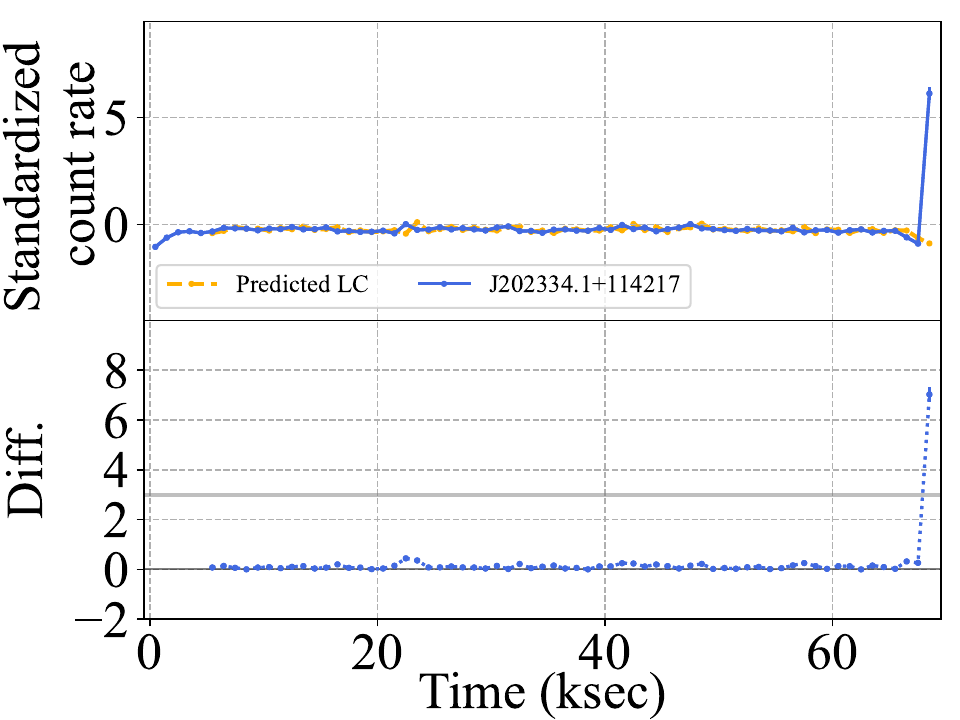}
  \includegraphics[width=3.9cm]{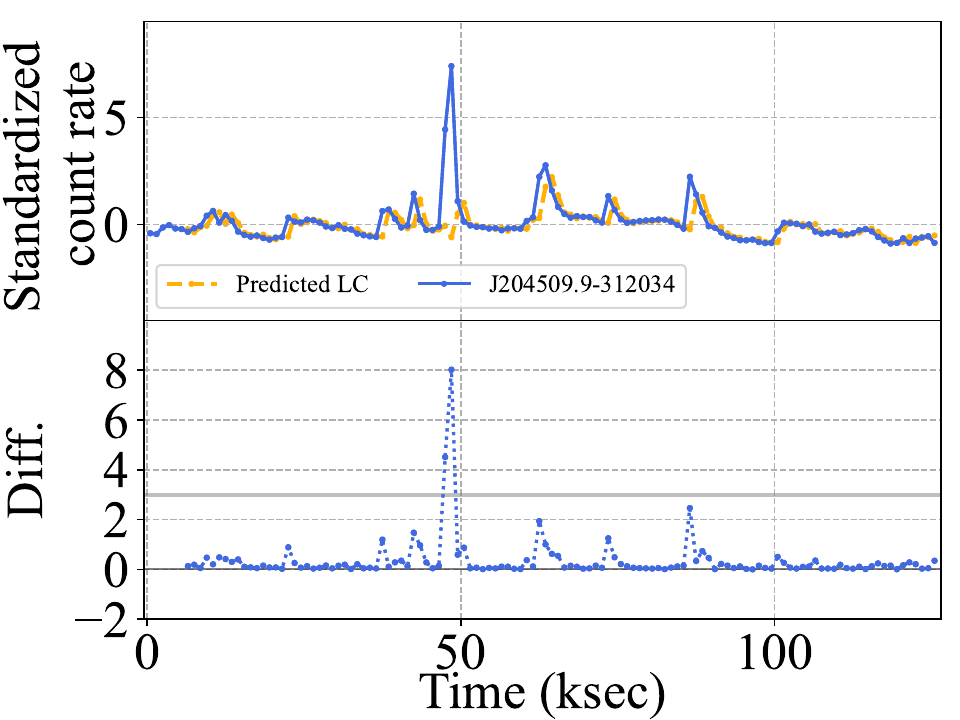}
  \includegraphics[width=3.9cm]{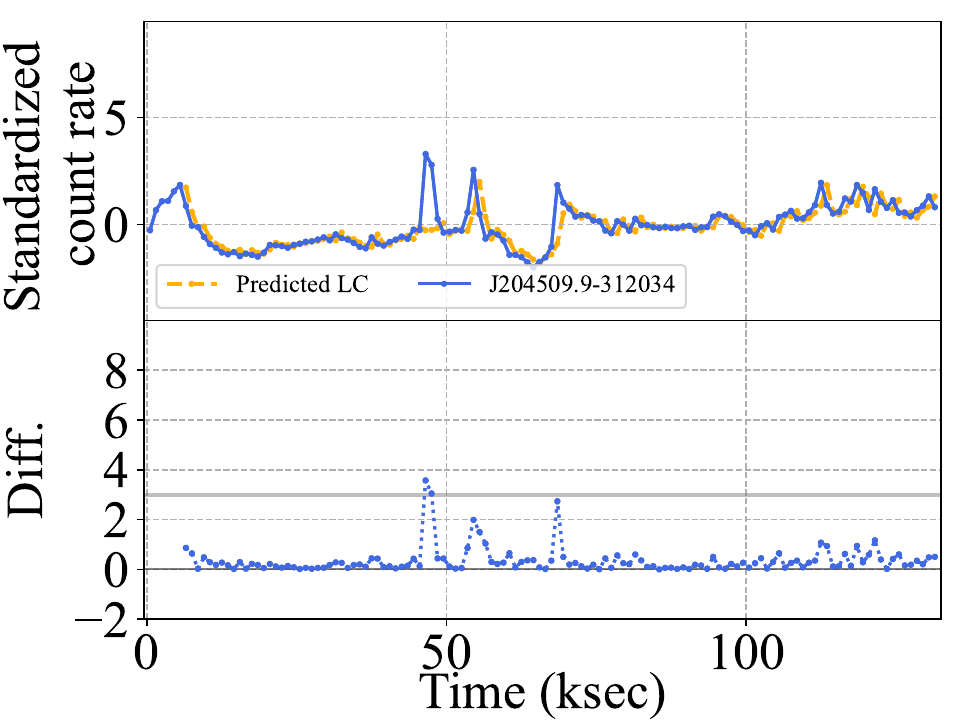}
  \includegraphics[width=3.9cm]{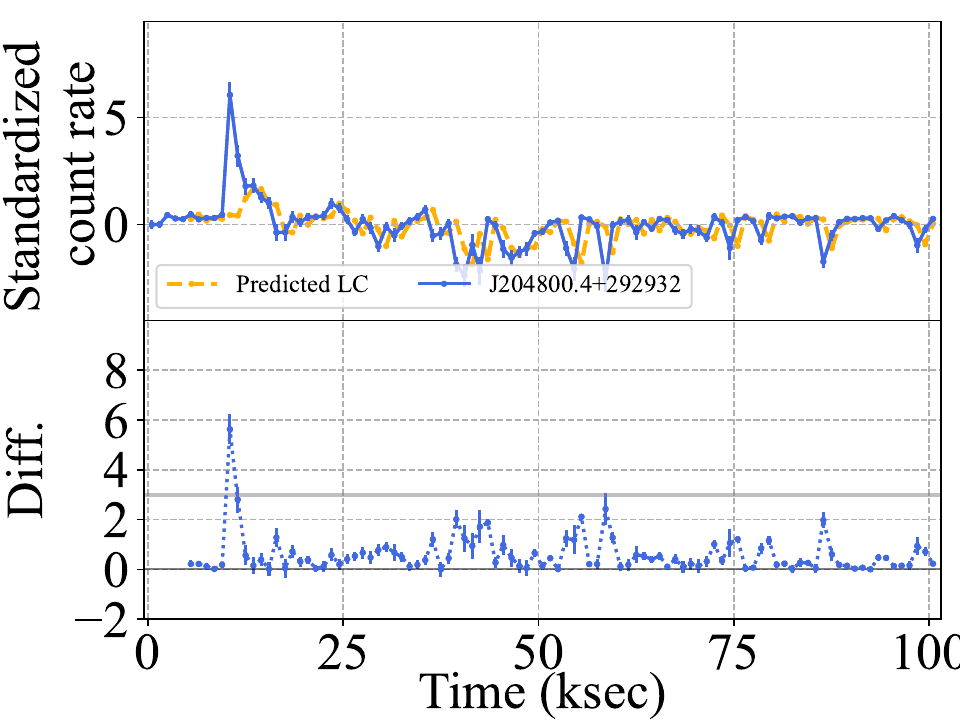}
  \includegraphics[width=3.9cm]{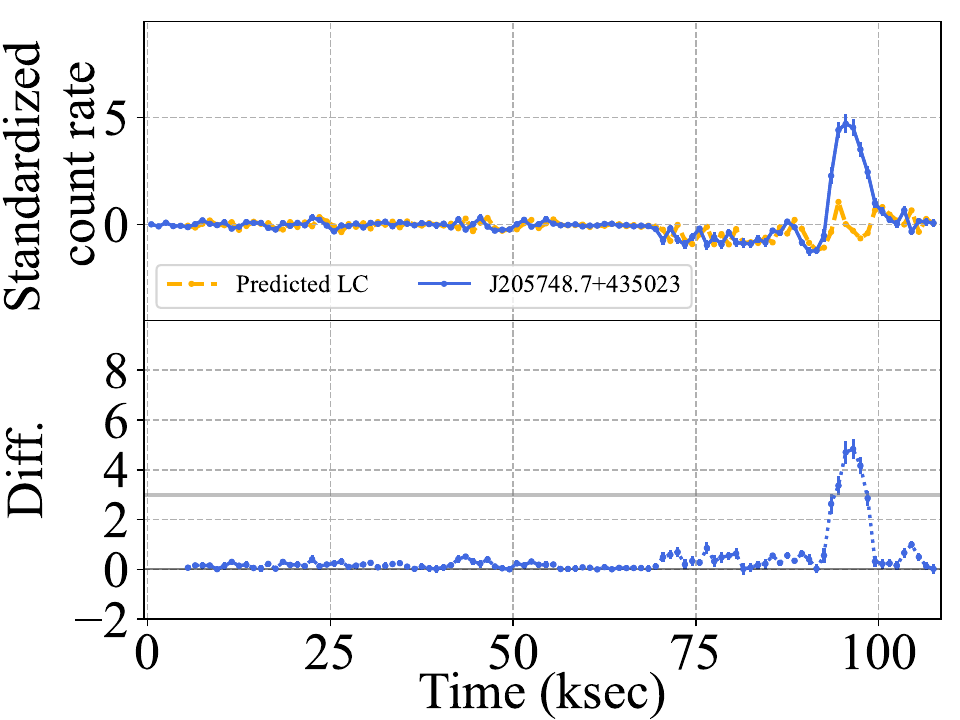}
  \includegraphics[width=3.9cm]{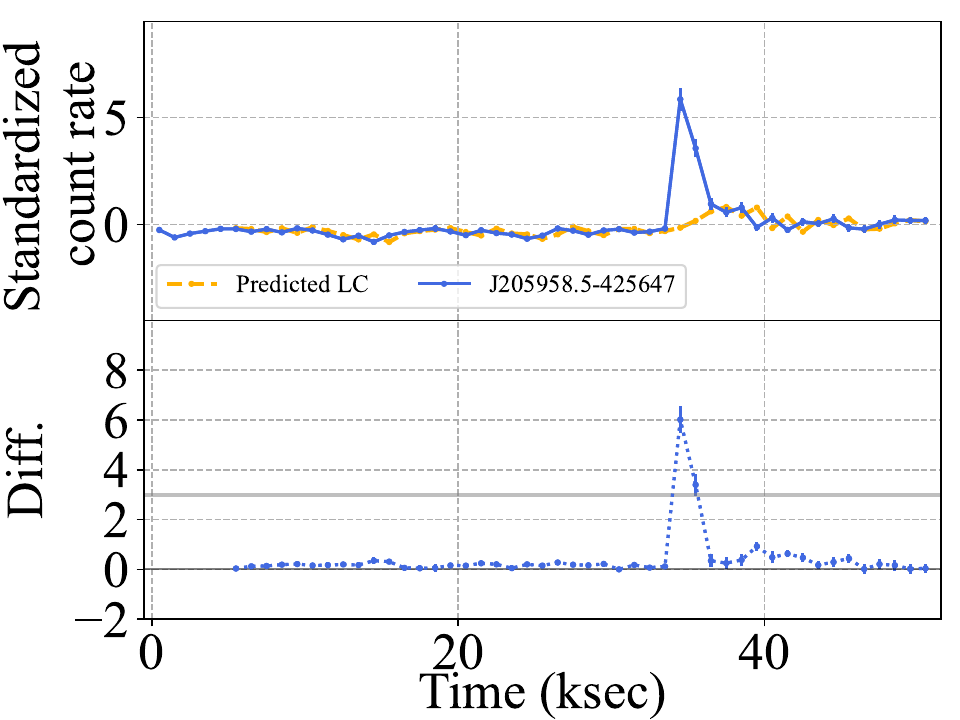}
  \includegraphics[width=3.9cm]{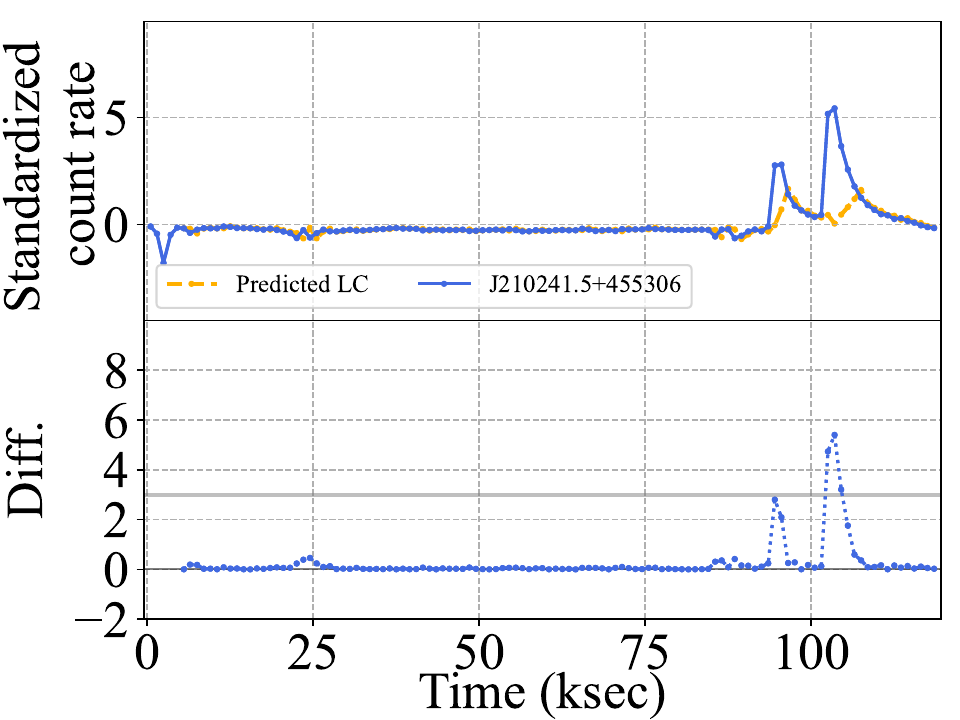}
  \includegraphics[width=3.9cm]{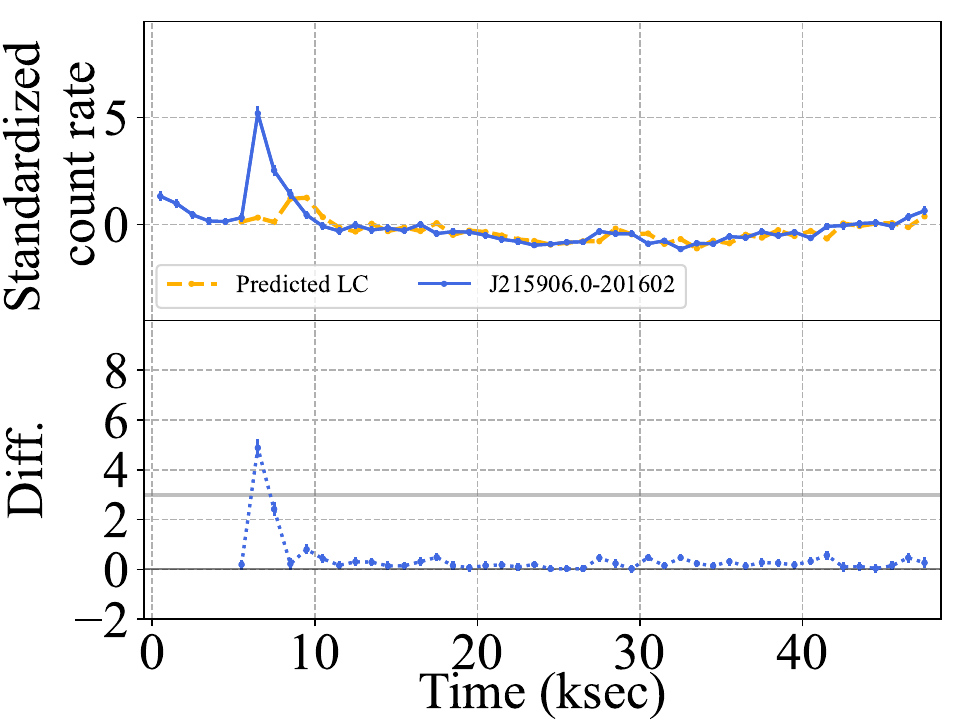}
  \includegraphics[width=3.9cm]{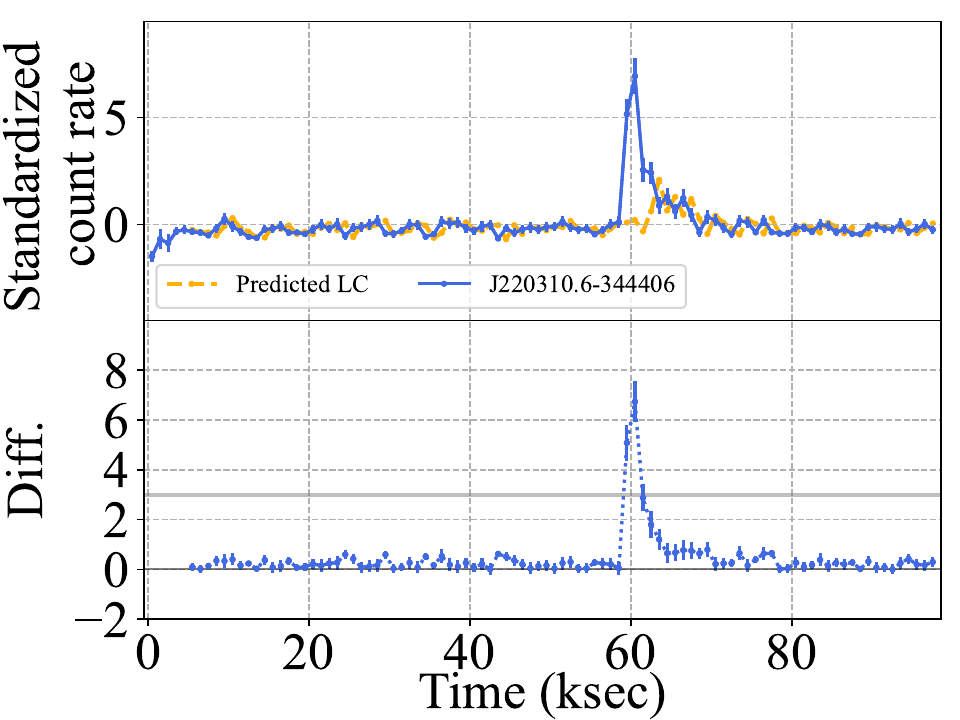}
  \includegraphics[width=3.9cm]{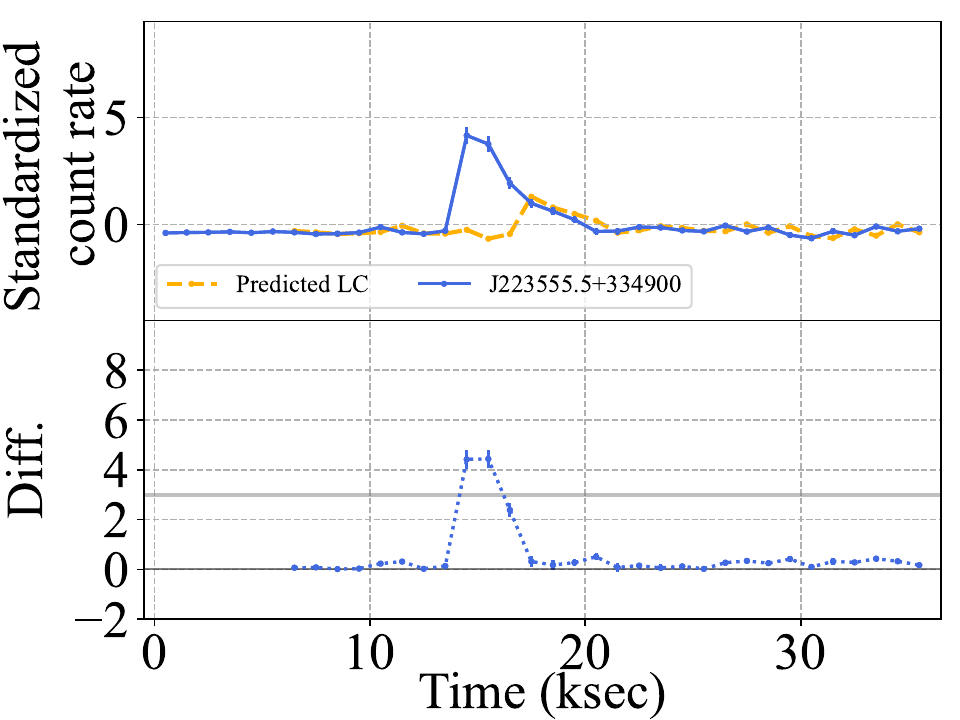}
  \includegraphics[width=3.9cm]{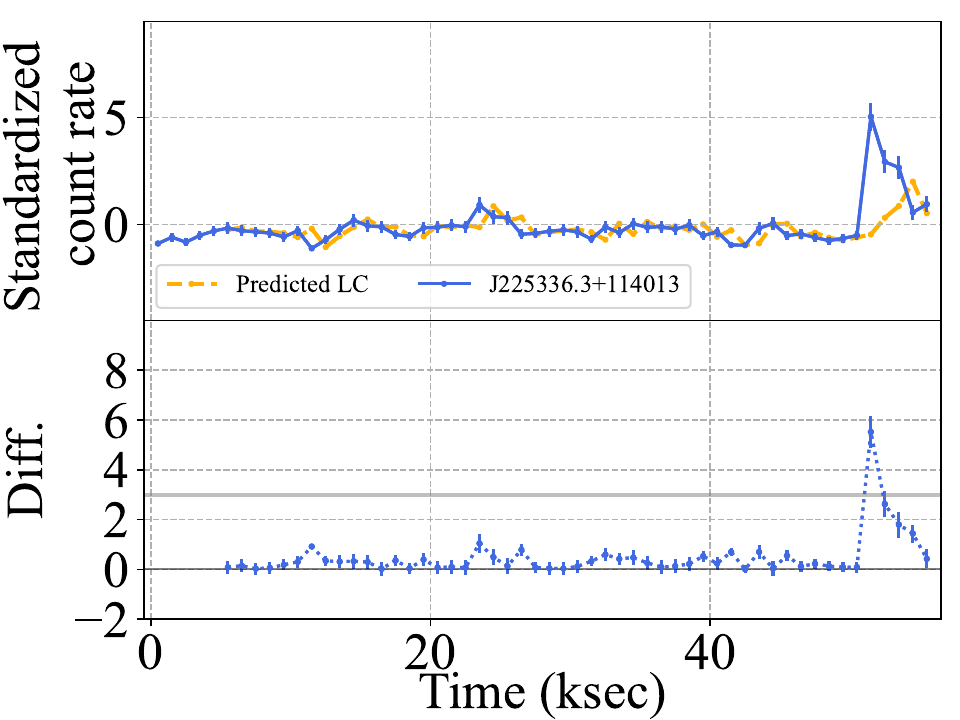}
 \end{center}
 \caption{
        Standardized XMM-Newton light curves in the 0.2--12 keV band (blue) and predicted ones (orange) based on the QLSTM model constructed with ZZFeatureMap and C20. The lower panels show 
        the absolute values of the differences between the real and predicted values. In each figure, the XMM-Newton name is denoted in the legend.
 }
 \label{fig:xmm_lc_supp3}
\end{figure*}
\begin{figure*}
 \begin{center}
  \includegraphics[width=3.9cm]{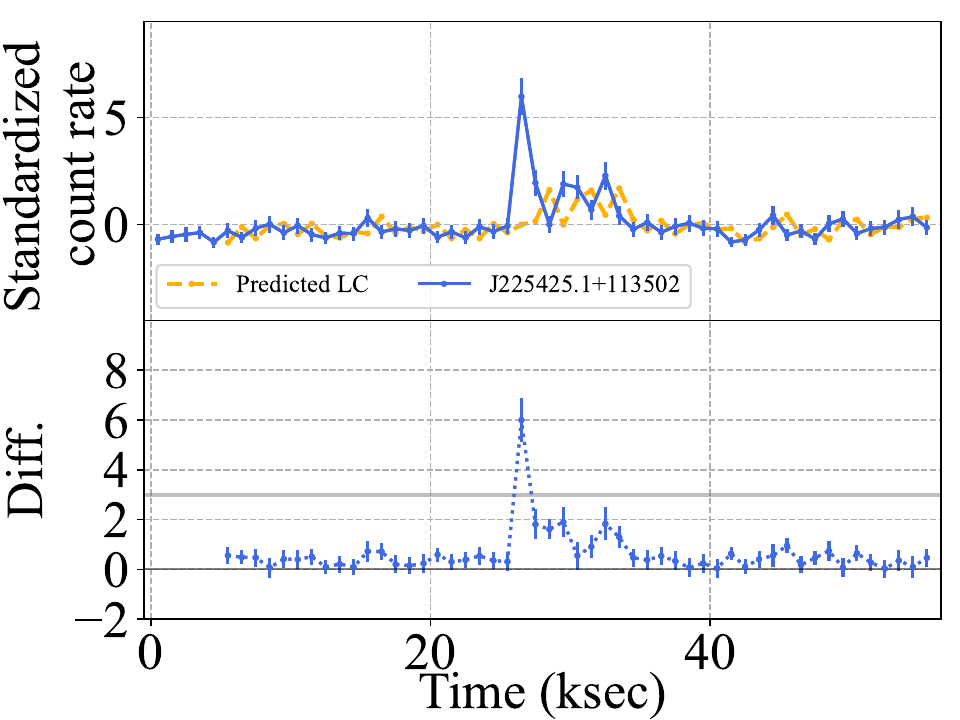}
 \end{center}
 \caption{
        Standardized XMM-Newton light curves in the 0.2--12 keV band (blue) and predicted ones (orange) based on the QLSTM model constructed with ZZFeatureMap and C20. The lower panels show 
        the absolute values of the differences between the real and predicted values. In each figure, the XMM-Newton name is denoted in the legend.
 }
 \label{fig:xmm_lc_supp4}
\end{figure*}

We present the light curves of 113 XMM-Newton detections for which anomalous behaviors are identified from their light curves using a QLSTM model (Figures~\ref{fig:xmm_lc_supp0},~\ref{fig:xmm_lc_supp1},~\ref{fig:xmm_lc_supp2},~\ref{fig:xmm_lc_supp3}, and ~\ref{fig:xmm_lc_supp4}) and their catalog (Table~\ref{tab:anom_catalog}). The model is constructed by adopting the feature map ZZFeatureMap and the ansatz C20.

\begin{longrotatetable}
\begin{deluxetable*}{lrrrrrrrrrrr}
\tablecaption{Catalog of detected transient event candidates\label{tab:anom_catalog}}
\tablewidth{700pt}
\tabletypesize{\scriptsize}
\tablehead{
\colhead{4XMM name} &
\colhead{Obs. ID} & 
\colhead{Det. ID} & 
\colhead{R.A.} & 
\colhead{Decl.} & 
\colhead{Counterpart name} & 
\colhead{type} & 
\colhead{R} & 
\colhead{SM} & 
\colhead{Stel.} & 
\colhead{PS} & 
\colhead{W} \\
\colhead{(1)} &
\colhead{(2)} &
\colhead{(3)} &
\colhead{(4)} &
\colhead{(5)} &
\colhead{(6)} &
\colhead{(7)} &
\colhead{(8)} &
\colhead{(9)} &
\colhead{(10)} &
\colhead{(11)} &
\colhead{(12)} 
} 
\startdata
J001427.0-303546 & 743850101 & 107438501010045 & 3.612859 & $-30.596339$ & ... & ... & ... & ... & ... & ... & ...  \\ 
J012642.1-011408 & 743700201 & 107437002010001 & 21.675742 & $-1.235603$ & TYC 4682-1697-1 & Star & \checkmark & \checkmark & 1.0 & \checkmark & \checkmark  \\ 
J013341.8+303849 & 650510101 & 106505101010010 & 23.423944 & $30.646651$ & [SB95] 14 & HighPM\* & ... & ... & ... & \checkmark & \checkmark  \\ 
J015709.1+373739 & 149780101 & 101497801010021 & 29.288018 & $37.627693$ & ... & ... & ... & ... & ... & ... & ...  \\ 
J022153.6+423025 & 780760501 & 107807605010017 & 35.474130 & $42.507134$ & 2MASS J02215361+4230262 & HighPM\* & ... & ... & ... & \checkmark & \checkmark  \\ 
J022304.5+421319 & 780760401 & 107807604010001 & 35.768779 & $42.221757$ & ... & ... & ... & ... & ... & \checkmark & \checkmark  \\ 
J025412.4+414303 & 605540101 & 106055401015017 & 43.551668 & $41.717575$ & ... & ... & ... & ... & ... & \checkmark & \checkmark  \\ 
J025643.0+001524 & 801610101 & 108016101010004 & 44.179322 & $0.256752$ & ... & ... & ... & \checkmark & 0.998 & \checkmark & \checkmark  \\ 
J033242.0-275702 & 555780601 & 105557806010004 & 53.175025 & $-27.950634$ & [LBX2017] 780 & HighPM\* & ... & \checkmark & 1.0 & \checkmark & \checkmark  \\ 
J033242.0-275702 & 604961201 & 106049612010006 & 53.175296 & $-27.950657$ & [LBX2017] 780 & HighPM\* & ... & \checkmark & 1.0 & \checkmark & \checkmark  \\ 
J033808.7-353418 & 781350101 & 107813501010003 & 54.536532 & $-35.572062$ & 2XMM J033808.7-353418 & X & ... & \checkmark & 0.996 & ... & \checkmark  \\ 
J035025.0+171446 & 844350101 & 108443501010001 & 57.604893 & $17.246247$ & V\* V471 Tau & SB\* & \checkmark & ... & ... & \checkmark & \checkmark  \\ 
J040304.9-154750 & 783880301 & 107838803010004 & 60.770738 & $-15.797497$ & ... & ... & ... & \checkmark & 0.983 & \checkmark & \checkmark  \\ 
J040312.3-752613 & 915391701 & 109153917010006 & 60.800971 & $-75.437116$ & ... & ... & ... & \checkmark & 0.986 & ... & \checkmark  \\ 
J042412.4+144529 & 101441501 & 101014415010001 & 66.052097 & $14.758201$ & V\* V895 Tau & BYDraV\* & \checkmark & ... & ... & \checkmark & \checkmark  \\ 
J051900.0+334454 & 600320101 & 106003201010001 & 79.750248 & $33.748427$ & V\* IQ Aur & alf2CVnV\* & ... & ... & ... & \checkmark & \checkmark  \\ 
J052049.2-454129 & 206390101 & 102063901010003 & 80.205182 & $-45.691431$ & ... & ... & \checkmark & ... & ... & ... & ...  \\ 
J052448.6+014627 & 554610101 & 105546101010016 & 81.202850 & $1.774254$ & CVSO 718 & TTauri\* & ... & \checkmark & 0.985 & \checkmark & \checkmark  \\ 
J052844.8-652654 & 123720201 & 101237202010001 & 82.186043 & $-65.448269$ & V\* AB Dor & TTauri\* & \checkmark & \checkmark & 1.0 & ... & \checkmark  \\ 
J053249.4-662212 & 142800101 & 101428001010001 & 83.207074 & $-66.370207$ & X LMC X-4 & HighMassXBin & \checkmark & \checkmark & 0.986 & ... & \checkmark  \\ 
J053502.2-052910 & 403200101 & 104032001010086 & 83.759010 & $-5.486061$ & V\* V1320 Ori A & Star & ... & \checkmark & 0.984 & \checkmark & ...  \\ 
J053512.1+095521 & 402050101 & 104020501010008 & 83.800532 & $9.922776$ & ... & ... & \checkmark & ... & ... & \checkmark & ...  \\ 
J053547.1-062910 & 89940301 & 100899403010004 & 83.946393 & $-6.486305$ & V\* V988 Ori & OrionV\* & ... & \checkmark & 0.998 & \checkmark & \checkmark  \\ 
J054837.0-510320 & 44740601 & 100447406010028 & 87.154440 & $-51.055574$ & ... & ... & ... & ... & ... & ... & ...  \\ 
J055630.7-331016 & 762750201 & 107627502010005 & 89.128018 & $-33.171543$ & ... & ... & ... & ... & ... & ... & \checkmark  \\ 
J055658.6-332036 & 762750201 & 107627502010006 & 89.244448 & $-33.343507$ & ... & ... & ... & \checkmark & 0.998 & \checkmark & \checkmark  \\ 
J055733.5-330117 & 824730201 & 108247302010032 & 89.389955 & $-33.021683$ & ... & ... & ... & \checkmark & 0.992 & \checkmark & \checkmark  \\ 
J063904.3+094307 & 694210101 & 106942101010001 & 99.768312 & $9.718617$ & ... & ... & \checkmark & \checkmark & 0.989 & ... & \checkmark  \\ 
J065423.9-240056 & 652250701 & 106522507010004 & 103.599894 & $-24.015732$ & ... & ... & ... & \checkmark & 0.981 & \checkmark & \checkmark  \\ 
J072639.5-260143 & 692550101 & 106925501010003 & 111.664772 & $-26.028876$ & ATO J111.6650-26.0290 & YSO\_Candidate & ... & \checkmark & 0.987 & \checkmark & \checkmark  \\ 
J074833.7-674507 & 160760101 & 101607601010001 & 117.141899 & $-67.751943$ & V\* UY Vol & LowMassXBin & \checkmark & ... & ... & ... & ...  \\ 
J081012.8-491409 & 501790101 & 105017901010002 & 122.553701 & $-49.235964$ & Cl\* NGC 2547    JND 13-98 & Star & ... & \checkmark & 1.0 & ... & \checkmark  \\ 
J083841.2+195946 & 721620101 & 107216201010018 & 129.671743 & $19.996256$ & 2MASS J08384128+1959471 & Eruptive\* & ... & \checkmark & 0.338 & \checkmark & \checkmark  \\ 
J084047.8-450329 & 506490101 & 105064901010001 & 130.199384 & $-45.058389$ & HD  74194 & HighMassXBin & ... & \checkmark & 1.0 & ... & \checkmark  \\ 
J084139.0-530923 & 112420101 & 101124201010010 & 130.412848 & $-53.156642$ & Gaia DR3 5318490476885356928 & Star & ... & ... & ... & ... & \checkmark  \\ 
J084143.2-445937 & 506490101 & 105064901010003 & 130.430316 & $-44.993887$ & ... & ... & ... & \checkmark & 0.978 & ... & ...  \\ 
J084224.7+193516 & 742570101 & 107425701010004 & 130.603115 & $19.587858$ & ... & ... & ... & \checkmark & 0.999 & ... & \checkmark  \\ 
J091058.0-585800 & 690200701 & 106902007010001 & 137.741972 & $-58.966918$ & \* a Car & SB\* & \checkmark & ... & ... & ... & \checkmark  \\ 
J101935.5+195211 & 891801301 & 108918013010001 & 154.898220 & $19.869861$ & ... & ... & \checkmark & ... & ... & \checkmark & ...  \\ 
J105126.8+355234 & 841481701 & 108414817010023 & 162.861798 & $35.876187$ & ... & ... & ... & ... & ... & \checkmark & \checkmark  \\ 

J105130.9+573439 & 147511001 & 101475110010002 & 162.878601 & $57.577711$ & [BCH2008] 444 & HighPM\* & \checkmark & ... & ... & \checkmark & \checkmark  \\ 
J105130.9+573439 & 147510801 & 101475108010001 & 162.878856 & $57.577631$ & [BCH2008] 444 & HighPM\* & \checkmark & ... & ... & ... & \checkmark  \\ 
J105130.9+573439 & 147510901 & 101475109010004 & 162.879091 & $57.577501$ & [BCH2008] 444 & HighPM\* & \checkmark & ... & ... & ... & ...  \\ 
J110204.8-011801 & 406541301 & 104065413010001 & 165.520054 & $-1.300460$ & ... & ... & ... & \checkmark & 0.999 & \checkmark & \checkmark  \\ 
J111154.0-761930 & 300270201 & 103002702010001 & 167.974964 & $-76.325277$ & Hn 15 & YSO & \checkmark & \checkmark & 0.996 & ... & \checkmark  \\ 
J120501.2-615131 & 109110101 & 101091101010014 & 181.255342 & $-61.858819$ & ... & ... & ... & \checkmark & 0.992 & ... & \checkmark  \\ 
J121358.3+141305 & 745110601 & 107451106010005 & 183.492902 & $14.218144$ & 2XMM J121358.3+141304 & Star & ... & \checkmark & 0.987 & \checkmark & \checkmark  \\ 
J122548.8+333248 & 824610101 & 108246101010001 & 186.453590 & $33.546883$ & NGC  4395 & Seyfert2 & ... & ... & ... & \checkmark & \checkmark  \\ 
J122944.8+075238 & 761630201 & 107616302010009 & 187.436773 & $7.877384$ & ... & ... & ... & ... & ... & ... & ...  \\ 
J124641.0-593143 & 761090201 & 107610902010003 & 191.670931 & $-59.528667$ & ... & ... & ... & ... & ... & ... & \checkmark  \\ 
J124901.5-410131 & 823580301 & 108235803010011 & 192.256367 & $-41.025385$ & ... & ... & ... & \checkmark & 0.991 & ... & \checkmark  \\ 
J125517.5-043210 & 801930801 & 108019308010017 & 193.823326 & $-4.536247$ & ... & ... & ... & ... & ... & ... & ...  \\ 
J130200.1+274657 & 904640301 & 109046403010002 & 195.500590 & $27.782729$ & 2MASX J13020015+2746579 & AGN & ... & ... & ... & \checkmark & \checkmark  \\ 
J130200.1+274657 & 851180501 & 108511805010001 & 195.500647 & $27.782786$ & 2MASX J13020015+2746579 & AGN & ... & ... & ... & \checkmark & \checkmark  \\ 
J130200.1+274657 & 864560101 & 108645601010001 & 195.500813 & $27.782451$ & 2MASX J13020015+2746579 & AGN & ... & ... & ... & \checkmark & \checkmark  \\ 
J132017.5-022416 & 800050101 & 108000501010001 & 200.073077 & $-2.404623$ & ... & ... & ... & \checkmark & 1.0 & ... & ...  \\ 
J132456.1-430258 & 863890201 & 108638902010013 & 201.234095 & $-43.049428$ & ... & ... & ... & ... & ... & ... & ...  \\ 
J142329.1+041016 & 861260301 & 108612603010002 & 215.871255 & $4.171259$ & ... & ... & \checkmark & ... & ... & ... & ...  \\ 
J142942.0-624044 & 49350101 & 100493501010001 & 217.425020 & $-62.679018$ & ... & ... & \checkmark & ... & ... & ... & ...  \\ 
J154829.6-534627 & 740640101 & 107406401010001 & 237.123407 & $-53.774421$ & 2MASS J15482965-5346272 & HighPM\* & ... & \checkmark & 0.983 & ... & ...  \\ 
J154910.4-534225 & 740640101 & 107406401010008 & 237.293388 & $-53.707116$ & ... & ... & ... & ... & ... & ... & ...  \\ 
J155717.3-373907 & 882060601 & 108820606010002 & 239.322565 & $-37.652185$ & UCAC4 262-089820 & YSO\_Candidate & ... & ... & ... & ... & ...  \\ 
J161527.4-223927 & 555650301 & 105556503010004 & 243.864207 & $-22.657662$ & 2MASS J16152743-2239275 & Low-Mass\* & ... & \checkmark & 0.998 & \checkmark & \checkmark  \\ 
J161534.5-224241 & 555650201 & 105556502010001 & 243.893957 & $-22.711568$ & V\* VV Sco & OrionV\* & \checkmark & ... & ... & \checkmark & ...  \\ 
J162438.9-233018 & 760900101 & 107609001010023 & 246.161702 & $-23.504818$ & ... & ... & ... & \checkmark & 0.981 & \checkmark & \checkmark  \\ 
J162522.1-233502 & 760900101 & 107609001010014 & 246.342206 & $-23.584008$ & ... & ... & ... & ... & ... & ... & ...  \\ 
J162534.7-233257 & 760900101 & 107609001010005 & 246.394899 & $-23.549196$ & ... & ... & \checkmark & ... & ... & ... & ...  \\ 
J162535.5-233253 & 760900101 & 107609001010035 & 246.398083 & $-23.548132$ & ... & ... & \checkmark & ... & ... & ... & ...  \\ 
J162544.8-232905 & 760900101 & 107609001010004 & 246.436601 & $-23.485782$ & ... & ... & ... & ... & ... & ... & ...  \\ 
J162554.4-233036 & 760900101 & 107609001010003 & 246.477007 & $-23.509411$ & ... & ... & \checkmark & ... & ... & ... & ...  \\ 
J162658.5-244535 & 800031001 & 108000310010008 & 246.743675 & $-24.759799$ & ... & ... & \checkmark & ... & ... & ... & ...  \\ 
J162719.5-244140 & 305540701 & 103055407010001 & 246.831323 & $-24.694604$ & EM\* SR   12 & OrionV\* & \checkmark & \checkmark & 1.0 & \checkmark & \checkmark  \\ 
J162721.8-244335 & 800031001 & 108000310010001 & 246.840819 & $-24.726551$ & [GY92] 253 & YSO & ... & ... & ... & ... & \checkmark  \\ 
J162722.2-244325 & 800031001 & 108000310010029 & 246.842528 & $-24.723749$ & ... & ... & ... & ... & ... & ... & ...  \\ 
J162724.5-242934 & 800031001 & 108000310010003 & 246.852473 & $-24.493151$ & [GY92] 259 & YSO & ... & ... & ... & ... & ...  \\ 
J162726.9-244050 & 800031001 & 108000310010004 & 246.862219 & $-24.680764$ & ... & ... & \checkmark & ... & ... & ... & \checkmark  \\ 
J162804.6-243456 & 305540701 & 103055407010014 & 247.019421 & $-24.582144$ & [GY92] 463 & TTauri\* & ... & ... & ... & ... & \checkmark  \\ 
J163216.7-672740 & 152620101 & 101526201010001 & 248.069713 & $-67.461190$ & Gaia DR3 5809528276749789312 & Star & \checkmark & \checkmark & 0.954 & ... & ...  \\ 
J170249.1-484723 & 760646701 & 107606467010001 & 255.705823 & $-48.789965$ & V\* V821 Ara & HighMassXBin & ... & \checkmark & 0.952 & ... & \checkmark  \\ 
J171208.7-230949 & 880280201 & 108802802010001 & 258.036323 & $-23.163996$ & CD-23 13197 & YSO & ... & \checkmark & 0.999 & \checkmark & \checkmark  \\ 

J171208.7-230949 & 880280501 & 108802805010001 & 258.036580 & $-23.163768$ & CD-23 13197 & YSO & ... & \checkmark & 0.999 & \checkmark & \checkmark  \\ 
J171208.7-230949 & 880280101 & 108802801010001 & 258.036824 & $-23.163946$ & CD-23 13197 & YSO & ... & \checkmark & 0.999 & \checkmark & \checkmark  \\ 
J173109.6-213015 & 842550101 & 108425501010019 & 262.790331 & $-21.504231$ & ... & ... & ... & ... & ... & \checkmark & ...  \\ 
J174301.2-362222 & 675040101 & 106750401010001 & 265.755295 & $-36.372783$ & SWIFT J1743.1-3620 & HighMassXBin & ... & \checkmark & 0.983 & ... & \checkmark  \\ 
J174327.0-295557 & 825140101 & 108251401010002 & 265.862772 & $-29.932628$ & ... & ... & ... & \checkmark & 0.986 & \checkmark & ...  \\ 
J174343.3-293645 & 862470101 & 108624701010002 & 265.931079 & $-29.613061$ & ... & ... & ... & \checkmark & 0.982 & \checkmark & ...  \\ 
J174502.3-285449 & 402430301 & 104024303010002 & 266.259801 & $-28.913680$ & ... & ... & ... & ... & ... & ... & ...  \\ 
J174503.7-291849 & 694641001 & 106946410010010 & 266.265762 & $-29.313852$ & ... & ... & ... & \checkmark & 0.982 & \checkmark & \checkmark  \\ 
J174505.3-291445 & 694640201 & 106946402010001 & 266.272306 & $-29.245869$ & ... & ... & ... & ... & ... & ... & ...  \\ 
J174610.7-290020 & 202670701 & 102026707010084 & 266.544859 & $-29.005734$ & ... & ... & ... & \checkmark & 0.983 & \checkmark & ...  \\ 
J174639.0-285351 & 762250301 & 107622503010003 & 266.662917 & $-28.897535$ & HD 316314 & gammaDorV\* & ... & \checkmark & 1.0 & \checkmark & \checkmark  \\ 
J175131.6-401533 & 763700301 & 107637003010019 & 267.881766 & $-40.259386$ & ... & ... & ... & ... & ... & ... & \checkmark  \\ 
J175425.2-261952 & 744600101 & 107446001010001 & 268.605308 & $-26.331288$ & IGR J17544-2619 & HighMassXBin & \checkmark & \checkmark & 0.999 & \checkmark & \checkmark  \\ 
J181410.4-324734 & 604860301 & 106048603010001 & 273.543695 & $-32.793100$ & HD 319139 & SB\* & \checkmark & \checkmark & 1.0 & ... & \checkmark  \\ 
J181422.0-324610 & 604860401 & 106048604010002 & 273.592108 & $-32.769659$ & ... & ... & ... & \checkmark & 0.991 & ... & \checkmark  \\ 
J184100.4-053546 & 604820301 & 106048203010001 & 280.251819 & $-5.596264$ & IGR J18410-0535 & HighMassXBin & ... & \checkmark & 0.989 & \checkmark & ...  \\ 
J184949.8-235012 & 601950101 & 106019501010001 & 282.457399 & $-23.836673$ & ... & ... & \checkmark & ... & ... & \checkmark & \checkmark  \\ 
J192637.8+012124 & 720173801 & 107201738010004 & 291.657538 & $1.356811$ & ... & ... & ... & \checkmark & 0.964 & \checkmark & \checkmark  \\ 
J195245.4+005725 & 840510201 & 108405102010004 & 298.189485 & $0.957056$ & ATO J298.1895+00.9569 & EclBin\_Candidate & ... & \checkmark & 0.995 & \checkmark & \checkmark  \\ 
J200043.8+224156 & 692290301 & 106922903010015 & 300.181685 & $22.702808$ & ... & ... & \checkmark & ... & ... & \checkmark & ...  \\ 
J200043.6+224234 & 692290301 & 106922903010001 & 300.181984 & $22.709620$ & HD 189733 & BYDraV\* & \checkmark & ... & ... & \checkmark & \checkmark  \\ 
J202334.1+114217 & 600690101 & 106006901010002 & 305.892171 & $11.704830$ & UCAC4 509-131194 & HighPM\* & \checkmark & ... & ... & ... & \checkmark  \\ 
J204509.9-312034 & 822740501 & 108227405010001 & 311.291186 & $-31.342853$ & ... & ... & \checkmark & \checkmark & 1.0 & ... & \checkmark  \\ 
J204509.9-312034 & 822740401 & 108227404010001 & 311.291237 & $-31.342583$ & ... & ... & \checkmark & \checkmark & 1.0 & ... & \checkmark  \\ 
J204800.4+292932 & 693400101 & 106934001015070 & 312.001695 & $29.492224$ & ... & ... & ... & ... & ... & \checkmark & \checkmark  \\ 
J205748.7+435023 & 781690101 & 107816901010004 & 314.452404 & $43.839849$ & V\* V2051 Cyg & OrionV\* & ... & ... & ... & \checkmark & \checkmark  \\ 
J205958.5-425647 & 691670101 & 106916701010011 & 314.993839 & $-42.946532$ & ... & ... & ... & \checkmark & 0.989 & ... & \checkmark  \\ 
J210241.5+455306 & 671930101 & 106719301010001 & 315.673153 & $45.885266$ & ... & ... & \checkmark & ... & ... & ... & ...  \\ 
J215906.0-201602 & 555220101 & 105552201010001 & 329.775264 & $-20.267354$ & ... & ... & ... & ... & ... & ... & \checkmark  \\ 
J220310.6-344406 & 722360301 & 107223603010022 & 330.794350 & $-34.735139$ & ... & ... & \checkmark & \checkmark & 0.999 & \checkmark & \checkmark  \\ 
J223555.5+334900 & 21140201 & 100211402010006 & 338.981545 & $33.816752$ & 2XMM J223555.5+334900 & Star & \checkmark & ... & ... & \checkmark & \checkmark  \\ 
J225336.3+114013 & 670880401 & 106708804010004 & 343.401616 & $11.670473$ & LSPM J2253+1140B & Star & ... & ... & ... & \checkmark & \checkmark  \\ 
J225425.1+113502 & 670880401 & 106708804010006 & 343.604867 & $11.583893$ & ... & ... & ... & \checkmark & 0.983 & \checkmark & \checkmark  

\enddata
\tablecomments{
    (1) IAU-style name in the 4XMM-DR14 catalog.
    (2) XMM-Newton observation ID.
    (3) Detection ID in the catalog.
    (4) Right ascension in units of degrees.
    (5) Declination in units of degrees.
    (6) Name of a counterpart determined based on SIMBAD.
    (7) Type of the counterpart. See http://vizier.u-strasbg.fr/cgi-bin/OType?\$1 for details.
    (8) Flag for a possible counterpart in the 2RXS catalog.     
    (9) Flag for a possible counterpart in the SkyMapper DR4 catalog. 
    (10) Maximum stellarity index from the photometry table (between 0 = no star and 1 = star). 
    (11) Flag for a possible counterpart in the Pan-STARRS1 catalog. 
    (12) Flag for a possible counterpart in the AllWISE catalog.  
}
\end{deluxetable*}
\end{longrotatetable}

\bibliography{ref}
\bibliographystyle{aasjournal}

\end{document}